\documentstyle[psfig,12pt]{report}
\newcommand{\be}{\begin{equation}}
\newcommand{\ee}{\end{equation}}

\newcommand{\mo}[1]{\sqrt{(#1/L)^{2} - 1 +2M/R}\,}
\newcommand{\mop}[1]{\sqrt{(#1/L)^{2} -1+2M_{+}/R}\,}
\newcommand{\moh}[1]{\sqrt{(#1/\hat{L})^{2} -1+ 2M/\hat{R}}\,}
\newcommand{\moph}[1]{\sqrt{(#1/\hat{L})^{2} - 1 +2M_{+}/\hat{R}}\,}
\newcommand{\smo}[1]{#1/L - \sqrt{(#1/L)^2 - 1 +2M/R\,}}
\newcommand{\smop}[1]{#1/L-\sqrt{(#1/L)^2-1+2M_{+}/R\,}}
\newcommand{\smoh}[1]{#1/\hat{L}-\sqrt{(#1/\hat{L})^{2}-1+2M/\hat{R}\,}}
\newcommand{\smohp}[1]{#1/\hat{L}-\sqrt{(#1/\hat{L})^{2}-1+2M/\hat{R}\,}}
\newcommand{\pip}{\pi_{L}(\rs+\epsilon)}
\newcommand{\pim}{\pi_{L}(\rs-\epsilon)}
\newcommand{\Rp}{R'(\rs+\epsilon)}
\newcommand{\Rm}{R'(\rs-\epsilon)}
\newcommand{\rmin}{r_{\mbox{{\tiny min}}}}
\newcommand{\sq}[1]{\sqrt{|1-#1|}}
\newcommand{\rs}{\hat{r}}
\newcommand{\rsd}{\dot{\hat{r}}}
 
\newcommand{\ak}{\hat{a}_{k}}
\newcommand{\akp}{\hat{a}_{k'}}
\newcommand{\akd}{\hat{a}^{\dagger}_{k}}

\newcommand{\bk}{\hat{b}_{k}}
\newcommand{\bkp}{\hat{b}_{k'}}
\newcommand{\bkd}{\hat{b}^{\dagger}_{k}}

\reversemarginpar
\begin{document}

\begin{titlepage}
\begin{center}
{\Large Non-Thermal Aspects of Black Hole Radiance}\\
%{\Large More Title}\\
\vspace{1.2in}
Per Kraus\\
\vspace{2in}
{\it A dissertation\\
presented to the faculty\\
of Princeton University\\
in candidacy for the degree\\
of Doctor of Philosophy\\
\vspace{.5in}
Recommended for acceptance\\
by the Department of Physics\\
}
\nopagebreak
\vspace{.8in}
November 1995
\end{center}
\end{titlepage}

\newpage
\vfill
\thispagestyle{empty}
\begin{center}
\copyright
Copyright by Per Kraus, 1995.  All rights reserved.
\end{center}

\newpage
\pagenumbering{roman}
\pagestyle{plain}
\setcounter{page}{3}
\begin{center}
{\bf Abstract}
\end{center}
\addcontentsline{toc}{chapter}{Abstract}
The phenomenon of black hole thermodynamics raises several deep issues 
which any proper theory of quantum gravity must confront: to what extent
does the inclusion of the back-reaction alter the thermal character of the
radiation, how can the entropy be understood from a microscopic standpoint,  
what is the ultimate fate of an evaporating black hole, and is the
outcome reconcilable  with unitary time evolution in quantum mechanics?  

In the first part of this thesis, 
we address the issue of determining what the actual emission
spectrum from a black hole is, once the gravitational field of the emitted
quanta is included in a quantum mechanical manner.  To make the problem
tractable, we employ two important approximations: we quantize only the s-wave
sector of the full theory, and we consider only single particle emission.
By proceeding in the framework of a Hamiltonian path integral description of
this system, we are able to integrate out the gravitational field, thereby
obtaining an effective action depending only on the matter degrees of freedom.
This effective action can then be second quantized in terms of new, corrected,
mode solutions thus enabling the calculation of the emission spectrum from
modified Bogoliubov coefficients. The results are particularly interesting in 
the case of emission from Reissner-Nordstrom black holes, since in
the extremal limit our results are dramatically different from what a naive,
and incorrect, semi-classical calculation would yield.

The other major topic which we discuss is the dynamics of quantum fields on
background geometries which undergo quantum tunneling.  An example of such a
system which has important implications for both cosmology and quantum gravity
in general, is the tunneling of a false vacuum bubble leading to the creation
of a new universe.  To determine what the state of a scalar field would be as a
result of such a process, we make a WKB approximation of the Wheeler-DeWitt
equation to obtain an imaginary time evolution equation for the scalar field.
The state after tunneling is then found by solving this equation on a portion 
of the Euclidean Kruskal manifold, the properties of which serve to ensure that
at late times thermal radiation emerges.

\newpage
\pagenumbering{roman}
\pagestyle{plain}
\setcounter{page}{4}
\begin{center}
{\bf Acknowledgements}
\end{center}
\addcontentsline{toc}{chapter}{Acknowledgements}
This thesis is a result of collaboration with my advisor, Frank Wilczek.
It is a pleasure to have this opportunity to thank him for sharing his
wisdom and insights, which have greatly enriched my understanding of physics.
I am also grateful to the many students and faculty who make Jadwin Hall such
an exciting place to do science.  Especially, I would like to thank my 
long-time office mate Joop Talstra for many memorable experiences, scientific
and otherwise.  

Finally, I am grateful to my parents for their support, 
and to Kate for many happy times, and for many more
to come.

\tableofcontents
\newpage
%\listoffigures
\newpage
\newpage
\setcounter{page}{0}
\pagenumbering{arabic}
\pagestyle{headings}

\chapter{Introduction}

Perhaps the most conspicuous deficiency in a physicist's current understanding
of nature is the absence of a usable quantum mechanical theory of gravity.
Standing in the way has been an utter lack of experimental data to serve as a
guide toward such a theory.  On one hand, the Standard Model of particle
physics, while not without unresolved issues of its own, provides a coherent
description of all phenomena which have been observed at existing accelerators;
on the other hand, string theory, while remaining a viable candidate for a 
theory of everything, is presently incapable of making further predictions
which can be tested by these same accelerators.  In the face of such a 
situation there is still reason to believe that progress can be made by
extracting clues from the structure of the theories that we know today.  The
remarkable quantum mechanical properties of black holes are an important 
example -- the phenomenon of black hole thermodynamics  is tantalizingly 
suggestive of a more comprehensive theory.  

Before discussing the relevance of
black holes to quantum gravity, which is the focus of the present work, it will
be useful to recall some of the development of the classical theory of black
holes.
Black holes are hard to see. For the experimentalist, this is true for obvious
reasons: the defining property of black holes as regions of spacetime
from which nothing can emerge means that their existence can only be inferred,
either through their gravitational effects on neighboring bodies, or by
viewing the characteristic glow of matter as it is sucked into a hole. 
Nevertheless, recent observations have essentially removed any doubt as to the
existence of black holes in our universe.  For many years, theorists were
similarly unable to see black holes in the equations of general relativity,
although in this case the difficulty was due to faulty vision rather than a 
lack of evidence.  Although Schwarzschild wrote down the metric describing the
geometry outside a spherically symmetric body,
$$
ds^2=-(1-2M/r)dt^2+\frac{dr^2}{1-2M/r}+r^2(d\theta^2 +\sin^2
\theta d\phi^2),
$$
immediately after the discovery of the field equations, an incomplete 
understanding of the geometry's global structure led to the erroneous 
conclusion that a singularity occurs at the horizon, $r=2M$, rendering the
solution for $r\le 2M$ unphysical.  Only much later was it realized that the
singularity was merely a coordinate artifact, and that physical quantities are
entirely well behaved at the horizon.  With this came the understanding that
there exists a {\em true} singularity at $r=0$, but it was suspected that 
deviations from spherical symmetry would cause the singularity to be smoothed
out in a realistic collapse process.  However, once Penrose \cite{penrose}
proved that singularities {\em do} occur in classical relativity under very
generic conditions, the modern age of black hole research was begun.

Two theorems in classical relativity served to presage the subsequent 
development of black hole thermodynamics.  First, it was found that the mass
and angular momentum of black holes obey the so-called ``first law of black
hole thermodynamics'':
$$
d\!M=\frac{1}{8\pi}\kappa d\!A + \Omega d\!J,
$$
$J$ being the hole's angular momentum, $A$ and $\Omega$ the horizon's area and 
angular velocity.  $\kappa$ is the surface gravity, defined as the force which
a person at infinity would need to exert in order to keep an object suspended
at the horizon.  For the Schwarzschild hole, $\kappa =1/4M$.  Second, it was
found that in any process, the total area of black hole horizons necessarily
increases --- the so-called ``second law of black hole thermodynamics''.
These results led Bekenstein \cite{bekenstein} to suggest that a black hole
possessed an entropy proportional to $A$, and that the conventional second
law of thermodynamics would be upheld when this entropy was taken into account.
While the formulae suggest that something proportional to $\kappa$ plays the
role of temperature, this point was entirely obscure from the standpoint of
classical relativity, since in this context a black hole can't emit anything,
much less thermal radiation.

At this point, quantum mechanics enters the game:  Hawking \cite{hawking}
showed that quantum effects cause a black hole to emit exactly thermal 
radiation with a temperature $T=\kappa/2\pi$.  This result immediately leads
to the identification $S=A/4$.  As we shall consider the derivation of this
effect in detail later, let us simply remark here that it is intimately
connected to the existence of the horizon as a surface of infinite redshift.
This allows particles of negative energy to exist inside the horizon, leading
to the possibility of particle-antiparticle creation, one particle flowing
into the hole, one flowing out, without violating conservation of energy.
Perhaps the most startling implication of this result is that, although it only
relies on fairly conservative assumptions about the interaction of quantum
fields with gravity, it leads to a fundamental conflict with the usual 
formulation of quantum mechanics.  For if the black hole continually emits
radiation it will eventually disappear altogether, and since thermal radiation
is uncorrelated, there will not be any trace left of the matter which 
originally formed the hole.   More concisely, the process is not described by
by unitary evolution since any initial state forming the black hole  leads to
the same thermal density matrix after the hole has evaporated.  If this is
true, then quantum mechanics will have to be revised  in order to be 
compatible with this phenomenon of information loss.  On the other hand, this
conclusion may be premature, as the approximation employed in computing the
radiance breaks down once the hole becomes sufficiently small, so that there
is in fact no compelling reason why the black hole has to disappear.  Various
alternatives to the information loss scenario have been the subject of much
discussion recently (two reviews, written from quite different viewpoints,
are \cite{strominger,banks}) but no consensus has been established.  Much of
the recent work has been done in the context of two dimensional models
\cite{CGHS} which, owing to conformal invariance, allow more analytical
progress than was previously possible.  Unfortunately, the key questions 
remain unanswered.  

The other closely related problem is to obtain a proper understanding of the
formula $S=A/4$.  It seems quite likely that a resolution of the information
loss puzzle is contingent upon understanding in what sense, if any, a black
hole has the number of states $e^{A/4}$.  One possibility is that a black hole
has this number of weakly coupled states localized near the horizon, proper 
accounting of which will lead to correlations in the outgoing radiation. 
Although the radiation may look approximately thermal, the correlations could
be sufficient to encode all details of the initial state.  Any computation
of such correlations must go beyond the free field approximation originally
employed by Hawking; with this goal in mind, a large part of the present work
is devoted to obtaining the leading corrections to the free field results.

In the course of this work we shall be discussing these problems mainly in the
context of the semiclassical approximation.  However, we should point out that
the validity of this approximation is by no means assured, due to the peculiar
properties of the horizon.  Indeed, it has been argued that strong coupling
effects \cite{schoutens} and quantum fluctuations \cite{ortiz} make this
description unreliable.

The remainder of this thesis is organized as follows.  In Chapter 2 we develop
the Hamiltonian formulation of gravity, which will play a key role in the 
material that follows.  We consider both classical and quantum aspects, paying
special attention to the spherically symmetric case.  Chapter 3 provides an
overview of quantum field theory in curved space, and a detailed discussion of
black hole radiance.  The discussion is phrased in terms of a nonstandard
coordinate system for the black hole geometry \cite{KW2}, 
which is particularly convenient
for doing computations near the horizon.  This coordinate system also provides
insights into the global structure of the geometry, as we discuss.  Finally,
the role of fluctuations in the stress-energy is considered, with emphasis on
the impact these effects have on the moving mirror model.  In Chapter 2 we
turn to the main focus of this thesis: the calculation of self-interaction
corrections to the black hole emission spectrum \cite{KW1,KW3}.  
Drawing upon material 
developed in the previous chapters, we show how these corrections can be
deduced by calculating the effective action for a gravitating thin shell.  We
then discuss the relation between the first and second quantized approaches to
this problem, and finally, the difficulty in obtaining multi-particle
correlations by a straightforward extension of the single particle calculation.
In Chapter 5 we describe some attempts at gaining a microscopic understanding
of the black hole entropy by counting fluctuations of the geometry.  As we will
see, the problem is not how to obtain a sufficiently large number of states,
but rather how to control the divergences which inevitably occur.  The topic
of Chapter 6 is the effect of quantum tunneling upon radiance phenomena 
\cite{bubble}.  The
main example concerns black hole formation via the tunneling of a false vacuum
bubble, a process which nicely illustrates the interplay between the 
Hamiltonian and Lagrangian approaches to quantum gravity.  We will show how
thermal radiation emerges from the black hole at late times, 
even though the
standard calculation of the radiance is inapplicable.

\chapter{Hamiltonian Formulation of Gravity}

\section{Classical Theory}

To proceed with the Hamiltonian approach to gravity we perform a $3+1$
decomposition of the spacetime manifold, singling out a time coordinate.
A Hamiltonian is then identified that propagates the three geometry along
the time direction.  The subtlety involved in this procedure is caused by
the reparameterization invariance of the Einstein-Hilbert action: due to
the existence of redundant variables, time evolution is not uniquely
defined unless a gauge is specified.  However, there is a well developed 
formalism for dealing with such systems \cite{dirac}, 
as will now be discussed. 

The key to separating out the physical and redundant degrees of freedom
is to write the metric in $3+1$ form as \cite{adm}
\be
ds^2 = -(N^t dt)^2 + h_{ij}(dx^i+N^i dt)(dx^j+N^j dt)
\ee
With this labelling, the Einstein-Hilbert action becomes:
\be
{\cal L}_{G} = \frac{1}{16\pi}\sqrt{-g}\,{\cal R} = \frac{1}{16\pi}
\sqrt{h}\,N^t\,[^{3}{\cal R} +K_{ij}K^{ij}-K^2],
\ee
where $^{3} {\cal R}$ is the Ricci scalar associated with $h_{ij}$, 
$K_{ab}$ is the extrinsic curvature of a constant $t$ hypersurface:
\be
K_{ab}=\frac{1}{2N^t}[\dot{h}_{ij}-N_{i|j}-N_{j|i}],
\ee
and $K\equiv K_{a}^{a}$.  Throughout, Latin indices are raised and lowered by
$h_{ij}$, and $|$ denotes covariant differentiation with respect to $h_{ij}$.

Now, it is seen that no time derivatives of $N^t$ or $N^i$ appear in the
gravitational action, nor will they when matter is included, so that the
canonical momenta conjugate to these variables vanish identically:
\be
\pi_{N^t}\equiv \frac{\partial {\cal L}}{\partial \dot{N}^t}=0
{\mbox \ \ \ \ ; \ \ \ \ } \pi_{N^i}\equiv \frac{\partial {\cal L}}{\partial
\dot{N}^i}=0.
\ee
These are constraints on the phase space.  $h_{ij}$, on the other hand, have
nonvanishing canonical momenta $\pi_{ij}$.  These can be computed and used 
to put ${\cal L}_{G}$ in canonical form: 
\be
{\cal L}_G = \pi_{ij}\dot{h}^{ij} - N^{t}{\cal H}_{t}-N_{i}{\cal H}^{i}
\ee
where
\be
{\cal H}_{t} = 8\pi h^{-\frac{1}{2}}(h_{ik}h_{jl}+h_{il}h_{jk}-h_{ij}h_{kl})
\pi^{ij}\pi^{kl} - \frac{1}{16\pi}h^{\frac{1}{2}} {\,^{3} {\cal R}},
\ee
\be
{\cal H}^{i}=-2\pi^{ij}_{|j}.
\ee
The constraints $\pi_{N^{t}}=\pi_{N^{i}}=0$ must hold at all times, which
means that their Poisson brackets with the Hamiltonian must vanish.  This 
condition yields the secondary constraints
\be
{\cal H}_{t}={\cal H}^{i}=0.
\ee
All the dynamics of gravity is contained in these constraints.  It should be
stressed that these constraints hold ``weakly'' --- they are to be imposed 
only after all Poisson brackets have been computed.

\subsection{Surface Terms}

In deriving the canonical form of the action we ignored all surface terms
which arose.  However, for gravity in asymptotically flat space surface terms
play an important role and cannot be neglected, so we write
\be
H_{G}=\int\! d^{3}\!x [N^t{\cal H}_{t}+N_i {\cal H}^i]+\mbox {surface terms}.
\label{Hnew}
\ee
Regge and Teitelboim \cite{regge}
showed that Einstein's equations are equivalent to
Hamilton's equations applied to $H_G$ only for a particular choice of
surface terms.  The point is that if Einstein's equations are written as
\be
\dot{h}_{ij}=A_{ij}(h,\pi) \mbox{ \ \ \ \ ; \ \ \ \ } \dot{\pi}_{ij}
=-B_{ij}(h,\pi)
\ee
then we need the variation of $H_G$ to be
\be
\delta H_G = A_{ij}\delta \pi^{ij} + B_{ij}\delta h^{ij}
\ee
to ensure consistency with Einstein's equations.  To put $\delta H_G$ in this
form we must integrate by parts, because space derivatives of $h_{ij}$ and
$\pi_{ij}$ appear in $H_G$, and then demand that the resulting surface term
cancels with the one we have included in (\ref{Hnew}).  In \cite{regge} the
surface terms were worked out assuming a particular rate of fall-off of
$h_{ij}$, $\pi_{ij}$, $N^t$, and $N^i$ at infinity.  This condition
can be relaxed \cite{ferraris}, and in general the full Hamiltonian is:
\be
H_G=\int_{\Sigma}\!d^3\!x[N^t {\cal H}_{t}+N_{i}{\cal H}^i]+\int_{\partial
 \Sigma} ds_{l}[{\cal H}_{1s}^{l}+{\cal H}_{2s}^{l}]
\ee
with
$$
H_{1s}^{l}=\frac{1}{16\pi}(N^t\sqrt{h}h^{ij}U_{ij}^{l}+2\pi^{li}N_{i})
$$
\be
H_{2s}^{l}=\frac{1}{16\pi}\left[\frac{1}{N^t}\sqrt{h}(N^{i}\partial_{i}
N^l-N^l\partial_{i}N^i)\right]
\label{surface}
\ee
where
\be
U_{jk}^{i} = \Gamma_{jk}^{i}-\delta^{i}_{(j}\Gamma_{k)}^l.
\ee
The result of Regge and Teitelboim corresponds to choosing fall-off 
conditions such that $H_{2s}^{l}$ vanishes at infinity.

Since ${\cal H}_{t}$ and ${\cal H}^i$ vanish when the constraints are
satisfied, the numerical value of the Hamiltonian is given solely by the
surface terms.  This numerical value is the ADM mass of the system.

\subsection{Spherical Symmetry}

We now specialize to the case of spherically symmetric geometries
\cite{bcmn,polch,thiemann}.  It is 
possible to proceed substantially further in this case because there are no
propagating degrees of freedom -- after solving the constraints there will
only be one free parameter remaining: the ADM mass.  The metric is written
\be
ds^2= - (N^t dt)^2 +L^2(dr+N^r dt)^2+R^2(d\theta^2+{\sin{\theta}}^2
d\phi^2)
\ee
and the action is 
\be
S=\frac{1}{16\pi}\int\! d^4\!x\sqrt{-g}\,{\cal R}=\int\!dt\,dr\,[\pi_{R}\dot{R}
+\pi_{L}\dot{L}-N^t{\cal H}_{t}-N^r{\cal H}_r]
\ee
with
\be
{\cal H}_t=\frac{L\pi_{L}^{2}}{2R^2}-\frac{\pi_{L}\pi_{R}}{R}+\left(
\frac{RR'}{L}\right)'-\frac{R'^{2}}{2L}-\frac{L}{2} \mbox{ \ \ \ \ ;
 \ \ \ \ } {\cal H}_r=R'\pi_{R}-L\pi_{L}'
\label{theconstraints}
\ee
where $'$ represents $d/dr$.

As before, the Hamiltonian is 
\be
H=\int\! dr[N^t{\cal H}_t+N^r{\cal H}_r] + M
\ee
and the constraints are
\be
{\cal H}_t={\cal H}_r=0.
\ee
The constraints can be solved as follows.  $\pi_R$ is eliminated by forming
the linear combination of constraints
$$
\frac{R'}{L}{\cal H}_t+\frac{\pi_{L}}{RL}{\cal H}_r=0.
$$
Defining
\be
{\cal M}(r)\equiv \frac{\pi_{L}^{2}}{2R}+\frac{R}{2}-\frac{RR'^2}{2L^2},
\ee
this constraint is equivalent to ${\cal M}'=0$. By comparing with 
(\ref{surface}) it can be shown that ${\cal M}(\infty)$ is the ADM mass.
The constraints can now be solved for the momenta:
\be
\pi_{L}=\eta R \sqrt{(R'/L)^2-1+2M/R} \mbox{ \ \ \ \ ; \ \ \ \ }
\pi_{R}=\frac{L}{R'}\pi_{L}'
\label{momsolve}
\ee
where $\eta=\pm 1$.

Since we have defined the new variable, ${\cal M}(r)$, it is natural to ask
what the momentum conjugate to ${\cal M}(r)$ is.  From the fundamental 
Poisson brackets,
\be
\{L(r),\pi_{L}(r')\}=\{R(r),\pi_{R}(r')\}=\delta(r-r')
\ee
it is straightforward to check that
\be
\pi_{{\cal M}}\equiv \frac{L\sqrt{(R'/L)^2-1+2{\cal M}/R}}{1-2{\cal M}/R}
\ee
satisfies 
\be
\{{\cal M}(r),\pi_{{\cal M}}(r')\}=\delta(r-r').
\ee
$\pi_{{\cal M}}$ has a simple physical interpretation. Consider the geometry
in Schwarzschild coordinates:
\be
ds^2=-(1-2M/r_s)dt_{s}^2+\frac{dr_{s}^{2}}{1-2M/r_s}+r_{s}^{2}(d\theta^2
+\sin^{2}\theta d\phi^2)
\ee
and then transform to new coordinates $t(t_s,r_s),\: r(t_s,r_s)$.  One 
finds that in the new coordinates,
\be
R=r_s(t,r) \mbox{ \ \ \ \ ; \ \ \ \ }
L^2=\frac{1}{1-2M/r_s}\left(\frac{\partial r_s}{\partial r}\right)^2
-(1-2M/r_s)\left(\frac{\partial t_s}{\partial r}\right)^2
\ee
which then yields
\be
\pi_{{\cal M}}(r)=\frac{\partial t_s}{\partial r}.
\ee
In other words, $\pi_{{\cal M}}$ measures the rate at which Schwarzschild time
changes along a hypersurface. $P_M \equiv \int\pi_{{\cal M}}(r)dr$ is conjugate
to the ADM mass: $\{M,P_M\}=1$.  This gives clear expression to the often heard
statement that the energy generates time translations at infinity.  

Since 
$P_M$ is related to the behavior at infinity, it is not surprising that it is
invariant only under small diffeomorphisms:
$$
\delta P_M  =\left\{\int\! dr[f(r){\cal H}_t+g(r){\cal H}_r]\,,P_M\right\}
$$
vanishes provided
$$
\left.
f(\infty), g(\infty) \le \frac{\pi_L}{RL}\right|_{r=\infty}
$$
as can be checked explicitly. For most coordinate choices, the inequality
reduces to $f(\infty)=g(\infty)=0$.

\section{Quantization}

Once the Hamiltonian formulation of gravity has been developed, the 
transition to the quantum theory is quite straightforward.  This is true
only in a formal sense, though,
 since operator ordering ambiguities will be left
unresolved, as will the problem of ultraviolet divergences.  Furthermore,
the interpretation  
of the resulting theory is not at all clear, as we will briefly discuss.
More detailed discussion of interpretational issues can be found in
\cite{unwald}.  

To quantize, we realize the phase space variables $h_{ij}$, $\pi_{ij}$ as
operators satisfying the commutation relations
\be
[h_{ij}(x^i),\pi^{mn}(x'^i)]=i\delta^{m}_{i}\delta^{n}_{j}\delta^{3}(x^i-x'^i)
\ee
We therefore make the replacements, 
$$
\pi^{ij} \rightarrow -i\frac{\delta}{\delta h_{ij}}.
$$
The states of the system are then described by functions of the three-geometry,
{\em i.e.} by the wave functionals $\Psi[h_{ij}(x^i)]$.  Physical states are
required to be annihilated by the constraints:
\be
{\cal H}^i \Psi[h_{ij}]=2i\left(\frac{\delta}{\delta h_{ij}}\right)_{|j}
\Psi[h_{ij}]=0
\ee
\be
{\cal H}_{t} \Psi[h_{ij}]=-\left[\frac{8\pi}{m_{p}^2}h^{\frac{1}{2}}G_{ijkl}
\frac{\delta}{\delta h_{ij}}\frac{\delta}{\delta h_{kl}}+\frac{m_{p}^{2}}
{16\pi}h^{\frac{1}{2}} {\,^{3}{\cal R}}\right]\Psi[h_{ij}]=0
\label{WD}
\ee
where $G_{ijkl}=h_{ik}h_{jl}+h_{il}h_{jk}-h_{ij}h_{kl}$, and we have restored
the Planck mass.  The first set of constraints simply says that physical states
should be invariant under reparameterizing the spacelike hypersurfaces; they
are analogous to Gauss' law as they are linear in derivatives.  The final
constraint is quadratic in derivatives and is known as the Wheeler-Dewitt
equation.  In asymptotically flat space, the wavefunction also satisfies a
non-trivial Schrodinger equation:
\be
H\Psi[h_{ij},t]=-i\frac{\partial \Psi[h_{ij},t]}{\partial t}.
\ee
It is then assumed that $\Psi^{*}[h_{ij}]\Psi[h_{ij}]$ can somehow be
interpreted as a probability density over three-geometries.

Even if we knew the correct operator ordering prescription, the full 
Wheeler-Dewitt equation would still be much too difficult to solve.  However,
it can be simplified greatly by employing the WKB approximation, which amounts
to taking $m_{p}\rightarrow \infty$.  In the WKB approximation the 
wavefunction is written in the form
\be
\Psi[h_{ij}]=e^{im_{p}^{2}S[h_{ij}]}.
\ee
Inserting this expression into (\ref{WD}) and keeping only terms of order 
$m_{p}^{2}$, we find that $S$ satisfies:
\be
16\pi G_{ijkl}\frac{\delta S}{\delta h_{ij}}\frac{\delta S}{\delta h_{kl}}
-\frac{1}{16\pi}h^{\frac{1}{2}} {\,^{3}{\cal R}}=0.
\ee
This is the Einstein-Hamilton-Jacobi equation, which means that $S[h_{ij}]$
is the classical action associated with a solution to Einsteins's equations.
Specifically, given some solution $g_{\mu \nu}(x^{\mu})$, $S[h_{ij}]$ is 
found by integrating $S_{G}=\frac{1}{16\pi}\int\!d^4\!x\sqrt{-g}\,{\cal R}$ in
the region between some reference hypersurface and the hypersurface
$h_{ij}(x^i)$.  Thus the WKB approximation reduces the task of solving a 
functional differential equation to solving a set of partial differential
equations.

As we remarked earlier, it is difficult to interpret the wavefunction 
$\Psi[h_{ij}]$. In particular, what does it mean to say that $|\Psi|^2$ is the
probability to find some three-geometry? It may well be that the wavefunction
only has meaning once matter is included.  Then, writing $\Psi[h_{ij},\phi]$,
we would interpret
$$
 \frac{|\Psi[h_{ij},\phi]|^2}{\int\! D\!\phi |\Psi[h_{ij},\phi]|^2}
$$
as the relative probability of finding various matter configurations on the
hypersurface $h_{ij}$.  We will return to this intepretation in later sections
when we couple a scalar field to gravity.

\subsection{Spherical Symmetry}

In the spherically symmetric case we have the operators $L$, $R$ and 
$\pi_{L}=-i\frac{\delta}{\delta L}$, $\pi_R=-i\frac{\delta}{\delta R}$, and
the wavefunction, $\Psi[L,R]$, which satisfies the constraints
\be
{\cal H}_{t}\Psi[L,R]={\cal H}_{r}\Psi[L,R]=0.
\ee
In this simplified theory the quadratic constraint is still too difficult to
solve so we again employ the WKB approximation and write
\be
\Psi[L,R]=e^{iS[L,R]}.
\ee
Since the WKB approximation amounts to the replacements $\pi_{L}\rightarrow
\frac{\delta S}{\delta L}$, $\pi_{R}\rightarrow \frac{\delta S}{\delta R}$,
and using (\ref{momsolve}), $S$ is found to satisfy
$$
\frac{\delta S}{\delta L} =\eta R\sqrt{(R'/L)^2-1+2M/R} 
$$
\be
  \frac{\delta S}{\delta R}=\frac{L}{R'}\frac{d}{dr}\left(\eta R
\sqrt{(R'/L)^2-1+2M/R}\right).
\ee
These expressions are most easily integrated as follows.  Start from some
arbitrary geometry $L$, $R$, and vary $L$ while holding $R$ fixed until
$(R'/L)^2-1+2M/R=0$.  The contribution to $S$ from this variation is
$$
S[L,R]=\eta R\int\! dr\, dL \sqrt{(R'/L)^2-1+2M/R}
$$
$$
=\eta\int\! dr \left[RL\sqrt{(R'/L)^2-1+2M/R}\right.
$$
\be
\left. \mbox{\ \ \ \ } +RR' \log{\left|\frac{R'/L
-\sqrt{(R'/L)^2-1+2M/R}}{\sqrt{|1-2M/R|}}\right|}\,\right].
\label{Ssolve}
\ee
Now vary $L$, $R$, while keeping $(R'/L)^2-1+2M/R=0$, to some set geometry.
There is no contribution to $S$ from this variation, so (\ref{Ssolve}) is the
complete solution.  The wavefunction, $\Psi$, either oscillates or decays
exponentially depending on the sign of $(R'/L)^2-1+2M/R$.  This is also the 
condition which determines whether the hypersurface $L$, $R$, can be embedded
in the classical Schwarzschild geometry.

The description of the state in terms of the wavefunction $\Psi[L,R]$ is, of
course, highly redundant since, according to the constraints, the wavefunction
must take the same value for many different $L$'s  and $R$'s. If we wish to 
eliminate this redundancy we should describe the state in terms of physical
observables, {\em i.e.} those which commute with the constraints.  In the 
present case there are only two such observables \cite{thiemann,kuchar}
: the ADM mass, M, and its
canonical momentum $P_M$.  To satisfy the commutation relation $[M,P_M]=i$
we make the replacement $M=i\partial/\partial P_M$, so that a state of
definite mass is written
\be
\Psi_M(P_M)=e^{-iMP_M}.
\ee
If the time dependence is also included we have
\be
\Psi_M(P_M,t)=e^{-iM(P_M+t)}.
\ee
It is seen that $P_M$ functions as an intrinsic time variable.

The emergence of an intrinsic time in this system is quite interesting in
light of the so-called ``problem of time'' in quantum gravity. In the 
asymptotically flat case this problem is relatively benign, since the 
wavefunction depends on the time $t$, which can be measured by clocks at
infinity.  However, for a closed universe there is no such asymptotic region
and no obvious variable to play the role of time.  This makes the wavefunction
of a closed universe especially difficult to interpret.  In quantum mechanics,
we interpret the wavefunction $\psi(x,t)$ by saying that at fixed time $t$ the
probability to find $x$ is $|\psi(x,t)|^2$.  To carry this procedure over to
the quantum theory of a closed universe we need to specify a variable which
can be fixed in order to determine probabilities.  In the spherically
symmetric, asymptotically flat case, $P_M$, like $t$, can play this role, but
no such variable is known in general.

\pagebreak
This completes our discussion of the
 quantization of the pure gravity theory.
In  what follows 
we turn to the effects of incorporating matter into the
system.

\chapter{Quantum Field Theory Near Black Holes}

\section{A Useful Coordinate System}

 Calculations of physical effects in the presence of black holes are greatly
simplified by employing appropriate coordinates. In this section, we will
describe the properties of a little known  set of coordinates 
\cite{KW2} which  will be
used repeatedly in the material that follows.  
 
Schwarzschild found his remarkable exact solution for
the geometry outside a star in general relativity
quite soon after Einstein
derived the field equations.  Further
study of this geometry over the course of several decades
revealed a series of surprises: the existence and physical
relevance of pure vacuum ``black hole'' solutions; the
incompleteness of the space-time covered by the
original Schwarzschild coordinates, and the highly
non-trivial global structure of its completion; and the
dynamic nature of the physics in this geometry despite its
static mathematical form, revealed perhaps most dramatically
by
the Hawking radiance \cite{hawking}.  Discussions
of this material can now be found
in advanced
textbooks \cite{birrel}, but they are hardly limpid.

In the course of investigating an improvement to the standard
calculation of this radiance to take into account its
self-gravity, as we detail in later sections,
we came upon a remarkably simple form for the line element of
Schwarzschild (and Reissner-Nordstrom) geometry.  
This line element has an interesting history~
\cite{israel}, but
as far as we know it has never been discussed from a modern
point of view.
We have found
that several of the more subtle features of the geometry become
especially easy
to see when this line element is used.
 
To motivate the form of the coordinates we reconsider the constraint
equations of spherically symmetric gravity (\ref{theconstraints}):
$$
{\cal H}_t =\frac{\pi_{L}^{2}}{2R^2}-\frac{\pi_L \pi_R}{R} +\left(\frac{
RR'}{L}\right)'-\frac{R'^2}{2L}=0,
$$
\be
{\cal H}_r=R'\pi_R -L\pi'=0.
\ee
The canonical momenta are given by
\be
\pi_L=\frac{N^rR'}{N^t}-\frac{R\dot{R}}{N^t} \mbox{ \ \ \ \ ; \ \ \ \ }
\pi_R=\frac{(N^rLR)'}{N^t}-\frac{\dot{(LR)}}{N^t}.
\ee
We can arrive at a particular set of coordinates by choosing a gauge and
solving the constraints.
Our choice is simply $L=1$, $R=r$.  With this choice the
equations simplify drastically, and
one easily solves to find
$\pi_L = \sqrt {2Mr }~,~ \pi_R = \sqrt {M\over 2r}$
and then $N^t = \pm 1,~ N^r = \pm \sqrt {2M\over r}$.
Thus for the line element we have
\be
ds^2~=~ -dt^2 + (dr \pm \sqrt {2M \over r} dt )^2 +
r^2 (d\theta^2  +  \sin^2 \theta d\phi^2 )~.
\label{nicemetric}
\ee
$M$, which appears as an integration constant, of course
is to be interpreted as the mass of the black hole described by
this line element.
 
For the Reissner-Nordstrom geometry, the same gauge choice leads
to a metric of the same form, with the only change that
$2M \rightarrow 2M - Q^2/r$.
 
These line elements are stationary -- that is, invariant under
translation of $t$,
but not static -- that is, invariant under reversal
of the sign of
$t$.  Indeed reversal of this sign interchanges the $\pm$
in (\ref{nicemetric}), a feature we will interpret further below.
Another peculiar feature is that each
constant time slice
$dt=0$ is simply flat Euclidean space!

We can obtain a physical interpretation of these coordinates by
comparing them to those of Lemaitre ~\cite{lemaitre}, in terms of 
which the line element reads
\be
ds^2~=~-d\tau^2 + {(2M)^{2/3} \over [{3 \over 2} (r_{L} + 
\tau)]^{2/3}}~dr_{L}^2~-~(2M)^{2/3} [{3 \over 2}(r_{L} -\tau)]^{4/3}
(d\theta^2 + \sin^2{\theta} d\phi^2).
\label{lemaitre}
\ee
As the Lemaitre coordinates are synchronous ($g_{\tau \tau}=-1$ , 
 $g_{\tau \rm i}=0$),~ a class of timelike geodesics is given by 
motion along the time lines ($r_{L},\theta, \phi =$constant), and
the proper time along the geodesics is given by the coordinate
$\tau$.  To arrive at (\ref{lemaitre}) we retain the Lemaitre time
coordinate, but now demand that the radial coordinate squared be
the coefficient multiplying $d\theta^2 + \sin^{2} \theta d\phi^2$.
In other words, we write
\be
t=\tau\mbox{ \ \ \ \ ; \ \ \ \ } r=(2M)^{1/3}[{3\over 2}(r_{L} -
\tau)]^{2/3}.
\label{coords}
\ee
A simple calculation then leads from (\ref{lemaitre}) to (\ref{nicemetric}) 
with the upper choice of sign; the lower choice is obtained by repeating
the same steps starting from Lemaitre coordinates with the sign of
$\rm \tau$ reversed. 

Finally, let us note that these coordinates are related to 
Schwarzschild coordinates,
\be
ds^{2}=-(1-\frac{2M}{r})dt_{s}^{2} + \frac{dr^{2}}{1-\frac{2M}{r}} +
r^{2}(d\theta^{2}
+\sin{\theta}^{2} d\phi^{2})
\ee
by a change of time slicing,
\be
t_{s}=t-2\sqrt{2Mr} - 2M \log{\left[\frac{\sqrt{r}-\sqrt{2M}}{\sqrt{r}+
\sqrt{2M}}\right]}.
\label{tdef}
\ee
In contrast to the surfaces of constant $t_{s}$, the constant $t$ surfaces pass
smoothly through the horizon and extend to the future  singularity 
free of coordinate singularities.
 
In terms of $r$ and $t$, the radially ingoing and outgoing null
geodesics are given by
$$
\hspace{-10mm}\mbox{ingoing: \ \ \ } t+r-2\sqrt{2Mr}+4M\log[{\sqrt{r}+\sqrt{2M}]}
=v = \mbox{ constant}
$$
\be
\mbox{outgoing: \ \ \ } t-r-2\sqrt{2Mr} -4M\log{[\sqrt{r}-\sqrt{2M}]}=u
=\mbox{ constant }.
\label{geo}
\ee
 
\subsection{Global Structure}

%xx line element continues through a horizon in non-singular fashion.
%on the other horizon, it poops out.
 
\begin{figure}[htb]
\centerline{\psfig{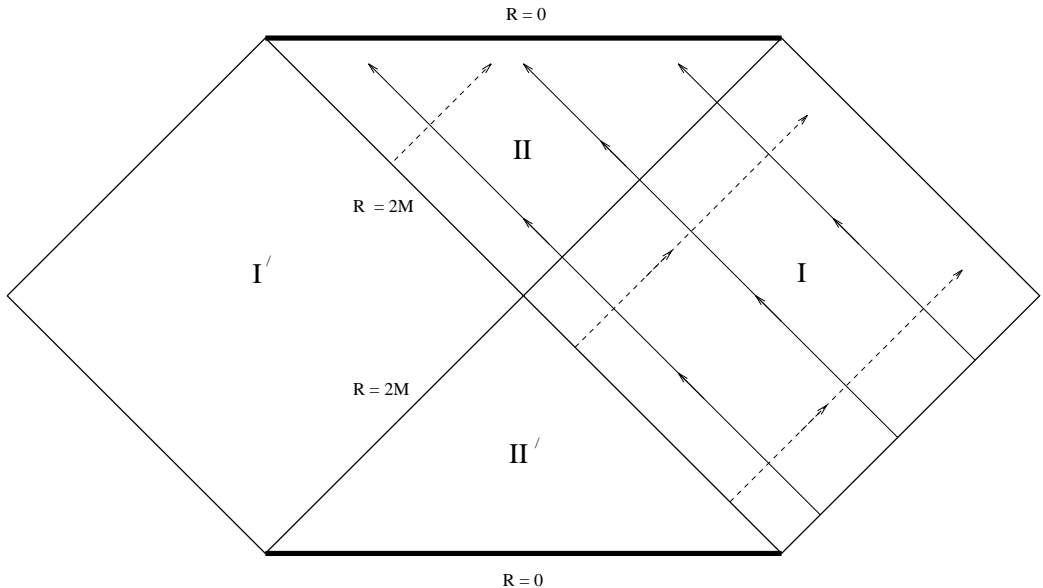}}
\caption{Penrose diagram for the Schwarzschild geometry.  As described in the
text, $r$ and $t$ in one coordinate patch, for the upper sign of the line
element, cover regions I and II.  As is clear from the diagram, the ingoing
light rays are captured in their entirety (into the singularity), whereas the
outgoing light rays cannot be traced back past the horizon.}
\label{metfig}
\end{figure}

Now let us discuss the global properties of our coordinate system.
Perhaps the clearest approach to such questions is via
consideration of the properties
of light rays.  Taking for definiteness the upper sign in (\ref{nicemetric}),
and without any essential
loss of generality restricting to the case
$d\theta = d\phi =0$ appropriate to the $\theta = \pi/2$ sections,
we find that $ds^2=0$ when
\be
{dr\over dt } ~=~ - \sqrt {2M\over r } \mp 1~.
\label{lightrays}
\ee
For the class of light rays governed by
the upper sign, we can cover the
entire range $0 < r < \infty $ as $t$ varies.  In particular one
meets
no obstruction, nor any special structure,
at the horizon $r=2M$.  For the class of light rays
governed by the lower sign there is structure at $r=2M$.
When $r > 2M$ one has a positive slope for ${dr\over dt}$, and
$r$ ranges over $2M < r < \infty$.  When $r < 2M$ one has a negative
slope for ${dr\over dt}$, and $r$ ranges over $0 < r < 2M$.
When $r = 2M$ it does not vary with $t$.  From these properties, one
infers that our light rays cover regions I and II in the Penrose
diagram, as displayed in Fig. ~\ref{metfig}.  Let us emphasize that the
properties of the Penrose diagram can be {\em inferred} from the
properties of the light rays, although we will not belabor that
point here.
 
%Here is the figure!!
\renewcommand\floatpagefraction{.9}
\renewcommand\topfraction{.9}
\renewcommand\bottomfraction{.9}
\renewcommand\textfraction{.1}

%xx to recover the complete Penrose diagram, one must include
%sectors where increasing $t$ is in the past light cone.

If one chooses instead the lower sign in (\ref{nicemetric}) , and performs
a similar analysis, one finds that regions I and II$^\prime$
are covered.
Patching these together with the sectors found previously,
one still does not have a complete space-time.
However our line element is not yet exhausted.  For in drawing
Figure 1 we have implicitly assumed that $t$ increases along light
rays which point up (``towards the future'').  Logically, and to
maintain symmetry, one should
consider also the opposite case, that the coordinate
$t$ increases towards the past.  By doing this, one generates
coordinate systems covering regions
I$^\prime$ and II$^\prime$
respectively I$^\prime$ and II$^\prime$,
for the upper and lower signs
in (\ref{nicemetric}) .   Thus the complete Penrose diagram is covered
with patches each governed by a stationary -- but not static --
metric, and with non-trivial regions of overlap.
 
%xx for Reissner-Nordstrom, note that
%the coordinate singularity occurs inside both horizons.  to get
%the complete space, build a tower.
 
In the Reissner-Nordstrom case the generalization of (\ref{nicemetric})
has
a coordinate singularity at $r=Q^2/2M$.  However this singularity
is inside both horizons, and does not pose a serious obstruction to
a global analysis.  One obtains the complete Penrose diagram also in
this case by
iterating
constructions similar to those just sketched.

%nature of time reversal
 
The usual
Schwarzschild line element appears to be time reversal symmetric,
but when the global structure of the space-time it defines is taken into
account one sees that this appearance is misleading.  The fully extended
light-rays in Figure 1 go from empty space to a singularity as $t$
advances (they pass from region I into region II),
which is definitely distinguishable from the
reverse process.  There is a
symmetry which relates these to the corresponding rays going from
region II$^\prime$ to region I$^\prime$, however it involves not merely
changing the sign of
$t$ in the Schwarzschild metric,
but rather going to a completely disjoint
region of the space-time.  This actual symmetry of the space-time is if
anything more obvious in our construction than in the standard one.
Thus by taking the line-element
in region I stationary rather than static
we have lost some false symmetry while making the true symmetry --
and its
necessary
connection with the existence of
region I$^\prime$ (constructed, as we have seen, by simultaneously
reversing the
sign of $t$ {\it and\/} interchanging the future with the past) --
more obvious.

\section{Quantum Field Theory in Curved Space}

In this section we briefly present some of the important results in the
theory of quantum fields in curved space.  Our goal is primarily to establish
notation and to write down some formulas which will be referred to in later
sections.  Comprehensive reviews of the subject can be found in 
\cite{birrel,fulling}.

Quantum field theory in curved space is a hybrid of quantum and classical
field theory, which one hopes reliably describes the behavior of matter in
regions of relatively low space-time curvature.  The approach is to couple
a quantum field to a metric tensor, which is treated as a classical variable.
For simplicity, we will consider a massless scalar field with action
\be
S=-\frac{1}{2}\int\! d^4\! x\sqrt{-g}\,g^{\mu \nu} 
\partial_{\mu}\phi\, \partial_{\nu}\phi.
\ee
The field satisfies the wave equation,
\be
(\sqrt{-g}\,g^{\mu \nu}\phi_{,\mu})_{,\nu}=0
\ee
Associated with this wave equation is the conserved inner product
\be
(\phi_1,\phi_2)\equiv -i\int_{\Sigma}d^3\! x\sqrt{h}\,n^{\mu}\phi_1(x^i,t)
\stackrel{\leftrightarrow}{\partial_{\mu}}\phi_{2}^{*}(x^i,t)
\ee
where $\phi_1$ and $\phi_2$ are two solutions of the wave equation.  Here
$\Sigma$ is a Cauchy surface, and $n^{\mu}$ is a future pointing unit vector
normal to $\Sigma$.  In order to second quantize the field we must write down
a complete set of ``positive frequency'' solutions, $u_{i}(x^{\mu})$, which
satisfy $(u_i,u_j)=\delta_{ij}$.  This is where the ambiguity in the 
quantization process occurs, since there are many different sets of $u_i$'s,
which will be shown to lead to inequivalent quantizations.  Once we have chosen
a set, the field operator is written in terms of creation and
annihilation operators:
\be
\phi=\sum_{i}\,[a_i u_i+a_{i}^{\dagger} u_{i}^{*}]
\label{phiu}
\ee
where $a$, $a^{\dagger}$ obey
\be
[a_i,a_{j}^{\dagger}]=\delta_{ij}.
\ee
The vacuum state is defined by $a_i |0_u \rangle=0$, and particle states
are created by applying $a^{\dagger}$'s to $|0_u \rangle$.

On the other hand, suppose that instead of $u_i$ we had chosen the modes
$v_i$ in which to expand the field:
\be
\phi=\sum_{i}\,[b_i v_i +b_{i}^{\dagger}v_{i}^{*}].
\label{phiv}
\ee
Then the vacuum would be defined by $b_i |0_v \rangle =0$, and particle
states would be created by applying $b^{\dagger}$'s to $|0_v \rangle$. 
The obvious question which arises is: what is the relation between the states
defined by $u_i$ and those defined by  $v_i$.  To answer this, we
note that by equating (\ref{phiu}) and (\ref{phiv}), and taking inner products,
 we obtain the ``Bogoliubov transformation'':
\be
b_j = \sum_{j}(\alpha_{ji}^{*} a_{i}-\beta_{ji}^{*} a_{i}^{\dagger})
\ee
where
\be
\alpha_{ij}=(v_i,u_j) \mbox{  \ \ \ \ ; \ \ \ \ } \beta_{ij}=-(v_i,u_{j}^{*})
\ee
The Bogoliubov coefficients obey the completeness relations,
$$
\sum_{k}(\alpha_{ik}\alpha_{jk}^{*}-\beta_{ik}\beta_{jk}^{*})=\delta_{ij}
$$
\be
\sum_{k}(\alpha_{ik}\beta_{jk}-\beta_{ik}\alpha_{jk})=0.
\label{complete}
\ee
It can then be shown that the states are related by
\be
|\psi_u \rangle =C :\!\exp{[\frac{1}{2}a(\alpha^{-1}\beta)a+a(\alpha^{-1}
-1)a^{\dagger}+
\frac{1}{2}a^{\dagger}(-\beta^{*}\alpha^{-1})a^{\dagger}]}\!:|\psi_v 
\rangle
\ee
where $C$ is a constant.  The average number of $v$ particles in the $u$
vacuum is
\be
\langle 0_u |b_{i}^{\dagger}b_{i}|0_{u}\rangle=\sum_{j}|\beta_{ji}|^2.
\ee
 
\section{Black Hole Radiance}

Now we turn to the quantization of a massless scalar field in the presence of
collapsing matter.   The goal is to compute the flux of particles on 
${\cal J}^+$ given some initial state defined on ${\cal J}^-$.  To completely
determine this flux on all of ${\cal J}^+$, one would have to solve 
the wave equation
mode by mode in the region bounded by ${\cal J}^-$ and $\Sigma$, 
where $\Sigma$ is a constant $t$ hypersurface which 
crosses the horizon to the future of the collapsing matter.
  This is clearly intractable, since the geometry
inside the collapsing matter may be very complicated, and the scalar field 
might interact with the matter in an arbitrarily complex fashion.  Fortunately,
all of the dependence of the particle flux on ${\cal J}^+$ on these factors
dies out at sufficiently late times, and the radiation becomes, at least to
leading order, completely independent of the details of the collapse process.
Thus we shift our goal to calculating this late-time radiation and to showing
that it is indeed universal.

The first important observation is that the geometry is entirely smooth as long
as one stays sufficiently far from the singularity at $r=0$.  In particular,
for a black hole much larger than $m_p$, the geometry is smooth near the
horizon --- in fact, the curvature can be made arbitrarily small by making the
black hole arbitrarily massive.  This strongly suggests the conlusion that 
however complicated the state of the field is after propagating through the
matter, it should certainly appear nonsingular to inertial observers near the
horizon.  Of course, this presupposes that the initial state on ${\cal J}^-$
is nonsingular.

Next, we examine the outgoing null geodesics in the region exterior to the
collapsing matter.  These obey:
\be
u \equiv t-2\sqrt{2Mr}-r-4M\log{(\sqrt{r/2M}-1)}=\mbox{constant}.
\ee
Let us consider two geodesics, labelled by $u_1$ and $u_2$, which are 
separated in time by $\Delta u= u_2 - u_1$.  On a constant $t$
surface, $\Sigma$, the geodesics have a radial separation given by
\be
\Delta u=r_1 -r_2 +2(\sqrt{2Mr_1}-\sqrt{2Mr_2})+4M\log{\left( \frac{\sqrt{r_1}
-\sqrt{2M}}{\sqrt{r_2}-\sqrt{2m}}\right)}.
\ee
For large $u_1$, $u_2$, which corresponds to geodesics which reach ${\cal J}^+$
at late times, the radial separation of the geodesics near the horizon is 
determined by
\be
\Delta u \approx 4M\log{\left(\frac{\sqrt{r_1}-\sqrt{2M}}{\sqrt{r_2}-\sqrt{2M}
}\right)}
\ee
or
\be
r_2 - 2M \approx (r_1 -2M) e^{-\Delta u/4M}.
\ee
In other words, the outoging geodesics pile up along the horizon.  
Alternatively, we can note that all the geodesics which reach ${\cal J}^+$
after $u=\hat{u}$, where $\hat{u}\gg M$, were contained on the surface $t=0$ in
the region between $r=2M$ and $r=2M+4Me^{-\hat{u}/4M}$.  
This has two important consequences.  First, we see that the late time
radiation is entirely determined by the state of the field at distances
arbitrarily close to the horizon.  Second, an outgoing wave suffers an 
arbitrarily large redshift in escaping from near the horizon to ${\cal J}^+$.
So, putting it all together, we see that to compute the late-time radiation
we need only consider nonsingular states in a region near the horizon.

It is easy to construct a nonsingular state if we first define a new time
coordinate.  The trouble with the coordinate $t$ used in (\ref{nicemetric}) 
is that its
flow is spacelike inside the horizon.  However, this can be rectified if we 
define a new time, $\tau$, as the value of $t$ along the curve 
$dr + \sqrt{2M/r}dt=0$.  $\tau$ has the virtue of being nonsingular and 
timelike.  We can choose a set of modes which are positive frequency with
respect to $\tau$, and expand the field in terms of them.  Then, by the
equivalence principle and standard quantum field theory in flat space, the
corresponding vacuum will have a nonsingular stress-energy as seen by inertial
observers.  Of course, this choice of state is not unique since other 
nonsingular states can be obtained by applying particle creation operators to
this state.  This is irrelevant if one is only interested in late-time 
radiation, since the excitations above the vacuum will eventually redshift
away.  

Now at spatial infinity (more accurately: conformal infinity
${\cal I}_+$) the vacuum state is defined
locally by the requirement that
modes having positive frequency with respect to the variable
$u = t_{\rm s} -r_*$ are unoccupied, where $t_{\rm s}$
is Schwarzschild
time and $r_* = r + 2M\ln (r -2M)$ is the tortoise coordinate.
We wish to find the relationship between this
requirement and the preceding one.
 
The relationship between $t$ and $t_{\rm s}$ is
\be
t = t_{\rm s} + 2 \sqrt{2Mr} +
  2M \ln {{\sqrt r - \sqrt{2M} } \over {\sqrt r + \sqrt{2M} }}
\label{times}
\ee
so that
\be
u = t_{\rm s} - r_* =
t - 2 \sqrt{2Mr} -r -4M \ln (\sqrt r - \sqrt{2M} )~,
\label{uexpression}
\ee
and thus one finds that along a curve with $dr+ \sqrt {2M\over r} dt =0$,
\be
{du\over dt}~=~ 2 + \sqrt {2M\over r} + {2M \over r - \sqrt{2Mr} }~.
\label{frequencyfactor}
\ee
Because the last term on the right hand side
is singular, the two definitions of positive
frequency -- with respect to $u$ or to $t$ -- do not coincide.
To remove the singularity, note that along any of
the curves of interest
$e^{-u/4M}$ has a simple zero at $r=2M$, but is otherwise positive.
Clearly then demanding positive frequency with respect to
$t$ along such curves requires positive frequency not with respect
to $u$ but rather with respect to
\be
U = - e^{-u/4M}~.
\label{kruskalU}
\ee
In this way we have arrived at the famous Unruh
boundary conditions \cite{unruh}. 
 
To summarize, the appropriate construction near the horizon is to expand the
field as
\be
\phi= \int\! \frac{d\omega}{\sqrt{2\pi 2\omega}}\,[b_{\omega}e^{i\omega U}
+b_{\omega}^{\dagger}e^{-i\omega U}]
\ee
and take the state resulting from collapse to be $|0_U\rangle$, where
$b_{\omega}|0_{U}\rangle =0$.  For simplicity, we have suppressed the 
field's dependence on the ingoing modes.  It remains to describe this state in
terms of particles defined by the modes $e^{-i\omega u}$. Since $u \rightarrow
\infty$ at the horizon, we need another set of modes in which to expand the 
field 
inside the horizon.  These will be referred to as $u_{\omega}^{in}(u)$; their
explicit form will not be needed.  We then have
\be
\phi = \int\! \frac{d\omega}{\sqrt{2\pi 2\omega}}\,[(a_{\omega}e^{-i\omega u}
+a_{\omega}^{\dagger} e^{i\omega u})\,\Theta\!\left(-U\right)
+\left(a_{\omega}^{in}u_{\omega}^{in}+a_{\omega}^{in \dagger} 
u_{\omega}^{in *}\right)\Theta\! \left(U\right)].
\ee
The relation between the modes for $U<0$ is given by
\be
e^{i\omega' U(u)}=\int\! \frac{d \omega}{2\pi}\,[\alpha_{\omega' \omega}
e^{-i\omega u}+\beta_{\omega' \omega}e^{i\omega u}] \mbox{ \ \ \ ; \ \ \ }
U<0
\ee
so
\be
\alpha_{\omega' \omega}=\int_{-\infty}^{\infty}\!du\, 
e^{i(\omega' U(u)+\omega u)}
\mbox{ \ \ \ ; \ \ \ } \beta_{\omega' \omega}=\int_{-\infty}^{ \infty}\!du\,
e^{i(\omega'U(u)-\omega u)},
\ee
or, since $U(u)=-e^{-u/4M}$,
\be
\alpha_{\omega' \omega}=4M\, \frac{\omega^{4\pi i \omega}}{\omega'}\, 
e^{2\pi M\omega}\, \Gamma(1-4iM \omega) \mbox{ \ \ ; \ \ }
\beta_{\omega' \omega}=4M\, \frac{\omega^{-4\pi i \omega}}{\omega'}\,
e^{-2\pi M \omega}\, \Gamma(1+4iM \omega).
\ee
So
\be
|\alpha_{\omega' \omega}|^2= \frac{(4M)^3 \pi \omega}{\omega'^2}
\frac{1}{1-e^{-8 \pi M\omega}} \mbox{ \ \ \ ; \ \ \ }
|\beta_{\omega' \omega}|^2 = \frac{(4M)^3 \pi \omega}{\omega'^2}
\frac{e^{-8\pi M \omega}}{1-e^{-8\pi M \omega}}.
\ee
If we are only interested in the radiation which flows to infinity it is
appropriate to form a density matrix by tracing over the states inside the
horizon.  To find the average number of particles radiated we should integrate
$|\beta_{\omega' \omega}|^2$.  But because the black hole radiates for an
infinite amount of time at a constant rate (in the current approximation),
this yields infinity.  To find the {\em rate} of emission, 
we can place the hole in a
large box and use the density of states $d\omega/2\pi$ for outgoing particles.
For normalized modes the completeness relation, (\ref{complete}),
 and the relation
\be
|\alpha_{\omega' \omega}|^2=e^{8\pi M \omega}|\beta_{\omega' \omega}|^2
\ee
imply
\be
\int\! d\omega' |\beta_{\omega' \omega}|^2 = \frac{1}{e^{8\pi M \omega}-1}.
\ee
The rate of emission of particles in the range $\omega$ to $\omega + d\omega$
is then:
\be
F(\omega)=\frac{d\omega}{2\pi}\frac{1}{e^{8\pi M \omega}-1}.
\ee
This is precisely the rate of emission from a black body, in one dimension,
at temperature $T=1/8\pi M$.  $F$ does not quite give the flux seen at infinity
since some fraction, $1-\Gamma(\omega)$, of the particles will be reflected by
the spacetime curvature back into the hole.  Thus, for the flux at infinity
we write,
\be
F_{\infty}(\omega)=\frac{d\omega}{2\pi}\frac{\Gamma(\omega)}{e^{8\pi M \omega}
-1}.
\ee

The preceding analysis shows that the average flux is that of a thermal body,
not that the full density matrix is thermal.  That the density matrix is
exactly thermal is, in fact, easy to show by utilizing a clever trick due to
Unruh \cite{unruh}.  
We have not chosen to use this method here, and instead have gone
through the rather laborious procedure of computing the Bogoliubov coefficients
directly, because only the latter method can be used when one wants to find
corrections to the spectrum.

\subsection{Reissner-Nordstrom}

A similar analysis can be carried out for the Reissner-Nordstrom geometry, the
metric for which is, in the $L=1$, $R=r$ gauge:
\be
ds^2=-dt^2+(dr+\sqrt{2M/r-Q^2/r^2\,}\,dt)^2 +
r^2(d\theta^2+{\sin}^2\theta \,d\phi^2).
\label{rnmetric}
\ee
The geometry has two horizons which are located at the two values of $r$ for 
which $t$ goes from being timelike to spacelike, and vice versa.  The outer
horizon radius is
\be
R_+=M+\sqrt{M^2-Q^2}
\ee
and the inner horizon radius is
\be
R_- =M-\sqrt{M^2-Q^2}.
\ee
It should be noted that the coordinates in (\ref{rnmetric}) fail to cover the
region inside the inner horizon. This will not pose any obstacle to determining
the radiance, though, as the calculation only depends in the geometry in the
vicinity of the outer horizon.

In the present case, the state resulting from collapse is specified by 
demanding the absence of particles positive in frequency with respect to $t$
along the curve $dr+\sqrt{2M/r-Q^2/r^2}$.  Now the calculation can proceed in
exact analogy to the uncharged case.  The resulting temperature is:
\be
T(M,Q)=\frac{1}{2\pi}\frac{\sqrt{M^2-Q^2}}{(M+\sqrt{M^2-Q^2})^2}.
\ee
The temperature vanishes for the extremal black hole $M=Q$.  If $Q>M$,
consideration of the global geometry reveals that there is no black hole at 
all, but rather a naked singularity, which we will refer to as the 
meta-extremal case.  Finally, if the radiated particles are themselves charged,
the factor governing the emission probability is not the Boltzmann factor,
$e^{-\omega/T}$, but rather:
\be
\exp{\left(-\frac{\omega-Qq/R_+(M,Q)}{T(M,Q)}\right)}
\ee
where the particles carry charge q.  The factor $Q/R_+$ is the electrostatic
potential at the outer horizon, and appears as a chemical potential causing
particles with the same sign of charge as the hole to be preferentially
emitted.

\subsection{Breakdown of the Semiclassical Approximation}

In the preceding treatment we have followed Hawking and considered the 
propagation of free fields on a classical background geometry, ignoring the
back reaction altogether.  It is usually assumed that the dominant effects of 
the back reaction can be incorporated by allowing the black hole to lose mass
quasistatically, so that the radiation at infinity is thermal but with a 
slowly varying temperature.  Indeed, this is how a normal thermal body is 
expected to behave during cooling.  While this scenario seems quite plausible 
for a wide range of configurations, both for black holes and normal objects,
there are notable instances where it is certain to be invalid, even 
qualitatively \cite{psstw}.  
In particular, if the emission of a single typical quantum
induces a relatively large change in temperature, a quasistatic description is
clearly inappropriate. Since a typical quantum has energy equal to $T$, the
temperature of the hole, we expect non-trivial back reaction effects to become
important when
\be
T\frac{\partial T}{\partial M}\approx T.
\ee

For a Schwarzschild black hole this happens when $M\approx m_p$, whereas for
a Reissner-Nordstrom hole it happens when $M\approx Q$. In the Schwarzschild
case, the breakdown occurs in a regime of high curvatures and large quantum
fluctuations, presumably inaccessible to semiclassical methods.  Not so,
however, for a large near-extremal Reissner-Nordstrom hole, $M\approx Q\gg 
m_p$, for which the curvature near the horizon remains small, suggesting that
meaningful calculations can be performed without the need for a full theory of
quantum gravity.  In the next chapter
 we perform such a calculation and see explicitly
that the radiance from the near extremal hole is markedly different from what
free field calculations indicate.

\section{Effect of Fluctuations}

The semiclassical approach to back reaction effects in quantum gravity is to
compute $<T_{\mu \nu}>$ for a field quantized on a classical background 
geometry, and then to use this quantity on the right hand side of Einstein's
equations,
\be
{\cal R}_{\mu \nu}-\frac{1}{2}g_{\mu \nu}{\cal R}=8\pi <T_{\mu \nu}>.
\label{semiclass}
\ee
The (generally intractable) problem then becomes to compute $<T_{\mu \nu}>$
for an arbitrary metric.  One expects this approach to provide a reasonable
description of the gross effects of back reaction provided that the 
fluctuations in $T_{\mu \nu}$ are not too large. For example, replacing
$<T_{\mu \nu}>$ by $\sqrt{<T_{\mu \nu}\,^2>}$ 
should not lead to any substantial
change in the geometry if (\ref{semiclass}) is to be physically relevant. 

In order to be quantitative, we shall consider the computation of stress
energy fluctuations in the moving mirror model, which is discussed in
\cite{Dav77,chung,hotta}.  We consider this model here
because it provides a vivid demonstration of just how misleading the
semiclassical approximation can be, and because we will referring to it later
in another context.  By ``moving mirror model'', we simply mean quantizing a
scalar field in two dimensions subject to the condition that the field vanish
along the timelike worldline $z_m(t)$.  We will use the coordinates
\be
u= t-z \mbox{ \ \ \ \ ; \ \ \ \ } v=t+z.
\ee
Without the mirror present the propagator is 
\be
G(1,2)\equiv \langle 0|T[\phi(1)\phi(2)]|0\rangle
=\frac{1}{4\pi} \log{[(u_2-u_1)(v_2-v_1)]}
\label{minkprop}
\ee
and for a mirror at rest, $z_m(t)=0$:
\be
G(1,2)=\frac{1}{4\pi}\log{\left[\frac{(u_2-u_1)(v_2-v_1)}{(u_2-v_1)
(v_2-u_1)}\right]}.
\ee
Now consider a general mirror trajectory, which we shall write in terms of
$u,v$ as $v_m(u)$.   It is easiest to proceed by defining new coordinates in
terms of which the mirror is at rest.  Working in the region to the right of
the mirror and defining 
\be
U(u)\equiv v_m(u) \mbox{ \ \ \ \ ; \ \ \ \ } V(v)\equiv v,
\ee
we see that the trajectory $v=v_m(u)$ corresponds to $U=V$, which is the
desired result.  In the new coordinates the metric is
\be
ds^2=\left(U'(u)\right)^{-1}dU dV,
\ee
and because of conformal invariance, the propagator has the same form as 
before:
\be
G(1,2)=\frac{1}{4\pi}\log{\left[\frac{\left(U(u_2)-U(u_1)\right)
\left(v_2-v_1\right)}{\left(U(u_2)-v_1\right)\left(v_2-v_1\right)}\right]}.
\label{mmprop}
\ee
The basic object we will be considering is the renormalized propagator defined
by subtracting (\ref{minkprop}) from (\ref{mmprop}):
\be
G_R(1,2)=\frac{1}{4\pi}\log{\left[\frac{U(u_2)-U(u_1)}{(u_2-u_1)(U(u_2)-v_1)
(v_2-U(u_1))}\right]}.
\ee
This prescription simply corresponds to normal ordering with respect to 
standard Minkowski space creation and annihilation operators.  $<T_{\mu \nu}>$
can now be found by operating on $G_R(1,2)$ according to
\be
T_{\mu \nu}=\partial_{\mu} \phi\, \partial_{\nu}\phi-\frac{1}{2}g_{\mu \nu}
\partial_{\rho}\phi\,\partial^{\rho}\phi,
\ee
so
\be
<T_{\mu \nu}>=\lim_{1\leftrightarrow 2}\,[\,
\frac{\partial}{\partial x_{1}^{\mu}}
\frac{\partial}{\partial x_{2}^{\nu}}-\frac{1}{2}g_{\mu \nu} g^{\alpha \rho}
\frac{\partial}{\partial x_{1}^{\rho}}\frac{\partial}{\partial x_{2}^{\alpha}}
\,]G_R(1,2).
\ee
In the present case the only nonzero element is
\be
<T_{uu}>=\lim_{1 \leftrightarrow 2} 
\frac{\partial}{\partial u_1}\frac{\partial}
{\partial u_2} G_R(1,2)
\ee
which, after some algebra, is found to be
\be
<T_{uu}>=\frac{1}{12\pi}\sqrt{v_m(u)}\frac{d^2}{du^2}\sqrt{\frac{1}{v_m(u)}}.
\ee
The result can be written in terms of the mirror trajectory $z_m(t)$ using
\be
\frac{dv_m(u)}{du}=\frac{1+\dot{z_m}}{1-\dot{z_m}},
\ee
yielding
\be
<T_{uu}>=-\frac{1}{12\pi}\frac{(1-\dot{z_m}^2)^{\frac{1}{2}}}{(1-\dot{z_m})^2}
\frac{d}{dt}\left[\frac{\ddot{z_m}}{(1-\dot{z_m}^2)^{\frac{3}{2}}}\right].
\ee
These expressions hold in the region to the right of the mirror.  The 
expressions to the left of the mirror are obtained by interchanging $u$ and
$v$.

These results can be used to find the radiation reaction force on the mirror.
The force four-vector is, by energy conservation:
\be
F_{\mu}=-\Delta T_{\mu \nu}v^{\nu}
\ee
where $\Delta T_{\mu \nu}=T_{\mu \nu}(+)-T_{\mu \nu}(-)$ is the difference in
$T_{\mu \nu}$ evaluated on the right and left hand sides of the mirror, and
$v^{\nu}$ is the mirror's velocity four-vector.  We obtain:
\be
F^{\mu}=\frac{1}{12\pi}\left(\frac{d^2v^{\mu}}{d\tau^2}-v^{\mu}v^{\nu}
\frac{d^2 v_{\nu}}{d\tau^2}\right)
\ee
where $\tau$ is proper time.  It is amusing to compare this to the 
radiation reaction force on a point charge in classical electrodynamics,
\be
F^{\mu}=\frac{2e^2}{3}\left(\frac{d^2v^{\mu}}{d\tau^2}-v^{\mu}v^{\nu}
\frac{d^2v_{\nu}}{d\tau^2}\right).
\ee
This expression, containing as it does third derivatives of position of with
respect to time, has some well known peculiar features such as runaway
solutions and pre-acceleration.  These carry over to the present case, although
we hasten to add that a realistic mirror has a high frequency cutoff
above which it becomes transparent; the worried reader need not be concerned
with the hazards of runaway mirrors.  

$<T_{\mu \nu}T_{\alpha \beta}>$ is found by operating on the renormalized four
point function, which by Wick's theorem is:
\be
G_R(1,2,3,4)=G_R(1,2)G_R(3.4)+G_R(1,3)G_R(2,4)+G_R(1,4)G_R(2,3).
\ee
We find, in the region to the right of the mirror:
$$
<(T_{uu})^2>=3<T_{uu}>^2 \mbox{ \ \ \ \ ; \ \ \ \ } <(T_{vv})^2>=0
$$
\be
<T_{uu}T_{vv}>=-\frac{1}{8\pi^2}\frac{(dv_m/du)^2}{[v-v_m(u)]^4}
\ee
The fluctuations become infinite at the position of the mirror, $v=v_m(u)$.
To see how this might affect the motion of the mirror we can consider the
fluctuations in the radiation reaction force.  Specifically, let us look at
the force normal to the mirror,
\be
F_N=-F_{\mu}n^{\mu}
\ee
where $n_{\mu}n^{\mu}=-1$ and $n_{\mu}v^{\mu}=0$.  The mean squared value
is found to be:
$$
<F_{N}^{2}>=n^{\mu}v^{\nu}n^{\alpha}v^{\beta}[<T_{\mu \nu}(-)T_{\alpha \beta}
(-)>+<T_{\mu \nu}(+)T_{\alpha \beta}(+)>
$$
\be
-2<T_{\alpha \beta}(-)><T_{\mu \nu}(+)>]\mbox{ \ \ \ \ }.
\ee
This is clearly infinite; for instance, the second term contributes
\be
\left. 
n^u n^v v^u v^v <T_{uu}(+)T_{vv}(+)>=\frac{1}{8\pi^2}\frac{(dv_m/du)^2}
{[v-v_m(u)]^4}\right|_{v=v_m(u)}.
\ee
The fluctuations diverge even for a stationary mirror.
In the spirit of the semiclassical approach we might have attempted to model 
the back reaction by solving
\be
ma_{\mu}=F_{\mu}.
\ee
However, in the case of the moving mirror this can not be interpreted as the
leading term of a fully quantum mechanical treatment, since the fluctuations
diverge.  Instead, it seems more likely that a fully quantum mechanical 
treatment simply does not exist.

While these effects are most important for the moving mirror, there is no
reason to believe that fluctuations are so violent in the case of a black
hole much larger then the Planck mass.  The point is simply that spacetime
is locally flat near the horizon on a scale much larger than  the Planck
length.  The divergences in the mirror model arose from the sharp boundary
condition at the mirror, a feature which has no analog in the black hole case.

\chapter{Self-Interaction Corrections}
 
Black hole radiance \cite{hawking} was originally derived in an approximation
where  the background geometry was given, by calculating the
response of quantum fields to this (collapse) geometry.  As we have seen,
in this
approximation the radiation is thermal, and much has been made both
of the supposed depth of this result and of the paradoxes that
ensue if it is taken literally.  For if the radiation is accurately
thermal there is no connection between what went into the hole and
what comes out, a possibility which is difficult to reconcile with
unitary evolution in quantum theory -- or, more simply, with the
idea that there are equations uniquely connecting the past with the
future.   To address such questions convincingly, one must
go beyond the approximation of treating the geometry as given, and
treat it too  as a quantum variable.  This is not easy, and as
far as we know no concrete correction to the original result
has previously
been derived in spite of much effort over more than twenty years.
Here we shall calculate what is plausibly the leading correction
to the emission rate of single particles in the limit of large
Schwarzschild holes, by a method that can be generalized in several
directions, as we shall outline.

There is a semi-trivial fact about the classic results
for  black hole radiation,
that clearly prevents the radiation from being accurately
thermal.  This is the effect that
the temperature
of the hole depends upon its mass, so that in calculating the
``thermal'' emission rate one must know what mass of the black hole
to use -- but the mass is different, before and after the radiation!
(Note that a rigorous identification of the
temperature  of a hot body from its radiation,
can only be made for sufficiently high frequencies,
such that the
gray-body factors approach unity.  But it is just in this limit
that the ambiguity mentioned above is most serious.)  As we
have emphasized, this problem is particularly
quantitatively
acute for near-extremal holes --- it is a general problem for
bodies with finite heat capacity, and in the near-extremal
limit the heat capacity of the black hole vanishes.
 
To resolve the above-mentioned
ambiguity, one clearly must allow the geometry to fluctuate,
namely to support black holes of different mass.  Another point
of view is that one must take into account the {\it self-gravitational
interaction\/} of the radiation.
 
\section{The Thin Shell Model}
 
%specification of the model.
 
To obtain a
complete description of a self-gravitating particle it would be necessary to
compute the action for an arbitrary motion of the particle and gravitational
field.  While writing down a formal expression for such an object is
straightforward, it is of little use in solving a concrete problem due to the
large number of degrees of freedom present.  To arrive at a more workable
description of the particle-hole system, we will keep only those degrees of
freedom which are most relevant to the problem of particle emission from
regions of low curvature.  The first important restriction is made by
considering only spherically symmetric field configurations, and treating the
particle as a spherical shell.  This is an interesting case
since black hole
radiation into a scalar field occurs primarily in the s-wave,
and virtual transitions to higher
partial wave configurations are formally suppressed by powers of
$\hbar$\begin{footnote}{Since
we do not address the ultraviolet problems of quantum gravity
these corrections are actually infinite, but one might anticipate that
in gravity theory with satisfactory ultraviolet behavior the virtual
transitions will supply additive corrections of order 
$\frac{\omega^2 \Lambda^2}{M^4}$,
where $\Lambda$ is the effective cutoff, and $M$ is the mass of the black hole,
 but will not alter the
exponential factors we compute.}\end{footnote}.
 
Before launching into the detailed calculation, which becomes rather
intricate, it seems
appropriate briefly to describe its underlying logic.
After the truncation to s-wave, the remaining dynamics
describes a
shell of matter interacting with a black hole of fixed mass and
with itself.  (The mass as seen from infinity is the
total mass, including that from the shell variable, and is allowed
to vary.  One could equally well have chosen the
total mass constant, and allowed the hole mass to vary.)
There is effectively
one degree of freedom, corresponding to the position of the
shell, but to isolate it one must choose appropriate
variables and solve
constraints, since the original action
superficially appears to contain much more than
this.  Having done that, one obtains an effective action for the
true degree of freedom.
This effective
action is nonlocal, and its full quantization would require one to
resolve factor-ordering ambiguities, which appears very difficult.
Hence we quantize it semi-classically, essentially by
using the WKB approximation.  After doing this one arrives at a
non-linear first order partial differential equation for the phase of
the wave function.  This differential equation may be solved by the
method of characteristics.  According to this method, one solves
for the characteristics, specifies the
function to be determined along a generic initial surface
(intersecting the
characteristics transversally), and evolves the function away from
the initial surface, by integrating the action along the characteristics.
(For a nice brief account of this,
see \cite{whitham}.)
 
When the background geometry is regarded as
fixed the characteristics for particle motion are simply the
geodesics in that geometry, and they are essentially
independent of
the particle's
mass or energy --- principle of equivalence --- except that
null geodesics are used for massless particles, and
timelike geodesics for massive particles.  Here we find that the
characteristics depend on the mass and energy in a highly non-trivial
way.  Also the action along the
characteristics, which would be zero for a massless particle
and proportional to the length for a massive particle, is
now a much more
complicated expression.
Nevertheless we can solve the equations, to obtain the
proper modes for our
problem.
 
Having obtained the modes, the final step is to identify the state
of the quantum field --- that is, the occupation of the modes ---
appropriate to the physical conditions we
wish to describe.  We do this
by demanding that a freely falling observer passing through the
horizon see no singular behavior, and that
positive frequency modes are unoccupied in the distant past.  This,
it has been argued,
is plausibly the appropriate prescription for the state of the
quantum field excited by collapse of matter into a black hole, at
least in so far as it leads to late-time radiation.  Using it, we
obtain a mixture of positive- and negative- frequency modes at
late times, which can be interpreted as a state of radiation from
the hole.  For massless scalar particles, we carry the explicit
calculation
far enough to identify the leading correction to the exponential
dependence of the radiation intensity on frequency.

\subsection{Effective Action}
 
We now derive the Hamiltonian effective action
for a self-gravitating particle
in the s-wave. 
First, we would like to explain why the Hamiltonian form of the
action is particularly well suited to our problem.
As explained above, our physical problem really contains just
one degree of freedom, but the original action appears to contain
several.  The reason of course is that
Einstein gravity is a theory with constraints and one
should only include a
subset of the spherically symmetric configurations in the
physical description, namely those satisfying the constraints.
In general, in eliminating constraints
Hamiltonian methods are more flexible than Lagrangian
methods.
This appears to be very much the case for our problem,
as we now discuss.
 
In terms of the variables appearing in the Lagrangian description,
the constraints have the form
$$
{\cal C}_{L} \left[ \rs, \rsd; g_{\mu \nu}, \dot{g_{\mu \nu}} \right] =0,
$$
where $\rs$ is the shell radius, and $ \,\dot{}\,$ represents $\frac{d}{dt}$.
When applied to the spherically symmetric, source free, solutions, one obtains
the content of Birkhoff's theorem -- the unique solution is the Schwarzschild
geometry with some mass, $M$.  Since this must hold for the regions interior
and exterior to the shell (with a different mass $M$ for each), and since $M$
must be time independent, we see that only those shell trajectories which are
``energy conserving'' are compatible with the constraints.  This feature makes
the transition to the quantum theory rather difficult, as one desires an
expression for the action valid for an arbitrary shell trajectory. This defect
is remedied in the Hamiltonian formulation, where the constraints are
expressed in terms of momenta rather than time derivatives,
$$
{\cal C}_{H} \left[ \rs, p; g_{ij}, \pi_{ij} \right] =0.
$$
At each time, the unique solution is again some slice of the Schwarzschild
geometry, but the constraints no longer prevent $M$ from being time dependent.
Thus, an arbitrary shell trajectory $\rs (t)$, $p(t)$, is perfectly consistent
with the Hamiltonian form of the constraints, making quantization much more
convenient.
 
As before, we begin by writing the metric in ADM form:
\be
ds^{2}=-N^{t}(t,r)^{2} dt^{2} + L(t,r)^{2}[dr +N^{r}(t,r)dt]^{2}
+R(t,r)^{2}[d\theta ^{2} + {\sin{\theta}}^{2} d\phi ^{2}]
\ee
In considering the above form, we have restricted ourselves to spherically
symmetric geometries at the outset.  With this choice of variables, the action
for the shell is written
\be
S^{s}=-m\int\! \sqrt{-\hat{g}_{\mu \nu} d\hat{x}^{\mu} d\hat{x}^{\nu}}
=-m\int\! dt\, \sqrt{\hat{N}^{t^{2}} -\hat{L}^{2}\left(\rsd+
\hat{N}^{r}\right)^{2}},
\ee
$m$ representing the rest mass of the shell,
and the carets instructing one to evaluate quantities at the
shell $\left(\hat{g}_{\mu \nu}=g_{\mu \nu}(\hat{t},\rs)\right)$.
 
The action for the gravity-shell system is then
\be
S= \frac{1}{16\pi} \int\!d^{4} x\, \sqrt{-g}\, {\cal R} - m \int\! dt\,
\sqrt{({\hat{N}}^{t})^{2} - {\hat{L}}^{2}(\rsd + {\hat{N}}^{r})^{2}}
+ \mbox{ boundary terms}
\ee
and can be written in canonical form as
\be
S=\int\! dt\, p\, \rsd\, + \int\! dt\, dr\, [\pi_{R}\dot{R} + \pi_{L}\dot{L}
-N^{t}({\cal H}_{t}^{s}+{\cal H}_{t}^{G}) - N^{r}({\cal H}_{r}^{s}+{\cal H}_{r}
^{G})] - \int\! dt\, M_{\mbox{{\scriptsize ADM}}}
\label{act}
\ee
with
\be
{\cal H}_{t}^{s}=\sqrt{(p/\hat{L})^{2}+m^{2}\,}\: \delta(r-\rs)
\ \ \ \ \mbox{;} \ \ \ \
{\cal H}_{r}^{s}=-p\, \delta(r-\rs)
\ee
\be
{\cal H}_{t}^{G}=\frac{L {\pi_{L}}^{2}}{2R^2} - \frac{\pi_{L} \pi_{R}}{R}
+\left(\frac{RR'}{L}\right)' - \frac{{R'}^{2}}{2L} - \frac{L}{2}
\ \ \ \ \mbox{;} \ \ \ \
{\cal H}_{r}^{G}=R'\pi_{R} - L \pi_{L}'
\label{gravcons}
\ee
where $'$ represents $\frac{d}{dr}$.
$M_{\mbox{{\scriptsize ADM}}}$ is the ADM mass of the system,
 and is numerically equal to the total mass of the
combined gravity-shell system.
 
We now wish to eliminate the gravitational degrees of freedom in order to
obtain an effective action which depends only on the shell variables.  To
accomplish this, we first identify the constraints which are
obtained by varying with respect to $N^{t}$ and $N^{r}$:
\be
{\cal H}_{t}={\cal H}_{t}^{s}+{\cal H}_{t}^{G}=0 \mbox{ \ \ \ ; \ \ \ }
{\cal H}_{r}={\cal H}_{r}^{s}+{\cal H}_{r}^{G}=0.
\label{con}
\ee
By solving these constraints, and inserting the solutions back into (\ref{act})
we can eliminate the dependence on $\pi_{R}$ and $\pi_{L}$.  We first consider
the linear combination of constraints
\be
0=\frac{R'}{L}{\cal H}_{t} + \frac{\pi_{L}}{RL} {\cal H}_{r}= -{\cal M}'
+\frac{\hat{R}'}{\hat{L}}{\cal H}_{t}^{s}+\frac{\hat{\pi}_{L}}{\hat{R}\hat{L}}
{\cal H}_{r}^{s}
\label{mcon}
\ee
where
\be
{\cal M} = \frac{{\pi_{L}}^{2}}{2R} + \frac{R}{2} - \frac{R{R'}^{2}}{2L^{2}}.
\label{mdef}
\ee
Away from the shell the solution of this constraint is simply ${\cal M}=$
constant.  By considering a static slice ($\pi_{L}=\pi_{R} = 0$), we see that
the solution is a static slice of the Schwarzschild geometry with $\cal{M}$ the
corresponding mass parameter.  The presence of the shell causes ${\cal M}$ to
be discontinuous at $\rs$, so we write
$$
{\cal M}=M \ \ \ \ r<\rs
$$
\be
{\cal M}=M_{+} \ \ \ \ r>\rs.
\ee
As there is no matter outside the shell we also have
$M_{\mbox{{\scriptsize ADM}}}=M_{+}$.
Then, using (\ref{mcon}) and (\ref{mdef}) we can solve the constraints to find
$\pi_{L}$ and $\pi_{R}$:
$$
\pi_{L}=R \mo{R'} \mbox{ \ \ \ ; \ \ \ } \pi_{R} = \frac{L}{R'} \pi_{L}'
\ \ \ \ \ \ \ \ \ \ \ r<\rs
$$
\be
\pi_{L}=R\mop{R'} \mbox{ \ \ \ ; \ \ \ } \pi_{R}=\frac{L}{R'} \pi_{L}';
\ \ \ \ \ \ \ \ \ \ \ r>\rs.
\label{mom}
\ee
The relation between $M_{+}$ and $M$ is found by solving the constraints at the
position of the shell.  This is done most easily by choosing coordinates such
that $L$ and $R$ are continuous as one crosses the shell, and $\pi_{R,L}$
are free of singularities there.  Then, integration of the constraints
across the shell yields
$$
\pip - \pim = -p/ \hat{L}
$$
\be
\Rp - \Rm = - \frac{1}{\hat{R}} \sqrt{p^{2} + m^{2} \hat{L}^{2}}
\label{cons}
\ee
 
Now, when the constraints are satisfied a variation of the action takes the
form
\be
dS= p\, d\rs + \int dr (\pi_{R} \delta\! R + \pi_{L} \delta\!L) - M_{+}\, dt
\label{var}
\ee
where $\pi_{R,L}$ are now understood to be given by (\ref{mom}),
and $M_{+}$ is determined by solving (\ref{cons}).
We wish to integrate the expression (\ref{var}) to find the action for an
arbitrary shell trajectory.  As discussed above, the
geometry inside the shell is taken to be fixed (namely, $M$ is held constant)
while the geometry outside the shell will vary in order to satisfy the
constraints.  It is easiest to integrate the action by initially varying the
geometry away from the shell.  We first consider starting from an arbitrary
geometry and varying $L$ until $\pi_{R}=\pi_{L}=0$, while holding
$\rs, p, R, \hat{L}$ fixed:
\be
\begin{array}{l}
\int\! dS = \int_{\rmin}^{\infty}\!dr \int_{\pi = 0}^{L}\! \delta\!L\, \pi_{L}
\vspace{4mm}
\\
=\int_{\rmin}^{\rs-\epsilon}\! dr \int_{\pi=0}^{L}\! \delta\!L\, R \mo{R'}
+\int_{\rs+\epsilon}^{\infty}\! dr \int_{\pi=0}^{L}\! \delta\!L\, R \mop{R'} 
\vspace{4mm}
\\
= \int_{\rmin}^{\rs-\epsilon}\! dr \left[ RL \mo{R'} + RR'
\log{\left|\frac{ \smo{R'}}{\sq{2M/R}}\right|}\right]
\vspace{1mm}
\\
\ \ \ + \int_{\rs+\epsilon}^{\infty}\! dr \left[ RL \mop{R'} +
RR' \log{\left|\frac{\smop{R'}}{\sq{2M_{+}/R}}\right|}\right]
\label{varyL}
\end{array}
\ee
where the lower limit of integration, $r_{{\scriptsize min}}$,
properly extends to the
collapsing matter forming the black hole; its precise value will not be
important.  We have discarded the constant arising from the lower limit
of the $L$ integration.  In the next stage we can vary $L$ and $R$,while
 keeping
$\pi_{R,L}=0$, to some set geometry.  Since the momenta vanish, there
is no contribution to the action from this variation.
 
It remains to consider nonzero variations at the shell.  If an arbitrary
variation of $L$ and $R$ is inserted into the final expression of (\ref{varyL})
one finds
\be
dS=\int_{\rmin}^{\infty}\! dr\, [\pi_{R} \delta\!R + \pi_{L} \delta\!L]
- \left [\frac{\partial S}{\partial \hat{R}'}(\rs+\epsilon) -
\frac{\partial S}{\partial \hat{R}'}(\rs-\epsilon)\right]d\!\hat{R}
+ \frac{\partial S}{\partial M_{+}} d\!M_{+}.
\label{try}
\ee
Since $R'$ is discontinuous at the shell,
$$
 \frac{\partial S}{\partial \hat{R}'}
(\rs+\epsilon) - \frac{\partial S}{\partial \hat{R}'}(\rs-\epsilon)
$$
is nonvanishing and needs to be subtracted in order that the relations
$$
\frac{\delta S}{\delta R} = \pi_{R} \mbox{ \ \ ; \ \ }
\frac{\delta S}{\delta L} = \pi_{L}
$$
will hold.  From (\ref{varyL}), the term to be subtracted is
\pagebreak
$$
-\left[\frac{\partial S}{\partial \hat{R}'}(\rs+\epsilon) -
\frac{\partial S}{\partial \hat{R}'}(\rs-\epsilon)\right] d\!\hat{R}
\vspace{4mm}
$$
$$
=-d\!\hat{R} \hat{R} \log{\left|\frac{\smoh{\Rm}}{\sq{2M/\hat{R}}}\right|}
$$
\be
\ \ \ + d\!\hat{R} \hat{R} \log{\left|\frac{\smohp{\Rp}}{\sq{2M_{+}/\hat{R}}}\right|}.
\ee
Similarly, arbitrary variations of $L$ and $R$ induce a variation of $M_{+}$
causing the appearance of the final term in (\ref{try}).  Thus we need to
subtract
\be
\frac{\partial S}{\partial M_{+}} d\!M_{+} =
-\int_{\rs+\epsilon}^{\infty}\! dr\, L \frac{\mop{R'}}{1-2M_{+}/R} d\!M_{+}.
\ee
Finally, we  consider variations in  $p, \rs$, and $t$.  $t$ variations
simply give $dS = - M_{+} dt$. We do not need to separately consider variations
of $p$ and $\rs$, since when the constraints are satisfied their variations
are already accounted for in our expression for $S$, as will be shown.
 
Collecting all of these terms, our final expression for the action reads
$$
S= \int_{\rmin}^{\rs-\epsilon}\!dr\,\left[ RL \mo{R'}
+RR'\log{\left|\frac{\smo{R'}}{\sq{2M/R}}\right|}\right]
\vspace{2mm}
$$
$$
+\int_{\rs+\epsilon}^{\infty}\! dr\, \left[ RL \mop{R'}
+RR' \log{\left|\frac{\smop{R'}}{\sq{2M_{+}/R}}\right|}\right]
\vspace{2mm}
$$
$$
-\int\! dt\, \dot{\hat{R}} \hat{R}\left [\log{\left|
\frac{\smoh{\Rm}}{\sq{2M/\hat{R}}}\right|}\right.
\vspace{2mm}
$$
$$
  +\left.  \log{\left|\frac{\smohp{\Rp}}{\sq{2M_{+}/\hat{R}}}\right|}\right]
\vspace{2mm}
$$
\be
+ \int\! dt \int_{\rs+\epsilon}^{\infty}\! dr\, \frac{L \mop{R'}}{1-2M_{+}/R}
\dot{M_{+}} - \int\! dt\, M_{+}.
\label{mess}
\ee
To show that this is the correct expression we can differentiate it; then
 it can
be seen explicitly that when the constraints are satisfied (\ref{var}) holds.
 
We now wish to write the action in a more conventional form as the time
integral of a Lagrangian.  As it stands, the action in (\ref{mess})
is given for
an arbitrary choice of $L$ and $R$ consistent with the constraints.  There is,
of course, an enormous amount of redundant information contained in this
description, since many $L$'s and $R$'s are equivalent to each other through a
change of coordinates.  To obtain an action which only depends on the truly
physical variables $p, \rs$ we make a specific choice for $L$ and
$R$, {\it ie.} choose a gauge.  In so doing, we must respect the condition
$$
\Rp - \Rm = -\frac{1}{\hat{R}} \sqrt{p^2 + m^2 \hat{L}^2}
$$
which constrains the form of $R'$ arbitrarily near the shell.  Suppose we
choose $R$ for all $r>\rs$; then $\Rm$ is fixed by the constraint, but we
can still choose $R$ for $r<\rs-\epsilon$, in other words, away from the
shell.  We will let $R_{<}'$ denote the value of $R'$ close to the shell but
far enough away such that $R$ is still freely specifiable.  We employ the
analogous definition for $R_{>}'$, except in this case we are free to choose
$R_{>}' = \Rp$.
 
In terms of this notation the time derivative of $S$ is
$$
L=\frac{dS}{dt}=\rsd\hat{R}\hat{L}\left[ \moh{R_{<}'} - \moph{R_{>}'}
\right]
$$
$$
 \ \ \ \ \ - \dot{\hat{R}} \hat{R} \log{\left| \frac{\smoh{\Rm}}{\smoh{R_{<}'}}
\right|}
$$
\be
 \ \ \ \ +\int_{\rmin}^{\rs-\epsilon}\! dr [\pi_{R} \dot{R} +\pi_{L}\dot{L}]
+ \int_{\rs+\epsilon}^{\infty}\! dr [\pi_{R} \dot{R} + \pi_{L} \dot{L}] -M_{+}.
\label{lag}
\ee
At this point we will, for simplicity, specialize to a massless particle
($m=0$) and define $\eta = \pm = \mbox{sgn($p$)}$.  Then the constraints
(\ref{cons}) read
$$
\Rm = \Rp + \frac{\eta p }{\hat{R}}
$$
\be
\moh{\Rm} = \moph{\Rp} + \frac{p}{\hat{L} \hat{R}}.
\label{newcon}
\ee
These relations can be inserted into (\ref{lag}) to yield
$$
L= \rsd \hat{R} \hat{L} \left[ \moh{R_{<}'} - \moph{R_{>}'}\right]
$$
$$
  -\eta \dot{\hat{R}} \hat{R}\log{\left|\frac{R_{>}'/\hat{L}
- \eta \sqrt{(R_{>}'/\hat{L})^{2} -1 +2M_{+}/\hat{R}}}
{R_{<}'/\hat{L} - \eta \sqrt{(R_{<}'/\hat{L})^{2}-1+2M/\hat{R}}}\right|}
$$
\be
 \ \ \ + \int_{\rmin}^{\rs-\epsilon}\! dr [\pi_{R}\dot{R} + \pi_{L}\dot{L}]
+\int_{\rs+\epsilon}^{\infty}\! dr [\pi_{R} \dot{R} +\pi_{L} \dot{L}] - M_{+}.
\label{nlag}
\ee
Now we can use the freedom to choose a gauge to make (\ref{nlag}) appear as
simple as possible.  It is clearly advantageous to choose $L$ and $R$ to be
time independent, so $\pi_{R} \dot{R} + \pi_{L} \dot{L}=0$.  Also, having
$R'=L$ simplifies the expressions further.  Finally, it is crucial that the
metric be free of coordinate singularities.  A gauge which conveniently
accommodates these features is
$$
L=1 \ \ ; \ \ \ R=r
$$
The $L=1, R=r$ gauge reduces the Lagrangian to
\be
L=\rsd [\sqrt{2M \rs} - \sqrt{2M_{+} \rs}] -\eta \rsd \rs
\log{\left|\frac{\sqrt{\rs}-\eta \sqrt{M_{+}}}{\sqrt{\rs}- \eta \sqrt{2M}}
\right|} - M_{+}
\label{laga}
\ee
where $M_{+}$ is now found from the constraints (\ref{newcon}) to be related to
$p$ by
\be
p= \frac{M_{+} - M}{\eta - \sqrt{2M_{+}/r}}.
\ee
The canonical momentum conjugate to $\rs$ obtained from \ref{laga} is
\be
p_{c}= \frac{\partial L}{\partial \rsd} = \sqrt{2M\rs} - \sqrt{2M_{+}\rs}
- \eta \rs \log{\left|\frac{\sqrt{\rs}-\eta \sqrt{2M_{+}}}{\sqrt{\rs}
-\eta \sqrt{2M}}\right|}
\label{pcan}
\ee
in terms of which we write the action in canonical form as
\be
S=\int\! dt [p_{c} \rsd - M_{+}]
\label{Scan}
\ee
which identifies $M_{+}$ as the Hamiltonian.  We should point out that $M_{+}$
is the Hamiltonian only for a restricted set of gauges.  If we look back at
(\ref{nlag}) we see that the terms $\pi_{R} \dot{R} + \pi_{L} \dot{L}$ will in
general contribute to the Hamiltonian.

\subsection{Quantization}
 
In this section we discuss the quantization of the effective action
 (\ref{Scan}).
First, it is convenient to rewrite the action in a form which explicitly
separates out the contribution from the particle.  We write
$$
M_{+}=M-p_{t}
$$
so
\be
S=\int\! dt [p_{c} \rsd +p_{t}]
\label{Act}
\ee
and the same substitution is understood to be made in (\ref{pcan}).  We have
omitted a
term, $\int\! dt M$, which simply contributes an overall constant to our
formulas.  In order to place our results in perspective, it
is useful to step back and consider the analogous expressions in flat space.
Our results are an extension of
\be
p=\pm\sqrt{{p_{t}}^{2} -m^{2}}
\label{pflat}
\ee
\be
S=\int\! dt [p\dot{r}+p_{t}].
\ee
Indeed, the $G \rightarrow 0$ limit of (\ref{pcan}), (\ref{Scan})
yields precisely these expressions  (with \linebreak $m=0$).
To quantize, one is tempted to
insert the substitutions $p \rightarrow -i\frac{\partial}{\partial r}$,
$p_{t} \rightarrow -i\frac{\partial}{\partial t}$ into (\ref{pflat}),
so as to satisfy the canonical commutation relations.
This
results in a rather unwieldy, nonlocal differential equation.  In this
trivial case we know, of course, that the correct description of the particle
is obtained by demanding locality and squaring both sides of (\ref{pflat})
before substituting $p$ and $p_{t}$.  So for this example  it is
straightforward to move from the point particle description
to the field theory
description, {\it i.e}. the Klein-Gordon equation.
Now, returning to (\ref{pcan})
we are again met with the question of how to implement the substitutions
$p \rightarrow -i \partial$.  In this case the difficulty is more severe;
we no longer have locality as a guiding criterion instructing us how to
manipulate (\ref{pcan}) before turning the $p$'s into differential operators.
This is because we expect the effective action (\ref{Act}) to be nonlocal on
physical grounds, as it was obtained by including the gravitational field of
the shell.
 
There is, however, a class of solutions to the field equations for which this
ambiguity is irrelevant to leading order, and which is sufficient to determine
the late-time radiation from a black hole.  These are the short-wavelength
solutions, which are accurately described by the geometrical optics, or WKB,
approximation.  Writing these solutions as
$$
\phi(t,r)=e^{i S(t,r)},
$$
the condition determining the validity of the WKB approximation is that
$$
|\partial S| \gg |\partial^{2}S|^{1/2}, \ |\partial^{3} S|^{1/3} \ \ldots
$$
and that the geometry is slowly varying compared to $S$.  In this regime,
derivatives acting on $\phi(t,r)$ simply bring down powers of $\partial S$,
so we can make the replacements
$$
p_{c} \rightarrow \frac{\partial S}{\partial r} \mbox{ \ \ \ ; \ \ \ }
p_{t} \rightarrow \frac{\partial S}{\partial t}
$$
and obtain a Hamilton-Jacobi equation for $S$.  Furthermore, it is well known
that the solution of the Hamilton-Jacobi equation is just the classical action.
So, if $\rs (t)$ is a solution of the equations of motion found by
extremizing (\ref{Act}), then
\be
S\left(t,\rs (t)\right)= S\left(0,\rs (0)\right) +
\int_{0}^{t}\! dt \left[p_{c}\left(\rs (t) \right) \rsd (t) + p_{t}  \right]
\label{phase}
\ee
where
\be
p_{c}\left(0,\rs \right) = \frac{\partial S}{\partial r}\left(0,\rs \right).
\label{pinit}
\ee
Since the Lagrangian in (\ref{Act}) has no explicit time dependence, the
Hamiltonian $p_{t}$ is conserved.  Using this fact, it is easy to verify that
the trajectories, $\rs (t)$, which extremize (\ref{Act}) are simply the null
geodesics of the metric
\be
ds^{2}=-dt^{2}+\left(dr+\sqrt{\frac{2M_{+}}{r}} dt\right)^{2}.
\ee
    From (\ref{geo}) the geodesics are:
$$
\mbox{ingoing: \ \ \ } t+\rs (t)+2\sqrt{2M_{+} \rs (t)} + 4M_{+}
\log{[\sqrt{\rs (t)}+\sqrt{2M_{+}}]}
$$
$$
\hspace{20mm} =\rs (0)+2\sqrt{2M_{+}\rs (0)} +4M_{+}\log{[\sqrt{\rs (0)}+\sqrt{2M_{+}}]}
\vspace{4mm}
$$
$$
\mbox{outgoing: \ \ \ } t-\rs (t) -2\sqrt{2M_{+}\rs (t)}-4M_{+}
\log{[\sqrt{\rs (t)}-\sqrt{2M_{+}}]}
$$
\be
\hspace{22mm}=-\rs (0)-2\sqrt{2M_{+} \rs (0)}-4M_{+}\log{[\sqrt{\rs (0)}-\sqrt{2M_{+}}]}.
\label{geodef}
\ee
$M_{+}$, in turn, is determined by the initial condition $S(0,r)$ according to
(\ref{pcan}) and (\ref{pinit}):
$$
\hspace{-7mm}\mbox{ingoing: \ \ \ } \frac{\partial S}{\partial r}(0,\rs (0))=
\sqrt{2M\rs (0)}-\sqrt{2M_{+}\rs (0)}+\rs (0)\log{\left|\frac{\sqrt{\rs (0)}
+\sqrt{2M_{+}}}{\sqrt{\rs (0)}+\sqrt{2M}}\right|}
$$
\be
\mbox{outgoing: \ \ \ } \frac{\partial S}{\partial r}(0,\rs (0))
=\sqrt{2M\rs (0)}-\sqrt{2M_{+}\rs (0)}-\rs (0)\log{\left|\frac{\sqrt{\rs (0)}
-\sqrt{2M_{+}}}{\sqrt{\rs (0)}-\sqrt{2M}}\right|}.
\label{mpdef}
\ee
Finally, we can use this value of $M_{+}$ to determine $p_{c}(t)$:
$$
\hspace{-7mm}\mbox{ingoing: \ \ \ } p_{c}(t)=\sqrt{2M\rs (t)}-\sqrt{2M_{+}\rs (t)}
+\rs (t)\log{\left|\frac{\sqrt{\rs (t)}+\sqrt{2M_{+}}}{\sqrt{\rs (t)}
+\sqrt{2M}}\right|}
$$
\be
\mbox{outgoing: \ \ \ } p_{c}(t)=\sqrt{2M\rs (t)}-\sqrt{2M_{+}\rs (t)} -\rs (t)
\log{\left|\frac{\sqrt{\rs (t)}-\sqrt{2M_{+}}}{\sqrt{\rs (t)}-\sqrt{2M}}
\right|}.
\label{pcdef}
\ee
These formulas are sufficient to compute $S(t,r)$ given $S(0,r)$.
 
As will be discussed in the next section, the relevant solutions needed to
describe the state of the field following black hole formation are those with
the initial condition
\be
S(0,r)=kr \ \ \ \ \ \ k>0
\label{sinit}
\ee
near the horizon.  Here, $k$ must be large ($\gg1/M$) if the solution is to be
accurately described by the WKB approximation.  In fact, the relevant $k$'s
needed to calculate the radiation from the hole at late times become
arbitrarily large, due to the ever increasing redshift experienced by the
emitted quanta as they escape to infinity.  We also show in the next section
 that to compute the emission probability of a quantum of frequency $\omega$,
 we are required to find the solution for all times in the region between
$r=2M$ and $r=2(M+\omega)$.  That said, we turn to the calculation of $S(t,r)$
in this region, and with the initial condition (\ref{sinit}).  The solutions
are determined from (\ref{phase}), (\ref{geodef})-(\ref{pcdef}).  Because of
the large redshift, we only need to keep those terms in these relations which
become singular near the horizon.  We then have for the outgoing solutions:
\be
S(t,r)=k\,\rs (0)-\int_{\rs (0)}^{r}\! d\rs\,\rs \log{\left[\frac{\sqrt{\rs}
-\sqrt{2M_{+}}}{\sqrt{\rs}-\sqrt{2M}}\right]} - (M_{+}-M)t
\label{phasecal}
\ee
\be
t-4M_{+}\log{[\sqrt{r}-\sqrt{2M_{+}}]}=
-4M_{+}\log{[\sqrt{\rs(0)}-\sqrt{2M_{+}}]}
\label{tcal}
\ee
\be
k=-\rs (0) \log{\left[\frac{\sqrt{\rs (0)}-\sqrt{2M_{+}}}
{\sqrt{\rs (0)} -\sqrt{2M}}\right]}.
\label{kcal}
\ee
 
To complete the calculation, we need to invert (\ref{tcal}) and (\ref{kcal})
to find $M_{+}$ and $\rs (0)$  in terms of $t$ and $r$, and then insert these
expressions into (\ref{phasecal}).  One finds that to next to leading order,
$$
\sqrt{2M_{+}}=\sqrt{2M}+(\sqrt{r}-\sqrt{2M})\frac{(e^{k/2M'}-1)e^{-t/4M'}}
{1+(e^{k/2M'}-1)e^{-t/4M'}}
$$
\be
\sqrt{\rs (0)}=\sqrt{2M}+(\sqrt{r}-\sqrt{2M})\frac{e^{(k/2M' -t/4M')}}
{1+(e^{k/2M'}-1)e^{-t/4M'}}
\label{Mr}
\ee
where
\be
M'=M+\sqrt{2M}(\sqrt{r}-\sqrt{2M})\frac{e^{(k/2M-t/4M)}}{1+e^{(k/2M-t/4M)}}.
\label{Mprime}
\ee
Plugging these relations into (\ref{phasecal}) and keeping only those terms
which contribute to the late-time radiation, one finds after some tedious
algebra,
\be
S(t,r)=-(2M^{2}-r^{2}/2)\log{\left[1+e^{(k/2M'-t/4M')}\right]}.
\label{sol}
\ee

\subsection{Results}
 
We will now discuss the application of these results to the problem of black
hole radiance. The procedure is a slight modification of the one we discussed
in the free field theory case.  As before, the point is that there are
 two inequivalent sets of modes which need to be considered:
those which are natural from the standpoint of an observer making measurements
far from the black hole, and those which are natural from the standpoint of an
observer freely falling through the horizon subsequent to the collapse of the
infalling matter.  The appropriate modes for the observer at infinity are those
which are positive frequency with respect to the Killing time, $t$.
Writing these modes as
$$
u_{k}(r) e^{-i\omega_{k} t},
$$
$\hat{\phi} (t,r)$ reads
\be
\hat{\phi}(t,r)=\int\! dk \left[ \hat{a}_{k}u_{k}(r) e^{-i\omega_{k} t}
+ \hat{a}_{k}^{\dagger}u_{k}^{*}(r) e^{i \omega_{k} t}\right].
\label{phik}
\ee
These modes are singular at the horizon,
$$
\frac{du_{k}}{dr} \ \rightarrow \infty \mbox{ \ \ as \ \ } r \rightarrow 2M.
$$
Symptomatic of this is that the freely falling observer would measure an
infinite energy-momentum density in the corresponding vacuum state,
$$
 \langle 0_{t} | T_{\mu \nu} | {0}_{t} \rangle \rightarrow \infty
\mbox{ \ \ as \ \ } r \rightarrow 2M
$$
where $\hat{a}_{k} | 0_{t}\rangle =0$. However, we do not expect this to be
the state resulting from collapse, since the freely falling observer is not
expected to encounter any pathologies in crossing the horizon, where the local
geometry is entirely nonsingular for a large black hole.
To describe the state resulting from collapse, it is more appropriate to use
modes which extend smoothly through the horizon, and which are positive
frequency with respect to the freely falling observer.
Denoting a complete set of
such modes by $v_{k}(t,r)$, we write
\be
\hat{\phi}(t,r) \int\! dk \left[\hat{b}_{k}v_{k}(t,r)+
\hat{b}_{k}^{\dagger} v_{k}^{*}(t,r)\right].
\label{phih}
\ee
Then, the state determined by
$$
\hat{b}_{k}|0_{v}\rangle=0
$$
results in a non-singular energy-momentum density at the horizon, and so is a
viable candidate.
The operators $\hat{a}_{k}$ and $\hat{b}_{k}$ are related by the Bogoliubov
coefficients,
\be
\hat{a}_{k}=\int\! dk' \left[\alpha_{kk'}\hat{b}_{k'}+\beta_{kk'}\hat{b}_{k'}
^{\dagger}
\right]
\label{bog}
\ee
where
$$
\alpha_{kk'}=\frac{1}{2\pi u_{k}(r)}\int_{-\infty}^{\infty}\! dt\,
e^{i\omega_{k} t} v_{k'}(t,r)
$$
\be
\beta_{kk'}=\frac{1}{2\pi u_{k}(r)}\int_{-\infty}^{\infty}\! dt\,
e^{i\omega_{k} t} v_{k'}^{*}(t,r).
\label{bcoef}
\ee
Note that here we compute the Bogoliubov coefficients by performing a $t$
integration, rather than an integration over a spatial coordinate, as is
conventional.  We are forced to do this in the present case since we do not
know the spatial dependence of the definite energy modes.
The flux seen at infinity is
\be
F_{\infty}(\omega_{k})=\frac{d\omega_{k}}{2\pi}\frac{\Gamma(\omega_{k})}
{|\alpha_{kk'}/\beta_{kk'}|^{2}-1}.
\label{flux}
\ee
 
Next, we consider the issue of determining the modes $v_{k}(t,r)$.  As stated
above, we require these modes to be nonsingular at the horizon.  Since the
metric near the horizon is a smooth function of $t$ and $r$, a set of such
modes can be defined by taking their behaviour on a constant time surface,
say $t=0$, to be
$$
v_{k}(0,r)\approx e^{ikr} \mbox{ \ \ as \ \ } r\rightarrow 2M.
$$
This is, of course, the initial condition given in (\ref{sinit}).  Now, the
integrals in (\ref{bcoef}) determining the Bogoliubov coefficients depend on
the values of $v_{k}(t,r)$ at constant $r$.  Since $v_{k}$ is evaluated in the
WKB approximation, the highest accuracy will be obtained when $r$  is as close
to the horizon as possible, since that is where $v_{k}$'s wavelength is short.
On the other hand, in calculating the emission of a particle of energy
$\omega_{k}$, we cannot take $r$ to be less than $2(M+\omega_{k})$, since the
solution $u_{k}(r)e^{-i\omega_{k} t}$ cannot be extended past that point.
Therefore, we calculate the integrals with $r=2(M+\omega_{k})$.
 
The results of the previous section give us an explicit expression
for $v_{k}$. From (\ref{sol}),
\be
v_{k}(t,2(M+\omega_{k}))=
e^{iS(t,2(M+\omega_{k}))}=
e^{i(4M\omega_{k}+2\omega^{2})\log{[1+e^{(k/2M'-t/4M')}] }}
\ee
where $M'$ is
\be
M'=M+\sqrt{2M}(\sqrt{2(M+\omega_{k})}-\sqrt{2M})\frac{e^{(k/2M-t/4M)}}
{1+e^{(k/2M-t/4M)}}\approx M+\omega_{k}
\frac{e^{(k/2M-t/4M)}}{1+e^{(k/2M-t/4M)}}.
\ee
Then, the integrals are,
\be
\int_{-\infty}^{\infty} dt e^{i\omega_{k} t}e^{\pm i(4M\omega_{k'}+2\omega_{k'}
^{2 })\log{[1+e^{(k'/2M'-t/4M')}]}},
\label{integral}
\ee
the upper sign corresponding to $\alpha_{kk'}$, and the lower to $\beta_{kk'}$.
 We
can compute the integrals using the saddle point approximation.
It is readily seen
that for the upper sign, the saddle point is reached when
$$
e^{(k'/2M'-t/4M')}
\rightarrow \infty,
$$
 so $t$ is on the real axis.  For the lower sign, the saddle
point is
$$
e^{(k'/2M'-t/4M')} \approx -1/2,
$$
which, to zeroth order in $\omega_{k}$, gives
$$
t=4i\pi M + \mbox{real}
$$
and to first order in $\omega_{k}$, gives
$$
t=4i\pi(M-\omega_{k})+ \mbox{real}.
$$
Inserting these values of the saddle point into the integrands gives for the
Bogoliubov coefficients,
\be
\left|\frac{\alpha_{kk'}}{\beta_{kk'}}\right|
= e^{4\pi(M-\omega_{k})\omega
 _{k}}.
\ee
The flux of radiation from the black hole is given by (\ref{flux}),
\be
F_{\infty}(\omega_{k})=\frac{d\omega_{k}}{2\pi}\frac{\Gamma(\omega_{k})}
{e^{8\pi(M-\omega_{k})\omega_{k}}-1}.
\ee
 
There is an alternative way of viewing the saddle point calculation, which
provides additional insight into the physical origin of the radiation.  Let us
rewrite the integral (\ref{integral}) as
\be
\int_{-\infty}^{\infty}\! dt e^{i\omega_{k} t \pm i S(t,2(M+\omega_{k'}))}.
\label{newint}
\ee
The saddle point is given by that value of $t$ for which the derivative of the
expression in the exponent vanishes:
$$
\omega_{k} \pm \frac{\partial S}{\partial t} (t,2(M+\omega_{k}))=0.
$$
But $\partial S/\partial t$ is just the negative of the Hamiltonian,
$$
\frac{\partial S}{\partial t}=p_{t}=M-M_{+}
$$
so the saddle point equation becomes
$$
M_{+}=M\pm \omega_{k}.
$$
To find the corresponding values of $t$, we insert this relation into
(\ref{tcal}) and (\ref{kcal}):
\be
t=4(M\pm \omega_{k})\log{\left[\frac{\sqrt{2(M+\omega_{k})+\epsilon\,}
-\sqrt{2(M+\omega_{k})\,}}
{\sqrt{\rs (0)}
-\sqrt{2(M\pm\omega_{k})}}\right]}
\label{tsad}
\ee
\vspace{2mm}
\be
k=-\rs (0)\log{\left[\frac{\sqrt{\rs (0)}-\sqrt{2(M\pm\omega_{k})}}
{\sqrt{\rs (0)} -\sqrt{2M}}\right]},
\label{ksad}
\ee
where we have written $\rs=2(M+\omega_{k})+\epsilon$ to make explicit that
$\rs$ must lie outside the point where the solutions $u_{k}(r)$ break down.
We desire to solve for $t$ as $k\rightarrow \infty$.  For the upper choice of
sign, we find from (\ref{ksad}) that
$$
\sqrt{\rs (0)}=\sqrt{2(M+\omega_{k})}+\mbox{O}\!\left(e^{-k/2M}\right),
$$
which, from (\ref{tsad}), then shows that the corresponding value of $t$ is
purely real.
 
For the lower choice of sign we have,
$$
\sqrt{\rs (0)}=\sqrt{2(M-\omega_{k})}-\mbox{O}\!\left(e^{-k/2M}\right).
$$
Continuing $t$ into the upper half plane, we find from (\ref{tsad}) that
$$
t=4i\pi (M-\omega_{k}) + \mbox{ real.}
$$
These results of course agree with our previous findings.
 
The preceding derivation invites us to interpret
the radiation as being due to
negative energy particles propagating in imaginary time.  The particles
originate from just inside the horizon, and cross to the outside in an
imaginary time interval $4\pi (M-\omega_{k})$.  This, perhaps, helps clarify
the analogy between black hole radiance and pair production in an electric
field, which, in an instanton approach \cite{affleck}, 
is also calculated by considering
particle trajectories in imaginary time.

 Finally, let us return to the question of thermality. One might have
guessed that the correct exponential suppression factor could be
the Boltzmann factor for nominal temperature corresponding
to the mass of the hole before the radiation, after the radiation,
or somewhere in between.  Thus one might have guessed that the
exponential suppression of the radiance could take the form
$e^{-\omega/T_{\rm before}}$, $e^{-\omega/T_{\rm after}}$, or
something in between.  Our result, to lowest order, corresponds
to the nominal temperature for emission being equal
to  $T_{\rm after}$.

\section{Corrections to Charged Black Hole Radiance}
 
In this section, two additional
things are done.  First, we extend the calculations
to include a charged black hole, and charged matter.  Although this step
does not present any significant formal difficulties,
the physical results we obtain are considerably richer
than what we found in our previous calculations involving neutral holes
and shells.  In the neutral case the final result could be
summarized as a simple replacement of the nominal temperature governing
the radiation by the Hawking temperature for the mass after
radiation, so that the ``Boltzmann factor'' governing emission of energy
$\omega$ from a hole of mass $M$ became
\be
e^{-\omega / T_{\rm eff.} } ~=~ e^{-\omega 8\pi (M -\omega )}~.
\label{teff}
\ee
Note that the argument of the exponential is {\it not\/}
simply proportional to the energy $\omega$, so that the spectrum is
not, strictly speaking, thermal.   While
the deviation from thermality is important in principle its structure,
in this case, is
rather trivial,
and one is left wondering whether that is
a general result.   Fortunately we find that for charged holes the
final results are much more complex.  We say ``fortunately'', not only
because this relieves us of the nagging fear that we have done a
simple calculation in a complicated way, but also for more physical
reasons.
For one knows on general grounds
that the thermal description of black
hole radiance breaks down completely for near-extremal holes.
One might anticipate, therefore, that something more drastic than a simple
modification of the nominal temperature will occur -- as indeed we
find.  A particularly gratifying consequence of the accurate formula is
a form of ``quantum cosmic censorship''.  Whereas a literal application
of the conventional thermal formulas for radiation yields a non-zero
amplitude for radiation past extremality -- that is, radiation leaving
behind a hole with larger charge than mass -- we find (within our
approximations) {\it vanishing\/} amplitude for such processes.
 
Second, we discuss in a more detailed fashion the relationship between
our method of calculation, which proceeds by reduction to an effective
particle theory, and more familiar approximations.  We show that it amounts
to saturation of the functional integral of the
underlying s-wave field theory with one-particle intermediate
states, or alternatively to neglect of vacuum polarization.  It is
therefore
closely related to conventional eikonal approximations.  We demonstrate
the reduction of the field theory to a particle theory
explicitly in the related problem of particle creation
by a strong spherically symmetric charge source,
which is a problem of independent
interest.
 
\subsection{Self-Interaction Correction}
 
Our system consists of a matter shell of rest mass m and charge q interacting
with the electromagnetic and gravitational fields.  The corresponding action is
\be
S=\int [ -m \sqrt{-\hat{g}_{\mu \nu} d\hat{x}^{\mu} d\hat{x}^{\nu}}
+ q \hat{A}_{\mu} d\hat{x}^{\mu}]
+\frac{1}{16\pi} \int\! d^4\! x \sqrt{-g}\, [{\cal R}-F_{\mu \nu}F^{\mu \nu}]
\label{acta}
\ee
The gravitational contribution to the Hamiltonian is the same as in 
(\ref{gravcons}), and the shell and electromagnetic contributions are
\be
{\cal H}_{t}^{s}=\left( \sqrt{(p/\hat{L})^{2}+m^{2}}-q\hat{A}_{t}
\right)\delta(r-\hat{r})
\mbox{ \ \ \ ; \ \ \ } {\cal H}_{r}^{s}=-p\,\delta(r-\hat{r})
\label{hshell}
\ee
\be
{\cal H}_{t}^{EM} = \frac{N^t L {\pi_{\!A_{r}}}^2}{2R^2} - A_t\,
 \pi_{\!A_{r}}'
\label{hem}
\ee
  To arrive at this
form we have chosen a gauge such that $A_t$ is the only nonvanishing component
of $A_\mu$.  Of course, we set $A_{r}=0$ only {\em after}
computing the canonical momentum $\pi_{\!A_{r}}$.
 
Constraints are found by varying the action with respect to $N^t$, $N^r$,
and $A_t$,
$$
{\cal H}_{t} \equiv {\cal H}_{t}^{s}+{\cal H}_{t}^{G}+{\cal H}_{t}^{EM}=0
\mbox{ \ \ \ ; \ \ \ } {\cal H}_{r}\equiv {\cal H}_{r}^{s} +{\cal H}_{r}^{G}=0
$$
\be
\pi_{A_r}'+q\, \delta(r-\hat{r})=0.
\ee
$\pi_R$ can be eliminated by forming the linear combination of constraints
\be
0=\frac{R'}{L}{\cal H}_{t} +\frac{\pi_L}{RL}{\cal H}_{r}
=-{\cal M}' + \frac{R'}{L}({\cal H}_{t}^{s}+{\cal H}_{t}^{EM})
+\frac{\pi_L}{RL}{\cal H}_{r}^{s}
\label{mcona}
\ee
where
\be
{\cal M}=\frac{{\pi_{L}}^2}{2R^2}+\frac{R}{2}-\frac{RR'^2}{2L^2}.
\label{mdefa}
\ee
We see from the Gauss' law constraint that $-\pi_{\!A_r}(r)$ is the charge
contained within a sphere of size $r$, so we define:
$Q(r) \equiv -\pi_{\!A_r}(r)$
 
Now, if the shell was absent  $ {\cal M}(r) $ and $Q(r)$ would be given by
\be
{\cal M}(r) = M-\int_{r}^{\infty}\! dr\, {R'(r){\cal H}_{t}^{EM}(r)\over L(r)}
\mbox{ \ \ \ ; \ \ \ } Q(r)=Q
\ee
with $M$ and $Q$ being the mass and charge of the black hole as seen from
infinity.  In the gauge $L=1, R=r$ these become
\be
{\cal M}(r)=M-Q^{2}/2r \mbox{ \ \ \ ; \ \ \ } Q(r)=Q.
\label{msola}
\ee
With the shell present we retain the expression (\ref{msola}) for the region
inside the shell, $r<\hat{r}$, whereas outside the shell we write
(with $L=1,\ R=r$),
\be
{\cal M}(r) = M_{+} - (Q+q)^{2}/2r \mbox{ \ \ \ ; \ \ \ } Q(r)=Q+q
\label{out}
\ee
where $M_+$ and $Q+q$ are the mass and charge of the hole-shell system as
measured at infinity.
 
By using the constraints we can determine $\pi_{R}$, $\pi_{L}$, and an
expression for $M_+$, in terms of the shell variables.  These relations can
then be inserted in the action to give an effective action
depending only on the shell variables.  The calculation for the present
case runs precisely parallel to the uncharged case, resulting in
$$
S=\int\! dt\left[\dot{\hat{r}}\left(\sqrt{2M\hat{r}-Q^{2}} -
\sqrt{2M_{+}\hat{r}
- (Q+q)^{2}\,}\right)\right.
$$
\be
\left.
- \eta \dot{\hat{r}}\hat{r} \log{\left|\frac{
\sqrt{\hat{r}}-\eta\sqrt{M_{+}-(Q+q)^{2}/2\hat{r}}}{\sqrt{\hat{r}}-\eta
\sqrt{M-Q^{2}/2\hat{r}}}\right|} - M_{+}\right]
\label{seff}
\ee
where $\eta \equiv$ sgn$\,(p)$, and we have now specialized to a massless shell
($m=0$).  The canonical momentum is then
\be
p_{c} = \sqrt{2M\hat{r}-Q^{2}}-\sqrt{2M_{+} \hat{r}- (Q+q)^{2}}
-\eta \hat{r} \log{\left|\frac{\sqrt{\hat{r}}-\eta \sqrt{M_{+}-(Q+q)^{2}/
2\hat{r}}}{\sqrt{\hat{r}}-\eta\sqrt{M-Q^{2}/2\hat{r}}}\right|}.
\label{pcana}
\ee
We need wish to find the short wavelength solutions which are accurately
described by the WKB approximation.  Writing these solutions as
$v(t,r)=e^{iS(t,r)}$ with $S$ rapidly varying, we can make the replacements
$$
p_{c} \rightarrow \frac{\partial S}{\partial r}
\mbox{ \ \ \ ; \ \ \ } M_{+}-M \rightarrow \frac{\partial S}{\partial t}.
$$
$S(t,r)$ satisfies the Hamilton-Jacobi equation, and so is found by computing
classical action along classical trajectories.  We first choose the initial
conditions for $S(t,r)$ at $t=0$:
\be
S_{k}^{q}(0,r)=kr.
\label{init}
\ee
We have a appended a subscript and a superscript to denote the initial
condition and charge of the solution.   The corresponding classical
trajectory has the initial condition $p_{c}=k$ at $t=0$.  $S_{k}^{q}(t,r)$ is
then given by
\be
S_{k}^{q}(t,r)=k\hat{r}(0) + \int_{\hat{r}(0)}^{r}d\hat{r}\,
 p_{c}(\hat{r}) -(M_{+}-M)t.
\label{sola}
\ee
To determine the radiance from the hole we will will only need to consider
the behaviour of the solutions near the horizon.  Furthermore, only the most
rapidly varying part of the solutions will contribute to the late-time
radiation.  With this in mind, we can write the momentum as (choosing
$\eta =1$ for an outgoing solution)
\be
p_{c}(\hat{r}) \approx -\hat{r} \log{\left| \frac{\hat{r}-R_{+}(M_{+},Q+q)}
{(\hat{r}-R_{+}(M,Q))(\hat{r}-R_{-}(M,Q))}\right|}
\label{pnew}
\ee
so that the initial condition becomes
\be
k=-\hat{r}(0) \log{ \left| \frac{(\hat{r}(0)-R_{+}(M_{+},Q+q))(\hat{r}(0)
-R_{-}(M_{+},Q+q))}{(\hat{r}(0)-R_{+}(M,Q))(\hat{r}(0)-R_{-}(M,Q))}\right|}.
\label{kinit}
\ee
Similarly, the classical trajectory emanating from $\hat{r}(0)$ is given by
approximately,
$$
t \approx \frac{2}{R_{+}(M_{+},Q+q)-R_{-}(M_{+},Q+q)} \left[
R_{+}(M_{+},Q+q)^{2} \log{\left|\frac{\hat{r}-R_{+}(M_{+},Q+q)}{\hat{r}(0)
-R_{+}(M_{+},Q+q)}\right|}\right.
$$
\be
\left. - R_{-}(M_{+},Q+q)^{2} \log{ \left| \frac{
\hat{r}-R_{-}(M_{+},Q+q)}{\hat{r}(0)-R_{-}(M_{+},Q+q)}\right|}\,\right].
\label{geoa}
\ee
These trajectories are in fact null geodesics of the metric
\be
ds^2 = -dt^2 + (dr+\sqrt{2M_{+}/r - Q^2}\, dt)^2.
\label{met}
\ee
The relations (\ref{kinit}) and (\ref{geoa}) allow us to determine $M_{+}$ and
$\hat{r}(0)$ in terms of the other variables, so that after integrating
(\ref{sola}) we can obtain an expression for $S_{k}^{q}(t,r)$ as a function
of $k$, $t$, and $r$.
 
We can now write down an expression for the field operator:
\be
\hat{\phi}(t,r)=\int\! dk\, [\hat{a}_{k}v_{k}^{q}(t,r)+\hat{b}_{k}^{\dagger}
v_{k}^{-q}(t,r)^{*}].
\label{phia}
\ee
The modes $v_{k}^{q}(t,r)$ are nonsingular at the horizon, and so the state
of the field is taken to be the vacuum with respect to these modes:
$$
\hat{a}_{k}\left|0_{v}\rangle = \hat{b}_{k} \left|0_{v}\rangle = 0.
\right.\right.
$$
Alternatively, we can consider modes which are positive frequency with
respect to the Killing time $t$.  We write these modes as
$u_{k}^{q}(r) e^{-i\omega_{k}t}$ where the $u_{k}^{q}(r)$ are singular at
the horizon, $r=R_{+}(M+\omega_{k},Q+q)$.  Then
\be
\hat{\phi}(t,r)=\int\! dk\, [\hat{c}_{k}u_{k}^{q}(r) e^{-i\omega_{k}t}
+\hat{d}_{k}^{\dagger}u_{k}^{-q}(r)^{*}e^{i\omega_{k}t}].
\label{phising}
\ee
The two sets of operators are related by Bogoliubov coefficients,
\be
\hat{c}_{k} = \int\! dk\, [\alpha_{kk'} \hat{a}_{k'} +\beta_{kk'}\hat{b}_{k'}
^{\dagger}].
\ee
  From
(\ref{phia},\ \ref{phising}) $\alpha_{kk'}$ and $\beta_{kk'}$ are found to be
$$
\alpha_{kk'} = \frac{1}{2\pi u_{k}^{q}(r)}\int_{-\infty}^{\infty}\! dt\,
e^{i\omega_{k}t}v_{k'}^{q}(t,r)
$$
\be
\beta_{kk'}=\frac{1}{2\pi u_{k}^{q}(r)}\int_{-\infty}^{\infty}\! dt\,
e^{i\omega_{k}t}v_{k'}^{-q}(t,r)^{*}.
\label{boga}
\ee
Here, $r$ is taken to be slightly outside the horizon,
$r=R_{+}(M+\omega_{k},Q+q) + \epsilon$.  These coefficients can be evaluated
in the saddle point approximation.  Recalling that $v_{k}^{q}(t,r)
=e^{iS_{k}^{q}(t,r)}$, the saddle point equation for $\alpha_{kk'}$ becomes
\be
\omega_{k}=-\frac{\partial S_{k'}^{q}}{\partial t} = M_{+}^{q} - M.
\label{alpha}
\ee
This leads to a purely real value of $t$ for the saddle point.
For $\beta_{kk'}$ we have
\be
\omega_{k}=\frac{\partial S_{k'}^{-q}}{\partial t}=M-M_{+}^{-q}.
\label{beta}
\ee
\mbox{}From (\ref{kinit},\ \ref{geoa}) we find that the saddle point
value for t has an
imaginary part given by
\be
\mbox{Im}(t_{s})=\frac{2\,R_{+}(M-\omega_{k},Q-q)^{2}}{R_{+}(M-\omega_{k},Q-q)
-R_{-}(M-\omega_{k},Q-q)}\,\pi=\frac{1}{2\,T(M-\omega_{k},Q-q)}.
\ee
Therefore,
\be
|{\beta_{kk'}/\alpha_{kk'}}|
= \frac{1}{|2\pi u_{k}(r)|}\exp{\left(\omega_{k}/T(M-
\omega_{k},Q-q) + \mbox{Im}[S_{k'}^{-q}(t_{s})^{*}]\right)}.
\ee
The terms in $S_{k'}^{-q}$ which contribute to the second term in the
exponent are
$$
\int_{\hat{r}(0)}^{r} d\hat{r}\, p_{c}(\hat{r}) +\omega_{k}\,\mbox{Im}(t_{s}).
$$
Using (\ref{pnew}-\ref{geoa}) this can be evaluated to give
\be
\mbox{Im}[S_{k'}^{-q}(t_{s})^{*}]=\frac{M\omega + \sqrt{M^{2}-Q^{2}\,}
\left(\sqrt{(M-\omega)^{2}-(Q-q)^{2}}-\sqrt{M^{2}-Q^{2}}\right)}
{2\,T(M-\omega,Q-q)\, R_{+}(M,Q)}
\ee
resulting in
\be
\left|\frac{\beta_{kk'}}{\alpha_{kk'}}\right|^{2}
=\exp{\left(-\frac{\sqrt{M^{2}-Q^{2}}\, [\omega - \sqrt{(M-\omega)^{2}
-(Q-q)^{2}}+\sqrt{M^{2}-Q^{2}\,}\,]}{T(M-\omega,Q-q)\,R_{+}(M,Q)}\right)}.
\label{bol}
\ee
This is the effective Boltzmann factor governing emission.
Sufficiently far from extremality, when $\omega$, $q \ll
\sqrt{M^{2}-Q^{2}}$, we can expand (\ref{bol}) to give
\be
\left|\frac{\beta_{kk'}}{\alpha_{kk'}}\right|^{2} \approx \exp{\left(-\frac{
\omega - \frac{Qq}{R_{+}(M,Q)} + \frac{M^{2}q^{2}+Q^{2}\omega^{2}
-2MQ\omega q}
{2(M^{2}-Q^{2})R_{+}(M,Q)}}{T(M-\omega,Q-q)}\right)}
\ee
as compared to the free field theory result \cite{hawking},
\be
\left|\frac{\beta_{kk'}}{\alpha_{kk'}}\right|^{2} =
\exp{\left(-\frac{\omega-\frac{Qq}{R_{+}(M,Q)}}{T(M,Q)}\right)}.
\label{free}
\ee
Near extremality, the self-interaction corrections cause the emission to
differ substantially from (\ref{free}).
 
We might ask whether it is possible to reach extremality after a finite
number of emissions.  Since $T(M-\omega,Q-q)$ appears in the denominator
of the exponent of (\ref{bol}), the transition probability to the extremal
state is in fact zero.  We can also ask whether there are transitions to a
meta-extremal ($Q>M$) hole.  This would have rather dramatic implications
as the meta-extremal hole is a naked singularity.  To address this question
we return to the saddle point equation (\ref{beta}). When $Q>M$, $R_{+}$
and $R_{-}$ become complex.  From (\ref{kinit}) we see that a saddle point
solution would require that $k$ be complex, but we do not allow this since a
complete family of initial conditions $S_{k}(0,r)=kr$ was defined with $k$
real.  Therefore, in the saddle point approximation the extremal hole is
stable.
 
Modes with $|\beta /\alpha | > 1$ formally require larger amplitudes
for higher occupation numbers, and thus require special interpretation.
Considering for simplicity the free field form of these
coefficients, (\ref{free}), we see that such modes occur when
$\omega < q Q/ R_+$, that is when the incremental energy gain from
discharging the Coulomb field overbalances the cost of creating
the charged particle.  Under these conditions one has dielectric
breakdown of the vacuum, just as for a uniform electric field in
empty space.  Since this physics is not our primary concern in the
present work, we shall restrict ourselves to a few remarks.
The occupation factor appearing in the formula for
radiation in these ``superradiant''
modes is negative, but the reflection probability
exceeds unity, so the radiation flux is positive as it should be.
And in general the formulas for physical quantities will appear sensible, 
although Fock space occupation
numbers are not.  We can avoid superradiance altogether by considering
a model with only
massive charged fundamental particles, and holes with a charge/mass ratio
small compared to the minimal value for fundamental quanta.
 
Another interesting variant is to consider a {\it magnetically\/} charged
hole interacting with neutral matter.  In that case, one simply
puts $q=0$ in the fomulae above (but $Q \ne 0$).  One could also consider
the interaction of dyonic holes with charged matter, and other variants
({\it e.g}. dilaton black holes) but we shall not do that here.
 
\section{Discussion}
 
We have arrived at our results by what may have appeared to be a
somewhat circuitous route.  Inspired by a field theory question, we
calculated the solutions of a single self-gravitating particle at the
horizon, and then passed back to field theory by interpreting the solutions
as the modes of a second quantized field operator.  In this section we hope
to clarify the logic of this procedure, and show that it is both correct and
efficient, by demonstrating how a single particle action emerges from the
truncation of a complete field theory.
 
We can illustrate this explicitly if we consider the simpler model of
spherically symmetric
electromagnetic and charged scalar fields interacting in flat space.  Our
goal is to show that the propagator for the scalar field can be expressed as a
Hamiltonian path integral for a single charged shell.
To achieve this, two important
approximations will be made.  The first is that the effects of vacuum
polarization will be assumed to be small, so we can  ignore scalar loop
diagrams.  The second is to assume that the dominant interactions involve
soft photons, so that the difference in the scalar particle's energy before
and after emission or absorption of a photon is small compared to the energy
itself.   Thus we expect that our expression will be valid for cases where
the scalar particle has a large energy, so that the energy transfer per
photon is relatively small, and is far from the origin, so that
the classical electromagnetic self energy of the particle is a slowly varying
function of the radial coordinate.  Field theory in this domain is in fact
well described by the eikonal approximation, which implements the same
approximations we have just outlined.  What follows is then essentially a
Hamiltonian version of the eikonal method.
 
We start from the action
$$
S=-\frac{1}{4\pi}\int \!d^{4}x\,\left[(\partial_{\mu}-iqA_{\mu})\phi^{*}
\, (\partial^{\mu}+iqA^{\mu})\phi +m^{2}\phi^{*}\phi
+\frac{1}{4}F_{\mu \nu}F^{\mu \nu}\right]
$$
$$
=\int\!dt\,dr\left[\pi_{\phi^{*}}\dot{\phi}^{*}+\pi_{\phi}\dot{\phi}
-\left(\frac{\pi_{\phi^{*}}\pi_{\phi}}{r^2}+r^{2}{\phi^{*}}'
\phi'+m^{2}r^{2}\phi^{*}\phi+\frac{{\pi_{\!A_{r}}}^{2}}{2r^2}
\right)\right.
$$
\be
\left.
-A_{t}\left(iq[\pi_{\phi^{*}}\phi^{*}-\pi_{\phi}\phi]-\pi_{\!A_{r}}'
\right)\right].\mbox{ \ \ \ \ }
\ee
Defining the charge density
\be
\rho(r)\equiv iq[\pi_{\phi^{*}}(r)\phi^{*}(r)-\pi_{\phi}(r)\phi(r)]
\ee
the solution of the Gauss' law constraint is
\be
Q(r)\equiv -\pi_{\!A_{r}}=\int_{0}^{r}\!dr\,\rho(r)
\ee
and so the scalar field Hamiltonian becomes
\be
H=\int_{0}^{\infty}\!dr\left[
\frac{\pi_{\phi^{*}}\pi_{\phi}}{r^2}+r^{2}{\phi^{*}}'\phi'+m^{2}\phi^{*}\phi
+\frac{Q(r)^2}{2r^2}\right].
\ee
The fields are now written as second quantized operators:
$$
\hat{\phi}=\int\!\frac{dk}{\sqrt{2\pi\,2\omega_k}}\,
\frac{[\ak e^{ikr}+\bkd e^{-ikr}]}{r}
$$
\be
\hat{\pi}_{\phi}=i\int\!\frac{dk}{\sqrt{2\pi}}\sqrt{\frac{\omega_k}{2}}\,r\,
[\akd e^{-ikr}-\bk e^{ikr}]
\ee
where $\omega_{k}=\sqrt{k^{2}+m^{2}}$, and we also have $\hat{\phi}^{*}
=\hat{\phi}^{\dagger}$ , $\hat{\pi}_{\phi^{*}}=\hat{\pi_{\phi}}^{\dagger}$.  To
ensure that the field is nonsingular at the origin we impose the conditions
$\hat{a}_{-k}=-\ak$ , $\hat{b}_{-k}=-\bk$, and take the limits of all $k$
integrals to be from $-\infty$ to $\infty$.
 
We now write the Hamiltonian in terms of the creation and annihilation
operators.  In doing so we shall normal order the operators, which corresponds
to omitting vacuum polarization since we do not allow particle-antiparticle
pairs to be created out of the vacuum.
 Also when evaluating $\phi'$ we shall use the geometrical optics
approximation, $(e^{ikr}/r)'\approx ike^{ikr}/r$, valid for $k\gg 1/r$.
Then the quadratic part of the Hamiltonian becomes,
\be
\int_{0}^{\infty}\!dr\left[\frac{\hat{\pi}_{\phi^{*}}\hat{\pi}_{\phi}}{r^2}
+r^{2}  \hat{\phi}^{*}{'} \hat{\phi}'+m^{2}\hat{\phi}^{*}\hat{\phi}\right]
=\frac{1}{2}\int\!dk\,\omega_{k}[\akd\ak +\bkd\bk].
\ee
Next we consider the interaction term.  When evaluating this there will arise
factors of $\sqrt{\omega_{k'}/\omega_{k}}$.  The essence of the soft photon
approximation is that we replace these factors by $1$, since we are assuming
that $\Delta \omega/\omega \ll 1$ for the emission or absorption of a single
photon.  Then, after normal ordering, we can evaluate the charge density to
be:
\be
\hat{\rho}(r)=q\int\!\frac{dk\,dk'}{2\pi}[\akd\akp-\bkd\bkp]e^{i(k-k')r}.
\ee
 
We now wish to calculate matrix elements of the Hamiltonian between one
particle states. A basis of one particle states labelled by position is given
by
\be
|r\rangle=\int\!\!\frac{dk}{\sqrt{2\pi}}\,e^{-ikr}\,\akd|0\rangle.
\ee
The free part of the Hamiltonian then has matrix elements
\be
\langle r_{2}|\,\frac{1}{2}\int\!dk\,\omega_{k}[\akd\ak+\bkd\bk]|r_{1}\rangle
=\int\!\frac{dk}{2\pi}\,\omega_{k}[e^{ik(r_{2}-r_{1})}-e^{ik(r_{2}+r_{1})}].
\ee
The second term in the brackets corresponds to the path from $r_1$ to $r_2$
which passes through the origin.  These paths will not contribute to local
processes far from the origin, so we drop this term.  The matrix elements
of the interaction term for closely spaced points $r_1$ and $r_2$ are:
\be
\langle r_{2}|\int_{0}^{\infty}\!\!dr\,\frac{\hat{Q}(r)^{2}}{2r^2}|r_{1}\rangle
=\frac{q^{2}}{2r_{1}}\int\!\frac{dk}{2\pi}e^{ik(r_{2}-r_{1})} .
\ee
Putting these together, we find the matrix elements of the Hamiltonian,
 
\be
\langle r_{2}|\hat{H}|r_{1}\rangle=
\int\! \frac{dk}{2\pi}\, e^{ik(r_{2}-r_{1})}
(\sqrt{k^{2}+m^{2}}+q^{2}/2 r_{1}).
\ee
Now we can follow the standard route which leads from matrix elements of
the Hamiltonian to a path integral expression for the time evolution
operator, with the result
\be
\langle r_{f}|e^{-i\hat{H}t}|r_{i}\rangle
=\int_{r(0)=r_{i}}^{r(t)=r_{f}}{\cal D}p\,{\cal D}r\,
e^{i\int_{0}^{t}\!dt'\,(p\dot{r}-\sqrt{p^{2}+m^{2}}-q^{2}/2r)}.
\ee
The action in the exponent is precisely that of a charged shell, with
$q^{2}/2r$ being the electromagnetic self energy.
 
We now discuss how this analysis can be applied to the case where
we include gravitational interactions.  The resulting field Hamiltonian
is much more complex, and so we will not be able to explicitly calculate the
effective shell action.  However, the preceding derivation allows us to argue
that were we to do so, we would simply derive the effective action obtained in
section 2. The nature of the black hole radiance calculation makes us believe
that the approximations used to arrive at a shell action are justified.  This
is so because for a large ($M\gg m_{p}$) hole the relevant field
configurations are short wavelength solutions moving in a region of
relatively low curvature, and these are the conditions which we argued make
the eikonal approximation valid.
 
For simplicity, we will consider an uncharged self-gravitating scalar field.
If we truncate to the s-wave we arrive at what is known as the BCMN model,
originally considered in \cite{bcmn} and corrected in \cite{unruh}. The
action is
$$
S=\frac{1}{4\pi}\int\! d^4\!x\, \sqrt{-g}\left[\frac{1}{4}{\cal R}
- \frac{1}{2}g^{\mu \nu}\partial_{\mu}\phi\partial_{\nu}\phi\right]
$$
\be
=\int\! dt\, dr\,\left[\pi_{\phi}\dot{\phi}+\pi_{R}\dot{R}+\pi_{L}\dot{L}
-N^{t}({\cal H}_{t}^{\phi}+{\cal H}_{t}^{G})-N^{r}({\cal H}_{r}^{\phi}
+{\cal H}_{r}^{G})\right] -\int\! dt\, M_{ADM}
\label{sphi}
\ee
with
\be
{\cal H}_{t}^{\phi}=\frac{1}{2}\left(\frac{{\pi_{\phi}}^{2}}{LR^2}
+\frac{R^2}{L}{\phi '}^{2}\right) \mbox{ \ \ \ ; \ \ \ }
{\cal H}_{r}^{\phi}=\pi_{\phi}\phi'.
\ee
The analog of (\ref{mcona}) is now
\be
{\cal M}'=\frac{R'}{L}{\cal H}_{t}^{\phi}+\frac{\pi_{L}}{RL}{\cal H}_{r}^{s}
=\frac{R'}{2L^2}\left(\frac{{\pi_{\phi}}^{2}}{R^2}+R^{2}{\phi'}^{2}\right)
+\frac{\pi_{L}\pi_{\phi}\phi'}{RL}
\label{mphi}
\ee
The Hamiltonian is
\be
H=M_{ADM}={\cal M}(\infty)=M+\int_{0}^{\infty}\!dr\left[\frac{R'}{L}
{\cal H}_{t}^{\phi}+\frac{\pi_{L}}{RL}{\cal H}_{r}^{s}\right].
\ee
 
To obtain an expression for $H$ which depends only on $\phi$ and $\pi_{\phi}$
we must choose a gauge and solve the constraints.  We can obtain an explicit
result if we choose the gauge $R=r$, $\pi_{L}=0$.  Then, defining
\be
 h(r)\equiv
\frac{1}{2}\left(\frac{{\pi_{\phi}}^{2}}{r^2}+r^{2}\phi'^{2}\right),
\ee
$L$ is determined from (\ref{mphi}),
\be
{\cal M}'(r)=\left(\frac{r}{2}-\frac{r}{2L^2}\right)'=\frac{h(r)}{L^2}
\ee
so
\be
\frac{1}{L^2}=-\frac{2M}{r}e^{-2\int_{0}^{r}\!dr'\,h(r')/r'}
+\frac{1}{r}e^{-2\int_{0}^{r}\!dr'\,h(r')/r'}
\int_{0}^{r}\!dr'\,e^{2\int_{0}^{r'}\!dr''\,h(r'')/r''}
\ee
which then leads to
\be
H=Me^{-2\int_{0}^{\infty}\!dr\, h(r)/r}
+\int_{0}^{\infty}\!dr\,h(r)e^{-2\int_{r}^{\infty}\!dr'\,
h(r')/r'}.
\label{phiham}
\ee
This generalizes the result of \cite{unruh} to include a nonzero mass $M$
for the pure gravity solution.  To make a direct comparison with our work in
the previous section, it would be preferable to obtain the Hamiltonian in
$L=1, R=r$ gauge.  This is more difficult and we do not know the explicit
expression.   For the moment, though, we are mainly interested in the
qualitative structure of the Hamiltonian, and (\ref{phiham}) will be
sufficient for our purposes.  The various nonlocal
terms contained in the Hamiltonian (\ref{phiham}) correspond to gravitons
attaching onto the particle's worldline.
If we expand the exponentials in (\ref{phiham}), we see that
there arise an infinite series of bi-local, tri-local,  \ldots, terms resulting
from the non-linearity of gravity. Now we could, in principle, repeat the
analysis which led to an effective shell action  for the charged field in
flat space.   In that case the calculation could be done with only modest
effort because there was only a single quartic interaction term. In the
present case we would have to sum the infinite series of terms that arise;
our point is that handling all of these terms is cumbersome,
to say the least, and that it is much simpler to proceed as in
section 2.

\section{Multi-particle Correlations}

We have seen how to obtain the self-interaction correction to the probability
of single particle emission.  The natural next step would be to obtain similar
results for multi-particle processes, involving some combination of ingoing
and outgoing particles.  The ultimate goal, of course, it to construct a
complete S-matrix relating arbitrary in and out states.  This brings to the
fore what is usually regarded as the central conceptual puzzle of black hole
physics: does such an S-matrix exist which unitarily relates states described
on ${\cal J}^+$ and ${\cal J}^-$, or is there ``information loss'' is the sense
that pure states on ${\cal J}^-$ can evolve into mixed states on ${\cal J}^+$?
While the latter possibility clearly violates a tenet of quantum physicists,
there is at present no satisfactory proposal as to how the former possibility
can be realized.

If information is preserved in gravitational collapse, and an S-matrix exists,
it will require the existence of intricate correlations between particles on
${\cal J}^+$ which encode the details of the state on ${\cal J}^-$.  An
understanding of how this situation might arise is currently precluded by the
fact that essentially nothing is known about how to compute {\em any}
correlations on ${\cal J}^+$, much less those which would preserve all 
information.

In free field theory, the state on ${\cal J}^+$ is described by an exactly
thermal density matrix, and no one knows how to go beyond the free field
approximation, except in the case of the single particle processes we have 
been discussing.  However, we can envision calculating the correlations between
two emitted particles by an extension of our previous methods, simply including
two shells instead of one.  Presumably, the correlations would be quite
complicated in the case of short time separation between the particles, but
upon going to to large time separations one would see the later particle being
emitted with a probability corresponding to a hole of mass $M-\omega$, where
$\omega$ is the energy of the earlier particle.  In addition, there is the 
possibility for the phases of the particles to be correlated even for large
time separation, owing to the fact that the two shells were closely spaced near
the horizon at early times.

There is no obstacle to writing fown the action for two shells,
\be
S=\int\! dt\, [p_1 \dot{r_1} +p_2 \dot{r_2} - H(r_1,r_2,p_1,p_2)].
\ee
$H$ is again given by the ADM mass, 
which can be expressed in terms of the shell
variables by solving the constraints.  The action $S_{k_1 k_2}(t,r_1,r_2)$ can
then be computed for the initial condition
\be
S_{k_1 k_2}(0,r_1,r_2)=k_1 r_1 +k_2 r_2
\ee
by integrating the Hamilton-Jacobi equation in precisely the same manner as for
the one particle case.  Presumably, all correlations between the particles are
encoded in this action; unfortunately, we do not know the code, for reasons 
that we now discuss.

In the single particle case it was possible to make progress because we know
how to pass from the first quantized description in terms of $S_k(t,r)$, to a
field description; namely by writing
\be
\phi(t,r)=\sum_{k}\,[a_k e^{iS_k(t,r)}+a_{k}^{\dagger} e^{-iS_k(t,r)}].
\label{field}
\ee
This then put the full power of Bogoliubov transformations at our disposal, 
which enabled us to determine the emission probabilities.  By constrast, in
the multi-particle case the relation between the first quantized description
and the field description is unclear --- we are not aware of any extension of
(\ref{field}) which takes into account the two particle solutions.  This
represents a major obstacle, since a proper interpretation of the theory
requires that we have a field description.  On the other hand, it seems most
likely that this problem is merely a technical one, as there is no reason to
believe that the theory is ill defined or inconsistent.

\chapter{Black Hole Entropy}

Almost all researchers agree that a black hole has an entropy equal to one
quarter the area of its event horizon, even though there is no consensus as
to what the entropy represents physically.  The most appealing possibility
is that the entropy counts the number of black hole microstates, {\em i.e.}
 the black 
hole Hamiltonian has $e^{S(M)}$ eigenvalues between $0$ and $M$.  Since it
may seem rather remarkable that this could be deduced from the study of free
field theory on a classical background geometry, we will attempt put the 
argument in perspective by recalling the corresponding derivation in ordinary 
thermodynamics.  Suppose we wish to determine the entropy of some body of 
matter.  To proceed, we would first perform the experiment of placing the
body in contact with a heat bath of temperature $T$, waiting for equilibrium
to occur, and then measuring $E(T)$, the energy of the body.  Then, assuming
that equilibrium occurs when the total number of states of the system is
maximized, we have
\be
\frac{\partial S_{\mbox{\scriptsize body}}(E)}{\partial E}=\frac{\partial
S_{\mbox{\scriptsize bath}}}{\partial E}.
\ee
Finally, since $\partial S_{\mbox{\scriptsize bath}}/\partial E = 
1/T$, we obtain
\be
S_{\mbox{\scriptsize body}}(E)=\int_{0}^{E}\frac{dE}{T(E)}.
\ee

Clearly, the derivation relies on two features: a) our ability to measure
$E(T)$, or $T(E)$, b) the assumption that in equilibrium the number of states
is maximized.  Now let is return to the black hole case.  Obviously, no one
has performed an experiment to see whether a black hole of mass $M$ is in 
equilibrium with a heat bath of temperature $1/8\pi M$.  The point is that
it appears that any nonsingular state of the black hole satisfies this
property.  We saw  that by requiring regularity of the stress energy
tensor at the horizon we were inevitably led to consider states which radiate
at the Hawking temperature.  The status of point (b) is much less clear.  Just
because the hole is in equilibrium with the heat bath we cannot conclude that
the black hole is making frequent transitions among its microstates, and that
the probability distribution of these states is Boltzmann.  A rather trivial 
example of this behaviour is provided by a perfectly reflecting mirror; 
it can certainly be in equilibrium with thermal radiation without satisfying
these other conditions.  In fact, we shall se shortly that it does not seem to
be possible to assign a Boltzmann probability distribution to the microstates
of the hole.  Nevertheless, let us assume for now that assumption (b) is
justified. Then,
\be
S_{\mbox{\scriptsize hole}}(M)=\int_{0}^{M}\!\frac{dM}{1/8\pi M}
=4\pi M^2=\frac{A}{4}.
\ee

The belief that the entropy does count microstates is bolstered by the 
existence of a completely unrelated derivation due to Gibbons and Hawking
\cite{gibbons}.
We will not go through the details but simply sketch the idea.  Their approach
relies on the fact that the partition function for a system can be expressed
as a path integral over configurations periodic in imaginary time:
\be
Z=\mbox{Tr}\,e^{-\beta H}=\int_{\mbox{\scriptsize periodic}}
{\cal D}g_{\mu \nu}e^{-S}.
\ee
Here, the path integral is over Euclidean metrics with periodicity $\Delta
\tau_{E}=i\beta$.  One then calculates the path integral in saddle point
approximation and notes that only the analytic continuation of the black
hole geometry with $\beta=8\pi M$ contributes, because other geometries have
a conical singularity at the horizon.  After calculating the action and
subtracting the flat space contribution, they obtain: 
\be
Z(\beta)=e^{A/4}e^{-\beta M}
\ee
leading to the identification $S=A/4$.  

This derivation also has serious problems, not least of which is the fact
that the Euclidean path integral for gravity may not even exist due to the
well known conformal instability.  Nevertheless, it has the virtue that the
answer agrees with the previous result.  In any event, it seems overwhelmingly
likely that the entropy has some sort of deep significance, and the most
natural interpretation is that it counts microstates. 

\section{State Counting}

If we are inclined to believe that a black hole has the enormous number of
states $e^{S}$, it behooves us to explain how these states can be understood
in terms of the underlying Hamiltonian.  It would be most satisfying if 
quantization of this Hamiltonian revealed a discrete set of states whose number
could be counted to yield the entropy.  In this section we will describe some
efforts along these lines, and the divergence problems 
\cite{thooft} which ensue.  In keeping
with the spirit of this thesis, the discussion will be confined to spherically
symmetric configurations.  While we have no particular reason to believe that
all of the states can be accounted for by considering only the s-wave, the
sorts of problems that arise seem to afflict the higher partial waves in the
same way. Pure gravity in the s-wave has no dynamical degrees of freedom, and 
so to have a nontrivial theory to quantize we include a massless scalar field.
It is a little unsettling that we are forced to include to matter in order to
calculate, since the expression for black hole entropy has no explicit 
dependence on the matter content of the world.  However, it is possible that
there {\em is} matter dependence, but it can be absorbed into the value of 
Newton's constant which appears in the entropy formula
\cite{susskind,larsen}.

As a first approach, we can try to proceed in the same way as if we were 
quantizing a soliton in flat space --- by quantizing the quadratic fluctuations
of the field about the background solution.  Considering only the quadratic
fluctuations amounts to doing free field theory.  For simplicity, we will
work in Schwarzschild coordinates and use modes of definite energy,
$e^{-i\omega(t-r_*)}$,  where $r_*=r+2m\log{(r/2M-1)}$, and write:
\be
\phi=\sum_{\omega}\,[a_{\omega}e^{-i\omega(t-r_*)}+a_{\omega}^{\dagger}
e^{i\omega(t-r_*)}].
\ee
We consider only the outgoing modes since they alone lead to the divergence
problems.  In order to count modes we must make the frequencies discrete in
some way.  The easiest thing to do is to impose periodic boundary conditions
at $r=2M+\epsilon$ and $r=L$.  The $\epsilon$ is included because the modes
become singular at the horizon, and $L$ is some arbitrarily chosen radius
outside the horizon.  We take $L\gg2M\gg\epsilon$.  
This leads to the allowed frequencies:
\be
\omega_n \approx\frac{\pi n}{M \log{(L/\epsilon)}}.
\ee
This means that there are 
\be
n=\frac{M\omega \log{(L/\epsilon)}}{\pi}
\ee
single particle states for the black hole with energies between $M$ and 
$M+\omega$.  The problem is that there is no physical reason to keep $\epsilon$
finite, but the number of states diverges as $\epsilon \rightarrow 0$.  The 
reason for this behavior is due to the fact that $r_* \rightarrow -\infty$
as $r\rightarrow 2M$, so the modes oscillate an infinite number of times before
they reach the horizon.

It seems quite possible that the infinite number of oscillations is simply a
result of using free field theory, and that once the proper self-interaction
corrections are included a finite result will be obtained.  To illustrate this,
let us recall the expression for the canonical momentum of a self-interacting
particle,
\be
p_c=\sqrt{2Mr}-\sqrt{2M_+ r}-r\log{\left|\frac{\sqrt{r}-\sqrt{2M_+}}
{\sqrt{r}-\sqrt{2M}}\right|}.
\ee
For an energy eigenstate $M_+=M+\omega$, and the corresponding mode is
\be
\psi(r,t)=e^{i\int^{r}\!p_c(r')dr'-i\omega t}.
\label{modes}
\ee
Thus the number of oscillations is finite or infinite depending on whether
$\int^{2(M+\omega)}p_c(r')dr'$ is finite or infinite.  The singular part of
$p_c$ as $r\rightarrow 2M$ is:
\be
p_c \approx -2(M+\omega) \log{[r-2(M+\omega)]}
\ee
which leads to,
\be
\int^{2(M+\omega)} p_c(r') dr' = \mbox{finite}.
\ee
The (incorrect) free field theory result is recovered by expanding in $\omega$.
Then $p_c$ has the singular part
\be
p_c=\frac{4M\omega}{r-2M}
\ee
and $\int^{2M}p_c dr$ is again infinite.  

Thus by including self-interaction we seem to have solved the divergence 
problem.  However, there is a major caveat.  The mode solutions (\ref{modes})
were obtained using the WKB approximation.  As we discussed earlier, 
the WKB approximation for a 
solution of the form $e^{iS}$ is only valid provided
\be
\left|\frac{\partial S}{\partial r}\right|\gg \left|\frac{\partial^2 S}
{\partial r^2}\right|^{\frac{1}{2}},\, 
\left|\frac{\partial^3 S}{\partial r^3}\right|
^{\frac{1}{3}} \ldots
\ee
In other words, $S$ should be rapidly oscillating but the rate of change of
oscillation must not be too large.  Applying this condition to (\ref{modes}) 
gives
\be
|p_c(r)|\gg \left|\frac{dp_c(r)}{dr}\right|^{\frac{1}{2}}
\ee
or:
\be
\left|2(M+\omega)\log{[r-2(M+\omega)]}\right|\gg 
\left|\frac{2(M+\omega)}{r-2(M+\omega)}
\right|^{\frac{1}{2}}.
\ee
This condition is not satisfied as $r\rightarrow 2(M+\omega)$.  Therefore, we
cannot trust the behavior of (\ref{modes}) near the horizon, including the
conclusion that it oscillates a finite number of times.  The most convincing
resolution would be to go beyond the WKB approximation and obtain the correct
result for $\psi$.  But, as we have discussed previously, this requires the
resolution of factor ordering problems which we do not know how to solve at 
the present time.  The most we can say is that the preceding analysis suggests
quite strongly that the divergences are connected with an improper treatment
of the self-interaction of the modes near the horizon.

The breakdown of the WKB approximation occurs when we  insist on using modes
of definite energy. Earlier, we saw how to derive a complete set of modes
which were nonsingular at the horizon and are accurately described by the
WKB approximation.  These were:
\be
u_k(t,r)=e^{iS_k(t,r)}
\label{smoothies}
\ee
where:
\be
S_{k}(t,r)=-(2M^2-r^2/2)\log{[1+e^{(k/2M' - t/4M')}]}
\ee
with $M'$ given by (\ref{Mprime}).  
The coordinates are those of (\ref{nicemetric}).  
Since these modes do
not have definite energy, there is no straightforward way to count states 
using the microcanonical ensemble.  However, we can imagine putting the black
hole in contact with a distant heat bath at the Hawking temperature.  If the
black hole behaved as an ordinary thermodynamic body we could proceed by
computing the partition function, and from that extract the entropy by 
standard thermodynamic formulas.  Since the partition function is a trace,
\be
Z=\mbox{Tr}\,(e^{-\beta H}),
\ee
it is independent of which basis we choose to describe the states, and so the
modes (\ref{smoothies}) are as good as any other. 
Our working assumption is that
we can describe all the states as fluctuations about a fixed black hole of mass
$M$.  Since in the s-wave there are no purely gravitational fluctuations, the
Fock space built on the modes (\ref{smoothies}) should provide a complete
description of these states. Therefore we can write
\be
Z=<0|e^{-\beta H}|0>+\int\! dk_1 <k_1|e^{-\beta H}|k_1>
+\int\! dk_1\, dk_2 <k_1 k_2|e^{-\beta H}|k_1 k_2> + \ldots.
\ee
$H$ is the total energy as measured at infinity, so $e^{-\beta H}$ generates
translations in imaginary time:
\be
e^{\beta H} \phi(t) e^{-\beta H}=\phi(t-i\beta).
\ee
This feature makes $Z$ easy to calculate.  The first term gives simply
\be
<0|e^{-\beta H}|0>=1
\ee
by definition of the vacuum.  The next term is more interesting,
\be
<k|e^{-\beta H}|k>=\frac{1}{2\pi}\int_{-\infty}^{\infty}\! dr\,
u_{k}^{*}(0,r) u_{k}(-i\beta,r)=\frac{1}{2\pi}\int_{-\infty}^{\infty}\!dr\,
e^{-iS_k(0,r)}e^{iS_k(-i\beta,r)}.
\ee
Let us concentrate on the behavior for large $k$.  For $k\gg M$ we have
\be
S_k(-i\beta,r) \approx 2M(r-2M)(k/2M' + i\beta/4M') \mbox{ \ \ \ \ ; \ \ \ \ }
M' \approx M+ \frac{1}{2}(r-2M).
\ee
This is strictly valid only near the horizon, but that is the only region in
which we need the solutions since we can always construct wavepackets 
localized near the horizon.  The point we now wish to stress is that
$<k|e^{-\beta H}|k>$ goes to a (nonzero) constant independent of $k$ as
$k \rightarrow \infty$.  The precise value of the constant depends on the form
of the wavepackets, but at any rate
\be
\int\! dk <k|e^{-\beta H}|k>=\infty
\ee
so that $Z$ cannot be defined.  

There is a simple physical explanation for this behavior. A nonsingular mode
has positive energy density for points outside the horizon, and negative 
energy density for points inside.  Now imagine being at fixed radius, $r$,
initially outside the horizon, and letting $k$ increase.  As $k$ increases, the
total mass inside radius $r$ increases.  However, the mass can never increase
past $M=2/r$, since if it did one would be inside the horizon, but the modes
have negative energy here and so cannot lead to an increase in mass.  So as
$k$ goes to infinity, the effect on the geometry simply goes to a constant,
which explains why the matrix element of $e^{-\beta H}$ also goes to a 
constant.  

The lesson to be learned from all of this is presumably that a black hole 
cannot be treated as an ordinary thermodynamic body in the sense that its
states are distributed according to a Boltzmann distribution when in 
equilibrium with a heat bath.  There are simply too many low energy states
localized near the horizon for this distribution to make sense.  Let us point
out, though, that this conclusion is based on the assumption that the states are
correctly described by a local quantum field theory.  This assumption may be
incorrect, and the divergences may disappear when the correct theory at short
distances, such as string theory, is taken into account. That string theory
plays a crucial role in black hole physics is argued in, for example,
Ref. ~\cite{susskind}.

\chapter{Black Holes and Quantum Tunneling}

The radiation of particles from matter evolving along a classical trajectory
has been heavily studied in recent years.  Less well studied is the radiation
accompanying quantum tunneling from one classically allowed trajectory to
another.  The following question is of interest: if a matter system impinges
upon a potential barrier with a radiation field in a certain state, what is
the state of the field given that the matter is subsequently observed to be
on the other side of the barrier?  A method to answer this question in the
context of false vacuum decay in flat space was developed by Rubakov
 \cite{Rub84} and has been generalized to include gravity as well as topology
changing processes \cite{Rub87,Kan89}.  The spectrum of radiation is
 found   by solving an imaginary time Schr\"{o}dinger equation, the occurrence
of which leads to novel features. Instead of solving field equations in real 
time, one is naturally led to consider propagation on the Euclidean solution 
interpolating between the two classical trajectories. As phase factors in real 
time are converted into exponential damping factors in imaginary time, the 
resulting particle creation can be distinctly different and is accompanied by 
the systematic supression of excitations present before tunneling.

Given this situation, it is natural to ask how the radiation from black holes
might be affected by the presence of tunneling.  If we consider a distribution
of matter, initally outside its Schwarzschild radius, which tunnels through a
potential barrier to form a black hole, the conventional calculation
 \cite{hawking} of the radiation does not apply. On the other hand, it would be
shocking if the same answer was not obtained for the radiation at late times,
as this is thought to depend only on the hole's late time geometry and not on
its history at early times.  Here we compute the radiation for this process
and show that while the Euclidean time evolution has an effect at early times,
it has none at late times so that the standard result is in fact obtained.  

In order illustrate the technique of Ref.~\cite{Rub84} in a simpler setting, we
 first study the effect of tunneling on another well known radiating system ---
the moving mirror \cite{Dav77}. We show in Sect.~(2)
 how an imaginary
 time Schr\"{o}dinger equation emerges from a Born-Oppenheimer
approximation, and use this result to calculate the shift in the spectrum of
radiated particles as a result of the tunneling.  It is shown that the
initial spectrum  is shifted to favor low energy excitations,
as is understood by realizing that the probability to tunnel is increased if
energy is transferred from the radiation to the mirror.   

In Sect.~(3) this approach is extended to include gravity in
asymptotically flat space.  A WKB approximation to the Wheeler-DeWitt equation,
as considered in \cite{Lap79,Banks85}, is used to obtain an imaginary time
Schr\"{o}dinger equation which can then be solved as before. In
 Sect.~(4) we use this result to examine the radiation from a
black hole which is formed by tunneling.  In particular, we consider the
tunneling of a false vacuum bubble, a system extensively studied in
 Refs.~\cite{Sato81} ---~\cite{Guth87}.  This example involves a
 complication due to the peculiar structure that arises;
 Refs.~\cite{Guth90,polch} show that the sequence of 
three-geometries encountered during tunneling can not be stacked together to
form a manifold.  Employing a slight modification of the
standard approach, we show how the behaviour of fields on the Euclidean
Schwarzschild manifold protects the late time radiation from being affected
by tunneling.  An intuitive reason for this is that the bubble's tunneling
 probability 
is unchanged by the presence of Hawking radiation, which involves the creation
of pairs of particles with zero total energy.  
     
\section{Tunneling Mirror}

\label{secmir}
Consider a mirror moving in a one dimensional potential in the presence of a
massless scalar field.  The Schr\"{o}dinger equation for this system is
\begin{equation}
  [\hat{H}_{m}+\hat{H}_{\phi}]\Psi[\phi,x_{m};t]=i\frac{\partial}{\partial
  t}\Psi[\phi,x_{m};t]
\end{equation}
where
\begin{equation}
\hat{H}_{m}=-\frac{1}{2m}\frac{\partial^{2}}{\partial x_{m}^{\;2}}+V(x_{m})
\end{equation}
and
\begin{equation}
\hat{H}_{\phi}=\frac{1}{2}\int_{x_{m}}^{\infty}dx\left[-\frac{\delta^{2}}
{\delta\phi(x)^{2}}+\left(\frac{d\phi}{dx}\right)^{2}\right].
\end{equation}
Note that $\Psi$ is a function of the mirror coordinate $x_{m}$, and a
functional of the field configuration $\phi(x)$.  The mirror boundary 
condition is imposed by demanding that the field vanish at $x_{m}$,
\begin{equation}
  \Psi[\phi,x_{m};t]=0\; \mbox{ \ if \ } \phi(x_{m})\neq0.
\end{equation}
The system is solved by assuming that the backreaction of the field on the
mirror is a small perturbation of the mirror's motion, and that the mass and
momenta of the mirror are large enough that it can be described by a well
localized wave packet.  In this domain the system admits a Born-Oppenheimer
approximation, which amounts to an expansion in $1/m$.  In particular, we seek
a solution to the time independent Schr\"{o}dinger equation
\begin{equation}
[\hat{H}_{m}+\hat{H}_{\phi}]\Psi[\phi,x_{m}]=E\Psi[\phi,x_{m}]
\label{e:ti}
\end{equation}
valid to zeroth order in $1/m$.  Following Refs.~\cite{Rub84,Banks85} the 
Born-Oppenheimer approximation is implemented by writing $\Psi$ in the form
\begin{equation}
\Psi[\phi,x_{m}]=\psi_{VV}(x_{m})\,e^{iS(x_{m})}\,\chi[\phi,x_{m}]
\end{equation}
where $\psi_{VV}$ is a slowly varying function to be identified with the 
Van Vleck determinant.  To lowest order in $1/m$, (\ref{e:ti}) reduces to
the Hamilton-Jacobi equation.
\begin{equation}
\frac{1}{2m}\left(\frac{dS}{dx_{m}}\right)^{2}+V(x_{m})=E
\label{e:hj}
\end{equation}
since $dS/dx_{m}$, $V(x_{m})$ and $E$ are all of order $m$. 

To zeroth order:
\begin{equation}
-\frac{i}{2m}\,\frac{d^{2}S}{dx_{m}^{\;2}}\,\psi_{VV}\,\chi[\phi,x_{m}]
-\frac{i}{m}\frac{dS}{dx_{m}}\frac{d\psi_{VV}}{dx_{m}}\chi[\phi,x_{m}]
\end{equation}
$$
-\frac{i}{m}\psi_{VV}\frac{dS}{dx_{m}}\frac{\partial}{\partial
x_{m}}\chi[\phi,x_{m}]
  +  \psi_{VV}\,\hat{H}_{\phi}\,\chi[\phi,x_{m}]=0.
$$
$\psi_{VV}$ is chosen so that the first two terms cancel, leaving
\begin{equation}
\hat{H}_{\phi}\, \chi[\phi,x_{m}]=\frac{i}{m}\frac{dS}{dx_{m}}
\frac{\partial}{\partial x_{m}}\chi[\phi,x_{m}].
\end{equation}
This can be put in a familiar form by defining the time variable $\tau(x_{m})$.
 In a classically allowed region, where $E-V(x_{m})>0$ and $dS/dx_{m}$ is real,
$ \tau $ is defined by
\begin{equation}
\frac{d\tau}{d x_{m}} = \frac{m}{dS/dx_{m}} 
\mbox{ \ \ \ allowed regions}
\end{equation}
whereas in a classically forbidden region with  $dS/dx_{m}$ imaginary,
\begin{equation}
\frac{d\tau_{E}}{dx_{m}} = i \frac{m}{dS/dx_{m}} 
\mbox{ \ \ \ forbidden regions.}
\end{equation}
The resulting zeroth order equations for $\phi$ are:
\begin{equation}
\hat{H}_{\phi} \,\chi[\phi,\tau] = i \frac{\partial}{\partial \tau}\chi[\phi,
\tau] \mbox{ \ \ \ allowed regions}
\end{equation}
\begin{equation}
-\hat{H}_{\phi} \,\chi[\phi,\tau_{E}] = \frac{\partial}{\partial \tau_{E}}
\chi[\phi,\tau_{E}] \mbox{ \ \ \ forbidden regions.}
\end{equation}
These are the fundamental equations governing the evolution of the scalar
field in the presence of the mirror.  In the allowed regions we have
recovered the time-dependent Schr\"{o}dinger equation with the postion of the
mirror playing the role of a clock, whereas in the forbidden regions we have
obtained a diffusion equation, which we shall refer to as the Euclidean
Schr\"{o}dinger equation, with the Euclidean time $\tau_{E}$ measuring the
position of the mirror in the potential barrier.

%Here is the figure!!
\renewcommand\floatpagefraction{.9}
\renewcommand\topfraction{.9}
\renewcommand\bottomfraction{.9}
\renewcommand\textfraction{.1}

\begin{figure}[htb]
\centerline{\psfig{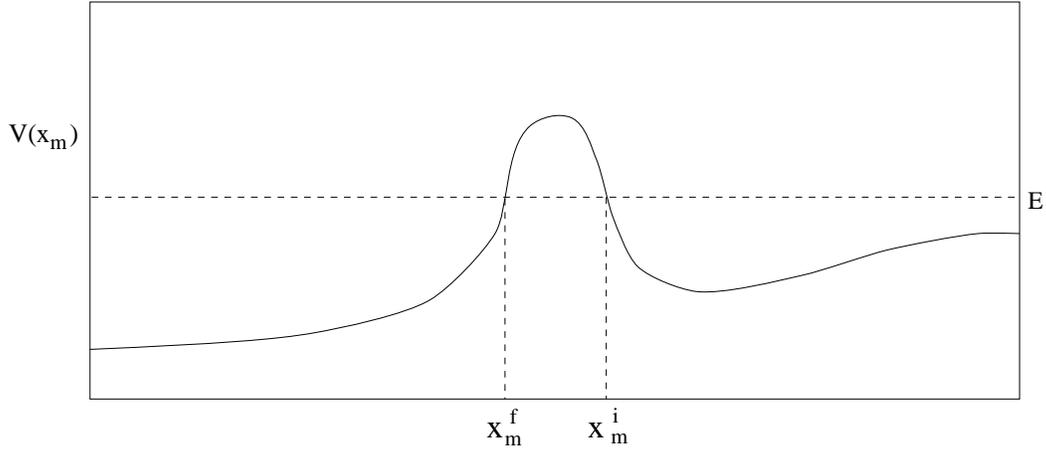}}
\caption{A generic mirror potential. The turning points for energy E are
indicated.}
\label{bubfig1}
\end{figure}

Now, choose the potential to be of the form illustrated in Fig. ~\ref{bubfig1}
 and let the
mirror come from right to left.  In the allowed region to the right of
 $x_{m}^{i}$
 the state $\chi[\phi,\tau]$ obeys the normal Schr\"{o}dinger equation, and so
standard methods can be used to find $\chi[\phi,\tau^{i}]$.  Between 
$x_{m}^{i}$ and $x_{m}^{f}$ the mirror is in a forbidden region, so the
state evolves according to
\begin{equation}
-\frac{1}{2}\int_{x_{m}(\tau_{E})}
^{\infty} dx \left[-\frac{\delta^{2}}{\delta \phi(x)^{2}} 
+ \left(\frac{d\phi}{dx}\right)^{2}\right] \chi[\phi,\tau_{E}]
=\frac{\partial}{\partial \tau_{E}} \chi[\phi,\tau_{E}]
\end{equation}
with $\chi[\phi,\tau_{E}^{i}]=\chi[\phi,\tau^{i}]$.
We wish to solve this equation  in order to find the state at the final turning
point $x_{m}^{f}$. It is useful to transform the mirror to rest by defining the
coordinate
\begin{equation}
y(x,\tau_{E})=x-x_{m}(\tau_{E})
\end{equation}
in terms of which the Euclidean Schr\"{o}dinger equation is
\begin{equation}
-\frac{1}{2} \int_{0}^{\infty} dy \left[ -\frac{\delta^{2}}{\delta \phi(y)
^{2}} + 2\,\frac{dx_{m}}{d\tau_{E}} \frac{d\phi}{dy}\frac{\delta}{\delta 
\phi(y)} + \left(\frac{d\phi}{dy}\right)^{2}\right] \chi[\phi,\tau_{E}]
=\frac{\partial}{\partial \tau_{E}} \chi[\phi,\tau_{E}]
\end{equation}
or
\begin{equation}
-\hat{H}_{\phi}^{E}(\tau_{E}) \chi[\phi,\tau_{E}] = 
\frac{\partial}{\partial \tau_{E}} \chi[\phi,\tau_{E}].
\end{equation}
The solution is
\begin{equation}
\chi[\phi,\tau_{E}]=T\exp\left[-\int_{\tau_{E}^{i}}^{\tau_{E}}
\hat{H}_{\phi}^{E}(\tau_{E}^{'}) d\tau_{E}^{'}\right] \chi[\phi,\tau_{E}^{i}]
=\hat{U}_{E}(\tau_{E},\tau_{E}^{i})\, \chi[\phi,\tau_{E}^{i}].
\end{equation}
Here T represents time ordering with respect to $\tau_{E}^{'}$.
  The crucial point
is that the Euclidean time evolution operator, $\hat{U}_{E}$, is non-unitary.
This is natural since we know that wavefunctions decay exponentially during
tunneling.  If $\hat{U}_{E}$ was unitary, the easiest way to calculate it 
would be to transform to the Heisenberg picture, solve the field equations
mode by mode, and compute Bogolubov coefficients.  However, as emphasized in
Ref.~\cite{Rub84} the non-unitarity of $\hat{U}_{E}$ implies that the 
Schr\"{o}dinger and Heisenberg pictures are inequivalent, making the standard
method inapplicable.  Instead, one can use the method developed in 
Ref.~\cite{Rub84} which closely resembles the standard one but is more general.
 We first describe the state right before tunneling.  For convenience, set
$x_{m}^{i}=\tau^{i}=\tau_{E}^{i}=0$.  Let $\xi_{\omega}(x,\tau)$ be a complete
set of positive norm solutions to the Klein-Gordon equation which vanish 
vanish at the mirror:
\begin{equation}
\left[-\frac{\partial^{2}}{\partial \tau^{2}} + \frac{\partial^{2}}
{\partial x^{2}}\right] \xi_{\omega}(x,\tau)=0
\end{equation}
\begin{equation}
i \int dx \left[\xi_{\omega}^{*}(x,\tau)\,\frac{\partial}{\partial \tau}
\xi_{\omega^{'}}(x,\tau)\,
-\frac{\partial}{\partial \tau}\xi_{\omega}^{*}(x,\tau)\xi_{\omega^{'}}
(x,\tau)\right]
=\delta_{\omega \omega^{'}}
\end{equation}
\begin{equation}
\xi_{\omega}(x_{m}(\tau),\tau)=0.
\end{equation}
The set of allowed frequencies $\omega$ is taken to be discrete, and 
$\sum_{\omega}$ represents summation over this set. The field operators can 
then be expanded in terms of these modes:
\begin{equation}
\hat{\phi}(x,\tau)=\sum_{\omega}\left[\hat{a}_{\omega} \xi_{\omega}(x,\tau)
+\hat{a}_{\omega}^{\dagger}\xi_{\omega}^{*}(x,\tau)\right]
\end{equation}
\begin{equation}
\hat{\pi}_{\phi}(x,\tau)=\frac{\partial}{\partial \tau}\hat{\phi}(x,\tau)
=\sum_{\omega}\left[
\hat{a}_{\omega}\frac{\partial}{\partial \tau}
\xi_{\omega}(x,\tau)+\hat{a}_{\omega}^{\dagger}
\frac{\partial}{\partial \tau}\xi_{\omega}(x,\tau)\right]
\end{equation}
with $\left[\hat{a}_{\omega},\hat{a}_{\omega'}^{\dagger}\right]=\delta_{
\omega \omega '}$.

 Now define Euclidean fields $\hat{\phi}^{E}(y,\tau_{E})$,
$\hat{\pi}_{\phi}^{E}(y,\tau_{E})$ which agree with $\hat{\phi}(x,\tau)$,
$\hat{\pi}_{\phi}(y,\tau)$ at $\tau=\tau^{E}=0$, but evolve according to
\begin{equation}
\hat{\phi}^{E}(y,\tau_{E})=\hat{U}_{E}^{-1}(\tau_{E},0)\,\hat{\phi}^{E}(y,0)\,
\hat{U}_{E}(\tau_{E},0)
\label{e:phi}
\end{equation}
\begin{equation}
\hat{\pi}_{\phi}^{E}(y,\tau_{E})=\hat{U}_{E}^{-1}(\tau_{E},0)\,
\hat{\pi}_{\phi}^{E}(y,0)\, \hat{U}_{E}(\tau_{E},0).
\label{e:pi}
\end{equation}
We will calculate $\hat{U}_{E}(\tau_{E},0)$ by first finding $\hat{\phi}
^{E}(y,\tau_{E})$, $\hat{\pi}^{E}_{\phi}(y,\tau_{E})$. 
 The field equations for these operators are
\begin{equation}
\frac{\partial \hat{\phi}^{E}}{\partial \tau_{E}} = -\left[\hat{\phi}^{E},
\hat{H}_{\phi}^{E}\right]=-i\hat{\pi}_{\phi}^{E}+\frac{dx_{m}}{d\tau_{E}}
\frac{\partial \hat{\phi}^{E}}{\partial y}
\end{equation}
\begin{equation}
\frac{\partial \hat{\pi}_{\phi}^{E}}{\partial \tau_{E}}=
-\left[\hat{\pi}_{\phi}^{E},\hat{H}_{\phi}^{E}\right]=
-i\frac{\partial^{2} \hat{\phi}^{E}}{\partial y^{2}}
+\frac{dx_{m}}{d\tau_{E}}\frac{\partial \hat{\pi}_{\phi}^{E}}{\partial y}.
\end{equation}
So
\begin{equation}
\hat{\pi}_{\phi}^{E}=i\left(\frac{\partial \hat{\phi}^{E}}{\partial \tau_{E}}
-\frac{dx_{m}}{d\tau_{E}}\frac{\partial \hat{\phi}^{E}}{\partial y}\right)
\end{equation}
and
\begin{equation}
\frac{\partial^{2} \hat{\phi}^{E}}{\partial \tau_{E}^{\;2}}
+\left[1+\left(\frac{dx_{m}}{d\tau_{E}}\right)^{2}\right]\frac{\partial^{2}
\hat{\phi}^{E}}{\partial y^{2}} - 2\frac{dx_{m}}{d\tau_{E}}
\frac{\partial^{2} \hat{\phi}^{E}}{\partial y \partial \tau_{E}}-\frac
{d^{2}x_{m}}{d\tau_{E}^{\;2}}\frac{\partial \hat{\phi}^{E}}{\partial y}=0.
\label{e;phi}
\end{equation}
Equation (\ref{e;phi}) can be obtained by varying the action
\begin{equation}
S=\frac{1}{2}\int dy\, d\tau_{E}\, \sqrt{g_{E}\,}\, g_{E}^{\mu \nu}
\partial_{\mu}\phi\partial_{\nu}\phi
\end{equation}
with the Euclidean metric 
\begin{equation}
ds_{E}^{2}=g^{E}_{\mu \nu}dx^{\mu}dx^{\nu}=d\tau_{E}^{\;2}+2\,\frac{dx_{m}
}{d\tau_{E}}\,dx\,d\tau_{E}+dx^{\;2}.
\end{equation}
$\hat{\phi}^{E}$, $\hat{\pi}_{\phi}^{E}$ can be expanded in terms of modes
$f_{\omega}$ which satisfy the Euclidean Klein-Gordon equation (\ref{e;phi})
and which vanish at $y=0$,
\begin{equation}
\hat{\phi}^{E}(y,\tau_{E})=\sum_{\omega}\hat{b}_{\omega}f_{\omega}(y,
\tau_{E})
\label{e:sta}
\end{equation}
\begin{equation}
\hat{\pi}_{\phi}^{E}(y,\tau_{E})=i\sum_{\omega}\hat{b}_{\omega}
\left(\frac{\partial}{\partial \tau_{E}}f_{\omega}(y,\tau_{E})
-\frac{dx_{m}}{d\tau_{E}}\frac{\partial}{\partial y}f_{\omega}(y,\tau_{E})
\right).
\end{equation}

 As the Euclidean Klein-Gordon equation is elliptic, one cannot in general
impose Cauchy boundary conditions at $\tau_{E}=0$  on $f_{\omega}$.  The
resulting solutions would not satisfy the mirror boundary condition.  With the
appropriate boundary conditions, either Dirichlet or Neumann, imposed at
$\tau_{E}=0$ and $\tau_{E}=\tau_{E}^{f}$, a detailed calculation is, of
course, required to find $f_{\omega}$ for a generic mirror trajectory. We shall
take the solutions as given and only use their specific forms in a region far
from the mirror, where they are simple.

Now, using the condition that the two sets of operators $\hat{\phi}$,
 $\hat{\pi}_{\phi}$ and $\hat{\phi}^{E}$, $\hat{\pi}_{\phi}^{E}$ are equal at
$\tau=\tau_{E}=0$, and taking inner products, the operators $\hat{b}_{\omega}$
can be expressed as a linear combination of $\hat{a}_{\omega}$, 
$\hat{a}_{\omega}^{\dagger}$:
\begin{equation}
\hat{b}_{\omega}=\sum_{\omega'}\left[\alpha_{\omega \omega'}\hat{a}_{\omega'}
+\beta_{\omega \omega'}\hat{a}_{\omega'}^{\dagger}\right].
\end{equation}
Then using
\begin{equation}
\hat{\phi}^{E}(y,\tau_{E}^{f})=\hat{U}_{E}^{-1}(\tau_{E}^{f},0)\,
\hat{\phi}_{E}(y,0)\,\hat{U}_{E}(\tau_{E}^{f},0)=\hat{U}_{E}^{-1}
(\tau_{E}^{f},0)\,
\hat{\phi}(y,0)\,\hat{U}_{E}(\tau_{E}^{f},0)
\end{equation}
and the analogous expression for $\hat{\pi}_{\phi}^{E}$, the following 
equations for $\hat{U}^{E}$ are obtained:\pagebreak
$$
\sum_{\omega}\sum_{\omega'}\left[\alpha_{\omega \omega'} \hat{a}_{\omega'}
+\beta_{\omega \omega'} \hat{a}_{\omega'}^{\dagger}\right]
f_{\omega}(y,\tau_{E}^{f})
$$
\begin{equation}
=\sum_{\omega}\left[\hat{U}_{E}^{-1}(\tau_{E}^{f},0)\,\hat{a}_{\omega}\,
\hat{U}_{E}(\tau_{E}^{f},0)\,\xi_{\omega}(y,0)
+\hat{U}_{E}^{-1}(\tau_{E}^{f},0)\,\hat{a}_{\omega}^{\dagger}\,
\hat{U}_{E}(\tau_{E}^{f},0)\,\xi_{\omega}^{*}(y,0)\right]
\end{equation}
and
$$
i\sum_{\omega}\sum_{\omega'}\left[\alpha_{\omega \omega'}\hat{a}_{\omega'}
+\beta_{\omega \omega'}\hat{a}_{\omega'}^{\dagger}\right]
\frac{\partial}{\partial \tau_{E}}f_{\omega}(y,\tau_{E}^{f})
$$
\begin{equation}
=\sum_{\omega}\left[\hat{U}_{E}^{-1}(\tau_{E}^{f},0)\,\hat{a}_{\omega}\,
\hat{U}_{E}(\tau_{E}^{f},0)\,\frac{\partial}{\partial \tau}\xi_{\omega}(y,0)
+\hat{U}_{E}^{-1}(\tau_{E}^{f},0)\,\hat{a}_{\omega}^{\dagger}\,
\hat{U}_{E}(\tau_{E}^{f},0)\,\frac{\partial}{\partial \tau}
\xi_{\omega}^{*}(y,0)
\right].
\end{equation}
Again taking inner products, this leads to relations of the form
\begin{equation}
\hat{U}_{E}^{-1}(\tau_{E}^{f},0)\,\hat{a}_{\omega}\,\hat{U}_{E}(\tau_{E}^{f},0)
=\sum_{\omega'}\left[u_{\omega \omega'}\hat{a}_{\omega'}
+v_{\omega \omega'}\hat{a}_{\omega'}^{\dagger}\right]
\label{e:tev}
\end{equation}
\begin{equation}
\hat{U}_{E}^{-1}(\tau_{E}^{f},0)\,\hat{a}_{\omega}^{\dagger}\,
\hat{U}_{E}(\tau_{E}^{f},0)=\sum_{\omega'}\left[w_{\omega \omega'}
\hat{a}_{\omega'}+z_{\omega \omega'}\hat{a}_{\omega'}\right].
\end{equation}
Then it can be shown that \cite{Rub84}
\begin{equation}
\hat{U}_{E}(\tau_{E}^{f},0)=\mbox{ const. }\times:\exp\sum_{\omega}
\sum_{\omega'}\left[\frac{1}{2}D_{\omega \omega'}\hat{a}_{\omega}^{\dagger}
\hat{a}_{\omega'}^{\dagger}+F_{\omega \omega'}\hat{a}_{\omega}\hat{a}
_{\omega'}+\frac{1}{2}G_{\omega \omega'}\hat{a}_{\omega}\hat{a}_{\omega'}
\right]:
\end{equation}
where the matrices $D$, $F$, and $G$ are defined by
\begin{equation}
D=vz^{-1}\mbox{ \ ; \ } F=\left(z^{T}\right)^{-1}-1\mbox{ \ ; \ }G=-z^{-1}w.
\end{equation}
The state after tunneling is then determined,
\begin{equation}
\left|\chi(\tau_{E}^{f})\right\rangle = \hat{U}_{E}(\tau_{E}^{f})
\left|\chi(0)\right\rangle
\label{end}
\end{equation}
and is expressed in terms of occupation numbers with respect to the modes
$\xi_{\omega}(y,0)$, where now $y=x-x_{m}^{f}$.  All of the information about
the final state is contained in the matrices $D$, $F$, and $G$, which are in
turn given in terms of inner products between the modes $f_{\omega}$ and
$\xi_{\omega}$.

As a simple application of these formul{\ae} we will calculate the shift in the
spectrum of outgoing particles which are far from the mirror at the time of
tunneling. It is assumed that the mirror was initially at rest and the field
in its ground state. The mirror subsequently accelerates in the potential
$V(x_{m})$ until it reaches the classical turning point $x_{m}^{i}$. It is well
known that as a result of the mirror's acceleration, a flux of outgoing 
particles is created whose spectrum is calculable by standard methods
\cite{Dav77}.  Outgoing particles far from the mirror are wavepackets composed
of superpositions of plane waves,
\begin{equation}
\xi_{\omega}(x,\tau)=\frac{1}{2\sqrt{\omega}}\,e^{-i\omega(\tau-x)}
\end{equation}
The spectrum of outgoing particles located at $x=\bar{x}\gg\omega^{-1}$
 at $\tau=0$ is written as
\begin{equation}
\sum_{\{n_{\omega}\}}S_{\bar{x}}\left(\{n_{\omega}\}\right)\left|\{n_{\omega}\}
\right\rangle
\end{equation}
where $\{n_{\omega}\}$ is a set of occupation numbers and $S_{\bar{x}}\left(
\{n_{\omega}\}\right)$ is the amplitude for the set to occur. 

Far from the mirror, the modes $f_{\omega}$ are easy to calculate since the
mirror boundary condition is irrelevant. They are of two types,
$$
f_{\omega}^{\mbox{--}}=\frac{1}{2\sqrt{\omega}}\,e^{-\omega \tau_{E}+i\omega x}
=\frac{1}{2\sqrt{\omega}}\,e^{-\omega \tau_{E}+i\omega \left(y+x_{m}(\tau_{E}
)\right)}
$$
\begin{equation}
f_{\omega}^{\mbox{+}}=\frac{1}{2\sqrt{\omega}}\,e^{\omega \tau_{E}+i\omega x}
=\frac{1}{2\sqrt{\omega}}\,e^{\omega \tau_{E}+i\omega \left(
y+x_{m}(\tau_{E})\right)}
\end{equation}
Then $\hat{\phi}$, $\hat{\pi}$ and $\hat{\phi}^{E}$, $\hat{\pi}_{\phi}^{E}$ 
are equal at $\tau=\tau_{E}=0$ if
\begin{equation}
\hat{b}_{\omega}^{\mbox{--}}=\hat{a}_{\omega} \mbox{ \ ; \ }
 \hat{b}_{\omega}^{\mbox{+}}=\hat{a}_{\omega}^{\dagger}.
\end{equation} 
Equation (\ref{e:tev}) gives:
$$
\hat{U}_{E}^{-1}(\tau_{E}^{f},0)\,\hat{a}_{\omega}\,\hat{U}_{E}(\tau_{E}
^{f}),0)
=e^{-\omega \tau_{E}^{f}+i\omega x_{m}^{f}}\,\hat{a}_{\omega}
$$
\begin{equation}
\hat{U}_{E}^{-1}(\tau_{E}^{f},0)\,\hat{a}_{\omega}^{\dagger}\,
\hat{U}_{E}(\tau_{E}^{f},0)=e^{\omega \tau_{E}^{f}+i\omega x_{m}^{f}}
\,\hat{a}_{\omega}^{\dagger}
\end{equation}
leading to
\begin{equation}
D=G=0 \mbox{ \ ; \ } F_{\omega \omega'}=\left(e^{-\omega \tau_{E}^{f}
-i\omega x_{m}^{f}}-1\right)\delta_{\omega \omega'}
\end{equation}
and
$$
\hat{U}_{E}(\tau_{E}^{f},0)=\mbox{ const. }\times :\exp\sum_{\omega}
\left[e^{-\omega \tau_{E}^{f}-i\omega x_{m}^{f}}-1\right]\hat{a}_{\omega}
^{\dagger}\hat{a}_{\omega}:
$$
\begin{equation}
=\mbox{ const. }\times:\exp\sum_{\omega}\left[e^{-i\omega x_{m}^{f}}
-1\right]\hat{a}_{\omega}^{\dagger}\hat{a}_{\omega}::\exp
\sum_{\omega}\left[e^{-\omega \tau_{E}^{f}}-1\right]\hat{a}_{\omega}^{\dagger}
\hat{a}_{\omega}:
\end{equation}
The first factor is a translation operator which expresses the state in terms 
of the modes $\xi_{\omega}(x,0)$ instead of $\xi_{\omega}(x+x_{m}^{f},0)$, and
the second factor acts on a state $\left|\{n_{\omega}\}\right\rangle$ to give
$e^{-E\left(\{n_{\omega}\}\right)\tau_{E}^{f}}\left|\{n_{\omega}\}
\right\rangle$, where $E\left(\{n_{\omega}\}\right)=\sum n_{\omega}\omega$
 is the energy of the state. Therefore, the state after tunneling is
\begin{equation}
\mbox{const. }\times\sum_{\{n_{\omega}\}}e^{-E\left(\{n_{\omega}\}\right)
\tau_{E}^{f}}\,S_{\bar{x}}\left(\{n_{\omega}\}\right)\left|\{n_{\omega}\}
\right\rangle.
\label{e:spec}
\end{equation}
The result of the tunneling is simply to shift the spectrum from $S_{\bar{x}}$
to $e^{-E({n_{\omega}})\tau_{E}^{f}}S_{\bar{x}}$.

It is not difficult to understand this result.  Since the total energy is fixed
, the state before tunneling is given by a superpostion of the various ways
of distributing the energy between the mirror and the radiation.  As the 
mirror's probability to tunnel depends exponentially on its energy, we expect
an inverse exponential correlation  between tunneling and energy in radiation.
Thus an observer measuring the spectrum of radiation, conditional on the mirror
tunneling, finds the result (\ref{e:spec}).
 Far from the mirror the shift in the
spectrum depends only on $\tau_{E}^{f}$, the amount of Euclidean time spent
during tunneling.  This is because the tunneling amplitude in the WKB 
approximation is $e^{-S}$, and the derivative of $S$ with respect to energy is
just the Euclidean time.

If we were to identify the Euclidean time with an inverse temperature, the 
shift would become a Boltzmann factor.  This makes it easy to generate thermal
distributions of radiation.  Specifically, if the distribution before tunneling
was a constant, then after tunneling tracing over the states of the mirror 
would yield a thermal density matrix for the radiation. A number of authors
have been led by this fact to seek a connection between the thermal radiation
that arises in the contexts of cosmology and black holes and an occurrence of
tunneling \cite{Kan89,Bro91,Cas92}.
 Such a connection relies upon assumptions about what is on the
other side of the barrier and what the spectrum of radiation is there. In this 
work we only consider situations
 where there is a well defined classical trajectory
on either side of the barrier;  we are interested in the case in which there is
collapsing matter on side of the barrier and a black hole on the other.  The
treatment of this process requires an extension of the previous method to
include gravity.  
      
\section{Application to Gravity}

\label{secgrav}
In this section we make a WKB approximation to gravity in a manner which
directly parallels that for the moving mirror. The action for gravity plus 
matter takes the form
$$
S=\frac{m_{p}^{\,2}}{16\pi}\int d^{4}x\sqrt{-g}\left({\cal R}-2\Lambda\right)
+S_{M}+\mbox{ boundary terms}
$$
\begin{equation}
=\int d^{4}x\left(\pi_{\phi_{i}}\dot{\phi}^{i}+\pi_{ij}\dot{h}^{ij}
-N^{t}{\cal H}_{t}-N_{i}{\cal H}^{i}\right)+\mbox{ boundary terms}.
\end{equation}
The Wheeler-DeWitt equation resulting from this action is
\begin{equation}
\hat{{\cal H}}_{t}\Psi=\left[-\frac{16\pi}{m_{p}^{\;2}}G_{ijkl}
\frac{\delta}{\delta h_{ij}}\frac{\delta}{\delta h_{kl}}-\frac{m_{p}^{\;2}}
{16\pi}h^{\frac{1}{2}}\left(^{3}{\cal R}-2\Lambda\right)+\hat{{\cal H}}_{t_{M}}
\right]\Psi=0
\end{equation}
 Proceeding as before, we seek a semiclassical
solution of the form
\begin{equation}
\Psi\left[h_{ij},\phi_{i}\right]=\psi_{VV}[h_{ij}]\,e^{im_{p}^{\;2}S[h_{ij}]}
\,\chi\left[\phi_{i},h_{ij}\right].
\end{equation}
At first order the Einstein-Hamilton-Jacobi equation is obtained:
\begin{equation}
\frac{16\pi}{m_{p}^{\;2}}G_{ijkl}\frac{\delta S}{\delta h_{ij}}\frac{\delta S}
{\delta h_{kl}}-\frac{m_{p}^{2\;}}{16\pi}h^{\frac{1}{2}}\left(^{3}{\cal R}-
2\Lambda \right)=0.
\label{e:ehj}
\end{equation}
Zeroth order yields
\begin{equation}
-\frac{16\pi}{m_{p}^{\;2}}i\,G_{ijkl}\frac{\delta S}{\delta h_{ij}}\frac{\delta
\chi}{\delta h_{kl}}+\hat{{\cal H}}_{t_{M}}\chi=0
\label{e:sc}
\end{equation}
provided $\psi_{VV}$ satisfies
\begin{equation}
G_{ijkl}\frac{\delta^{2}S}{\delta h_{ij} \delta h_{ij}}\psi_{VV}
+G_{ijkl}\frac{\delta S}{\delta h_{ij}}\frac{\delta \psi_{VV}}{\delta h_{kl}}
=0.
\end{equation}
The momentum constraints at first order are
\begin{equation}
\left(\frac{\delta S}{\delta h_{ij}}\right)_{|j}=0
\end{equation}
and at zeroth order are
\begin{equation}
2i\left(\frac{\delta \chi}{\delta h_{ij}}\right)_{|j}+\hat{{\cal H}}_
{M}^{i}\,\chi=0.
\label{e:mo}
\end{equation}

Equations (\ref{e:sc}) and (\ref{e:mo}) describe how the matter wave function
evolves as the spatial geometry changes. Quantum field theory in curved space
can be recovered by writing $\chi$'s dependence on $h_{ij}$ in terms of a 
time functional $\tau[x;h_{ij}]$, and by reintroducing a lapse $N^{\tau}$ and
 shift $N_{i}$, demanding that they obey
\begin{equation}
G_{ijkl}\frac{\delta S}{\delta h_{ij}}=\frac{m_{p}^{\,2}}{16\pi N^{\tau}}
\left(\int dy \frac{\delta h_{kl}}{\delta \tau[y;h_{ab}]}
-N_{i|j}-N_{j|i}\right).
\label{e:con}
\end{equation}
Then
\begin{equation}
-i\frac{16\pi}{m_{p}^{\;2}}\int\left(N^{\tau}G_{ijkl}\frac{\delta S}
{\delta h_{ij}} \frac{\delta \chi}{\delta h_{kl}}+2iN_{i}\left(\frac{\delta
 \chi}{\delta h_{ij}}\right)_{|j}\right)=-i\int\frac{\delta h_{ij}}{\delta
 \tau}\frac{\delta \chi}{\delta h_{ij}}
\end{equation}
so that the equation for $\chi$ becomes
\begin{equation}
\int d^{3}x
\left[N^{\tau}\hat{{\cal H}}_{t_{M}}+N_{i}\hat{{\cal H}}_{M}^{i}\right]
\chi[\phi_{i};\tau]= i\frac{\partial}{\partial \tau}\chi[\phi_{i};\tau].
\label{e:scg}
\end{equation}
The condition (\ref{e:con}) agrees with the classical relation between 
$\pi_{ij}$ and $h_{ij}$, demonstrating that $\tau[x;h_{ij}]$ is the classical
time and that (\ref{e:scg}) is the Schr\"{o}dinger picture version of quantum
field theory in curved space.  

As with the mirror example, $\tau$ becomes imaginary during tunneling so we
define a Euclidean time $\tau_{E}$ along with a Euclidean lapse $N^{\tau_{E}}
=iN^{\tau}$, in terms of which $\chi$ obeys
\begin{equation}
-\int dx [N^{\tau_{E}}\hat{{\cal H}}_{t_{M}}+iN_{i}\hat{{\cal H}}_{M}^{i}]\chi
[\phi_{i},\tau_{E}]=\frac{\partial}{\partial \tau_{E}}\chi[\phi_{i},\tau_{E}].
\end{equation}
For a massless scalar field with action
\begin{equation}
S=-\frac{1}{2} \int d^{4}x \sqrt{-g\,}g^{\mu \nu}\,\partial_{\mu}\phi
 \partial_{\nu}\phi,
\end{equation}
we have
\begin{equation}
\hat{{\cal H}}_{t_{M}}=\frac{1}{2}\left(h^{-\frac{1}{2}}\hat{\pi}_{\phi}^{2}
+h^{\frac{1}{2}} h^{ij} \partial_{i}\hat{\phi}\, \partial_{j} \hat{\phi}\right)
\end{equation}
\begin{equation}
\hat{{\cal H}}_{i_{M}}=\partial_{i}\hat{\phi}\, \hat{\pi}_{\phi}.
\end{equation}
To evolve $\chi$ through the tunneling region one is required to calculate the
Euclidean time evolution operator
\begin{equation}
\hat{U}_{E}(\tau_{E}^{f},\tau_{E}^{i})=T \exp\left[-\int_{\tau_{E}^{i}}
^{\tau_{E}^{f}} \hat{{\cal H}}^{E}_{\phi} d\tau_{E}\right]
\end{equation}
with
\begin{equation}
\hat{H}^{E}_{\phi}=\int d^{3}x\, [N^{\tau_{E}}(-\frac{1}{2}h^{-\frac{1}{2}}
\frac{\delta^{2}}{\delta \phi^{2}} +\frac{1}{2}h^{\frac{1}{2}}h^{ij}
\partial_{i}\phi\, \partial_{j} \phi)+N_{i} \partial_{i} \phi \frac{\delta}
{\delta \phi}].
\end{equation}
As before, one proceeds by defining Euclidean fields obeying (\ref{e:phi},\ref{e:pi}). In the present case the resulting field equations are:
$$
\left(\sqrt{g_{E}}\,g_{E}^{\mu \nu}\partial_{\mu}\hat{\phi}^{E}\right)_{,\nu}=0
$$ 
\begin{equation}
\hat{\pi}_{\phi}^{E}=i\frac{h^{\frac{1}{2}}}{N^{\tau_{E}}}\left(
\frac{\partial \hat{\phi}_{E}}{\partial \tau_{E}}-N^{i}\partial_{i}\phi\right)
\end{equation}
with
\begin{equation}
ds_{E}^{\;2}=g_{\mu \nu}^{E}\,dx^{\mu} dx^{\nu}=
\left(N^{\tau_{E}}d\tau_{E}\right)^{2}
+h_{ij}\left(dx^{i}+N_{i}d\tau_{E}\right)\!\left(dx^{j}+N^{j}d\tau_{E}\right).
\end{equation}
The evolution operator, and therefore the state after tunneling, is determined 
by solving the field equations mode by mode, and repeating the steps leading
from (\ref{e:sta}) to (\ref{end}).
\label{s:grav}  

\section{Black Hole Radiance in the Presence of Tunneling}

\label{sechole}

We can now apply this method to determine how the radiation from a black hole
is affected by tunneling. It is well known that a black hole formed
classically from collapsing matter radiates in a complicated manner at early 
times due to the time dependent geometry, but at late times will inevitably 
radiate as a black body at the Hawking temperature. Is this scenario altered 
if the black hole is formed while tunneling?  We shall show that it is not. 
The form of the late time radiation is insensitive to the hole's 
unconventional history in a way that is consistent with the intuitive picture 
of Hawking radiation being caused by pair production near the horizon.

\begin{figure}[htb]
\centerline{\psfig{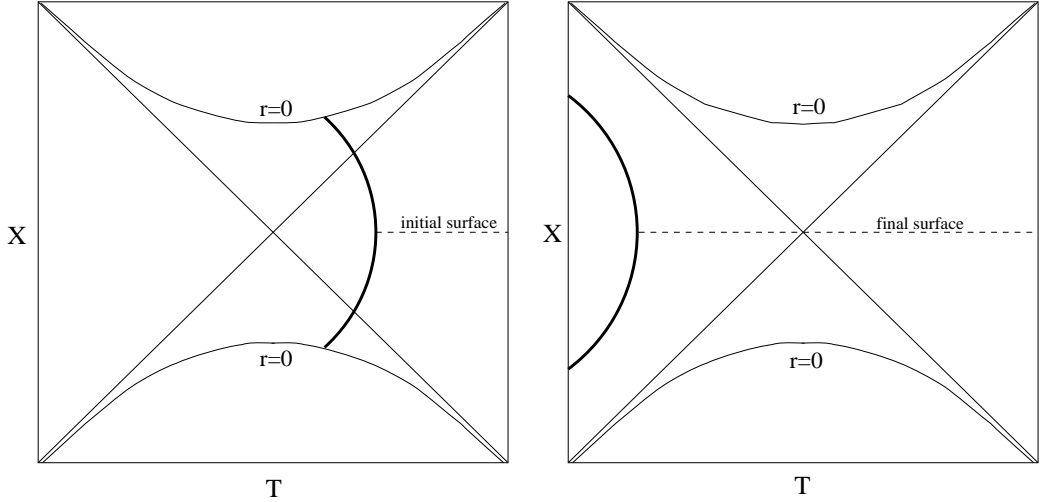}}
\caption{The type (a) and (b) solutions.  The heavy lines represent the bubble
trajectory, and the dashed lines are the initial and final surfaces of the
tunneling solution.  In these figures, only the regions to the right of the
trajectory are of interest, as they are outside of the bubble.}
\label{bubfig2}
\end{figure}

We consider the behaviour of a scalar field on the background of a false 
vacuum bubble which tunnels leading to the formation of a black hole.  The 
action for a false vacuum bubble in the thin wall approximation is
\begin{equation}
S=\frac{1}{16\pi} \int d^{4}x \sqrt{-g\,}\,{\cal R} -\frac{\Lambda_{I}}
{8\pi}\int_{bubble} \! d^{4}x \sqrt{-g\,}\, -\frac{\mu}{4\pi}\int_{wall}d^{3}A
\end{equation}
where $\Lambda_{I}$ is the cosmological constant of the false vacuum, and 
$\mu$ is the energy density of the bubble wall.  The classical solutions for
this action have been derived in Refs.~\cite{Sato81}-\cite{Guth87}.
  In what follows
we refer to the treatment of Ref.~\cite{Guth87}.  The spherically symmetric 
solutions are characterized by three parameters: $\Lambda_{I}$, $\mu$, and
the total mass $M$.  In addition, for given $\Lambda_{I}$ and $\mu$ there is
a critical mass $M_{cr}$ below which there are two solutions: type (a),
where the bubble emerges from a singularity with zero  radius, subsequently
expands to a maximum radius, and then recollapses; type (b), where the bubble
initially collapses from infinite radius, reaches a minimum radius, and then
reexpands.  Using the results of Refs.~\cite{Guth90,polch}, we focus on an 
expanding solution of type (a) which tunnels to an expanding solution of 
type (b). 
 We confine our interest to the region outside the bubble where
the metric, written in terms of Schwarzschild time $t$ and $r_{*}=r+2M
\ln({r}/{2M}-1)$, is
\begin{equation}
ds^{2}=\left(1-\frac{2M}{r}\right)\left(-dt^{2}+dr_{*}^{2}\right)+r^{2}
d\Omega^{2}.
\end{equation}
As $t$ and $r_{*}$ cover only part of the complete manifold, we introduce
Kruskal-Szekeres coordinates,
\begin{equation}
ds^{2}=\frac{32M^{3}e^{-r/2M}}{r}(-dT^{2}+dX^{2}) +r^{2}d\Omega^{2}.
\end{equation}
The two sets of coordinates are related by
$$
\left(\frac{r}{2M}-1\right)e^{r/2M}=X^{2}-T^{2}
$$
\begin{equation}
t=\left\{\begin{array}{ll}
4M\tanh^{-1}(T/X) & \mbox{if \, $|T/X|<1$}\\
4M\tanh^{-1}(X/T) & \mbox{if \, $|T/X|>1$}
\end{array}\right.
\end{equation}
Using these cordinates the type (a) and (b) solutions of interest are depicted
in Fig. ~\ref{bubfig2}.

%Here is the figure!!
\renewcommand\floatpagefraction{.9}
\renewcommand\topfraction{.9}
\renewcommand\bottomfraction{.9}
\renewcommand\textfraction{.1}

The tunneling amplitude for this process has been computed by two different
methods.  In Ref.~\cite{polch} the solution to the Wheeler-DeWitt equation 
is found in the
WKB approximation by solving the Einstein-Hamilton-Jacobi equation
 (\ref{e:ehj}). Since the solution behaves as $e^{-S}$, and the tunneling
amplitude is given by the ratio of the wavefunction evaluated at the initial
and final geometries, the tunneling amplitude is 
\begin{equation}
\exp\left(S[h_{ij}^{\mbox{\scriptsize{initial}}}]-S[h_{ij}^
{\mbox{\scriptsize{final}}}]\right)
\end{equation}
No difficulties arise in this approach; the calculation of tunneling
amplitude proceeds in a straightforward fashion.

\begin{figure}[htb]
\centerline{\psfig{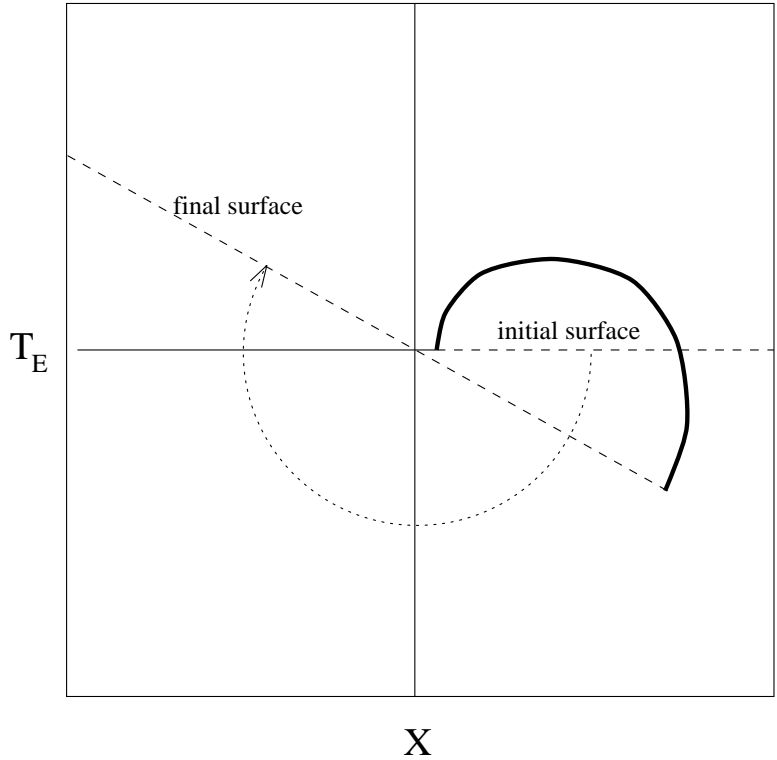}}
\caption{Bubble trajectory in Euclidean Schwarzschild space.}
\label{bubfig3}
\end{figure}

In Ref.~\cite{Guth90} the calculation is performed using the functional
 integral.
In this formalism one looks for a manifold which interpolates between the 
initial and final surfaces and which is a solution to the Euclidean Einstein
equations. The tunneling amplitude is $e^{-S}$, where S is the action of the
solution.  It is found, however, that solving the field equations leads to a
sequence of three geometries which do not form a manifold.  To see this, first
note that the geometry outside the bubble is Euclidean Schwarzschild space,
obtained by $t\rightarrow it_{E}$, $T\rightarrow iT_{E}$,
\begin{equation}
ds_{E}^{2}=\left(1-\frac{2M}{r}\right)\left(dt_{E}^{2}+dr_{*}^{2}\right)
+r^{2}d\Omega^{2}
=\frac{32M^{3}e^{-r/2M}}{r}\left(dT^{2}+dX^{2}\right)+r^{2}d\Omega^{2}
\end{equation}
with
\begin{equation}
\left(\frac{r}{2M}-1\right)e^{r/2m}=X^{2}+T_{E}^{2} \mbox{\ ;\ }
t_{E}=4M\tan^{-1}(T_{E}/{X}).
\end{equation}
It remains to describe the motion of the bubble wall.  Solving the equations
of motion leads to the trajectory in Fig. \ref{bubfig3}.

%Here is the figure!!
\renewcommand\floatpagefraction{.9}
\renewcommand\topfraction{.9}
\renewcommand\bottomfraction{.9}
\renewcommand\textfraction{.1}

 It is seen that the bubble wall
crosses the initial surface during the course of its motion, creating a 
situation in which it is impossible to identify a region which is swept out
by the evolving hypersurface.  Some regions of the manifold are crossed
twice by the hypersurface, some once, and some not at all.  The authors of
Ref.~\cite{Guth90} call
this object a pseudomanifold and give a prescription to calculate its action
by assigning covering numbers to the various regions, but this is not needed
for what follows.

With these results in hand, the technique of Sect.~(3) can be
used to calculate the state of the scalar field after tunneling. It was seen
that once the solution of the Einstein-Hamilton-Jacobi equation is given, the
field wave functional $\chi$ is fully determined by equations (\ref{e:sc}) and
(\ref{e:mo}).  Since $S[h_{ij}]$ is calculated in Ref.~\cite{polch},
 we have all
 that we need to find $\chi$.   This would, however, require finding the
 solution
to an unfamiliar functional differential equation. To cast it in in the form
of the Schr\"{o}dinger equation a lapse $N^{\tau_{E}}$, shift $N_{i}$
 and time $\tau_{E}$ were reintroduced leading to the appearance of the
Euclidean metric $g_{\mu \nu}^{E}$.  In the present case there is no true
interpolating Euclidean manifold, so that any choice of $N^{\tau_{E}}$ and
$N_{i}$ which define a well behaved $g_{\mu \nu}^{E}$ will lead to a bubble
trajectory that is a multivalued function of time.  Alternatively, a choice
of time functional which gives a single valued bubble trajectory will
necessarily lead to a Euclidean metric with vanishing determinant at some
point.  In either case, it is not clear that the resulting Schr\"{o}dinger
equation is well defined.  This is apparent from Fig. \ref{bubfig3}, 
where it can be seen
that boundary conditions imposed on the initial surface and on the bubble wall
may contradict each other.  These difficulties arise as a result of trying to
compute the final state of the field in one step, which requires a Euclidean
manifold interpolating all the way from the initial surface to the final
surface, and can be avoided by calculating the state on a series of 
intermediate hypersurfaces.  In this approach, it does not matter that the 
bubble wall eventually crosses the initial surface since once the state is
 calculated at some intermediate point we can forget about what preceded it.  

For simplicity, we will consider only the s-wave component of the scalar field
and frequencies high enough such that the geometrical optics approximation is
valid.  This means that the field equation is taken to be
\begin{equation}
g_{E}^{\mu \nu}\partial_{\mu} \partial_{\nu} \phi =0.
\end{equation}
The state of the field on the initial surface, $t=T=0$, is most conveniently 
expressed in terms of the coordinates $r_{*}$ and $t$.  We divide the modes
into ingoing and outgoing,
$$
\xi_{\omega}^{\mbox{\scriptsize{in}}}(r_{*},t)=C_{\omega}\,
e^{-i\omega(t+r_{*})}
$$
\begin{equation}
\xi_{\omega}^{\mbox{\scriptsize{out}}}(r_{*},t)=
C_{\omega}\,e^{-i\omega(t-r_{*})} 
\end{equation}
and write the field operator as
\begin{equation}
\hat{\phi}(r_{*},t)=\sum_{\omega}\left[\hat{a}_{\omega}^
{\mbox{\scriptsize{in}}}
\xi_{\omega}^{\mbox{\scriptsize{in}}}+\hat{a}_{\omega}^
{\mbox{\scriptsize{in}}\dagger}\xi_{\omega}^{\mbox{\scriptsize{in}*}}
(r_{*},t) + \mbox{ in} \rightarrow \mbox{out}\right].
\end{equation}
$C_{\omega}$ are normalization constants whose values will not be important. 
We shall only consider the {\em in} modes as the treatment of the {\em out}
 modes is exactly the same. We also suppress the {\em in} superscript.   

In the first stage of the evolution the hypersurface is pivoted around 
$r_{*}=r_{*}^{b}$ by $180^{\circ}$, where $r_{*}^{b}$ is the position
 of the bubble wall on the
initial surface. The solutions to the Euclidean field
equations are most conveniently obtained by choosing Cauchy boundary conditions
on the initial surface, (clearly a valid procedure in this case)
$$
f_{\omega}^{+}(r_{*},0)=\xi_{\omega}(r_{*},0) \mbox{ \ ; \ }
\frac{\partial}{\partial t_{E}}f_{\omega}^{+}(r_{*},0)
=-i\frac{\partial}{\partial t}\xi_{\omega}^{*}(r_{*},0)
$$
\begin{equation}
f_{\omega}^{-}(r_{*},0)=\xi_{\omega}^{*}(r_{*},0) \mbox{ \ ; \ }
\frac{\partial}{\partial t_{E}}f_{\omega}^{-}(r_{*},0)
=-i\frac{\partial}{\partial t}\xi_{\omega}^{*}(r_{*},0)
\end{equation}
It is also easiest to use the $X$, $T$ coordinates as they are well behaved 
everywhere.  Since the evolution of the hypersurface is simply a reflection 
about the point $X=X^{b}$, a mode which has the form $f(X,T_{E})$ on the
 initial
surface has the form $f(-X+2X^{b},T)$ on the new surface. Using the relations
\begin{equation}
r_{*}=4M\ln\sqrt{X^{2}+T_{E}^{2}\,} \mbox{ \ ; \ } t_{E}=4M\tan^{-1}(T_{E}/X)
\end{equation}
and that
\begin{equation}
f_{\omega}^{\pm}(r_{*},t_{E})=C_{\omega}\,e^{\pm \omega t_{E} -i\omega r_{*}}
\end{equation}
near the initial surface, one sees that near the new surface,
\begin{equation}
f_{\omega}^{\pm}(X,T_{E})=C_{\omega}\exp\left(\frac{\mp 4M\omega 
T_{E}}{-X+2X^{b}}
-4iM\omega\ln(-X+2X^{b})\right).
\end{equation}
Since on the new surface,
 $f_{\omega}^{+}=(f_{\omega}^{-})^{*}$ and $\partial f_
{\omega}^{+}/\partial t_{E}=-(\partial f_{\omega}^{-}/\partial t_{E})^{*}$,
 the
evolution operator $\hat{U}_{E}$ is unitary.  This means that the state on the
new surface has the same form as it did on the initial surface, but is now
expressed in terms of the modes
\begin{equation}
\xi_{\omega}(X,T)=C_{\omega}\exp\left(\frac{-4iM\omega T}{-X+2X^{b}}
+4iM\omega \ln(-X+2X^{b})\right).
\end{equation}
These modes can be approximated near $T=0$ as
\begin{equation}
\xi_{\omega}= \left\{\begin{array}{ll}
C_{\omega}\, e^{i\omega (t-r_{*})} & \mbox{ if $ \;  |X|\gg X^{b}$} \\
C_{\omega}\,e^{-(2iM\omega/X^{b})(T-X)} & \mbox{ if $ \;  |X|\ll X^{b}$}
\end{array} \right.
\end{equation}
Now it is useful to express the state in terms of modes which are nonzero only
inside or outside the horizon,
$$
\eta_{\omega}^{<}=\left\{\begin{array}{ll}
D_{\omega}\,e^{i\omega(t-r_{*})} & \mbox{ if $\; X<0$}\\
0 & \mbox{ if $ \; X>0$}
\end{array} \right.
$$
\begin{equation}
\eta_{\omega}^{>}=\left\{\begin{array}{ll}
0 & \mbox{ if $\; X<0$} \\
D_{\omega}\,e^{-i\omega(t+r_{*})} & \mbox{ if $\; X>0$.}
\end{array} \right.
\end{equation}
A fundamental result \cite{hawking,unruh} in the derivation of black hole 
radiance is that the vacuum state with respect to modes which have a time
dependence $e^{-i\omega T}$ is the state
\begin{equation}
\mbox{const.}\times \sum_{\{n_{\omega}\}}e^{-E(\{n_{\omega}\})/2T_{H}}
 \left| \{n_{\omega}\}\right\rangle_{<}\left| \{n_{\omega}\}\right\rangle_{>}
\label{e:dm}
\end{equation}
with respect to the modes $\eta_{\omega}^{<}$ and $\eta_{\omega}^{>}$.  The 
sum runs over all sets of occupation numbers, $E=\sum n_{\omega}\omega$, and
$T_{H}=1/8\pi M$ is the Hawking temperature.  Further, near the horizon, any
deviation of $\left|\chi \right \rangle$ from the 
 vacuum state can be ignored because
of the arbitrarily large redshift as $r_{*}\rightarrow -\infty$.  Far from the
horizon $\xi_{\omega}$ and $\eta_{\omega}^{<}$ agree so the form of the state
is unchanged there.  

Now the hypersurface can be evolved the remainder of the way.  If we restrict
our attention to the region $X<X^{b}$, then the motion of the hypersurface
 is simply a translation, $t_{E}\rightarrow t_{E}-\Delta t_{E}$. 
 This causes states with time dependence $e^{i\omega t}$ to be damped by a
 factor $e^{-\omega \Delta t_{E}}$, and states with time dependence
$e^{-i\omega t}$ to be amplified by a factor $e^{\omega \Delta t_{E}}$.
  Near the
horizon, the state $\left|\chi \right\rangle$ consists of pairs of positive and
negative frequency states according to (\ref{e:dm}).  One member of the pair
is damped but the other is amplified by a compensating amount so as to leave
the state $\left|\chi \right\rangle$ unchanged. 
The final state of the field can then be summarized as follows.
  Far from the hole, where there is no pairing, the initial state is damped:
\begin{equation}
\sum_{\{n_{\omega}\}} S(\{n_{\omega}\})\left|\{n_{\omega}\}\right\rangle
\longrightarrow \mbox{const.}\times \sum_{\{n_{\omega}\}}e^{-E(\{n_{\omega}\})
\Delta t_{E}}\,S(\{n_{\omega}\})\left|\{n_{\omega}\}\right\rangle.
\end{equation}
Near the horizon the final state is given by (\ref{e:dm}). This is true for 
both the in and out modes, so an observer stationed on either side of the 
horizon would observe a thermal distribution of both ingoing and outgoing
particles.  As time passes, all of the ingoing particles will eventually
cross the horizon and be swallowed by the hole, whereas the outgoing particles
will propagate out to infinity where they can be detected at arbitarily late
times as a flux of thermal radiation at the Hawking temperature.  

\section{Comments}

It was shown that the standard picture of black hole radiance is unchanged by
tunneling.  At late times, the hole radiates just as it would have had it been
formed from a classical collapse.  This makes sense if one thinks of Hawking
radiation as pair production.  The probability of tunneling is not affected
by the creation of a pair, since the pair has zero total energy. From this
point of view it is also clear that what happens at early times cannot
possibly affect the late time radiation, since the produced pairs only see the 
late time geometry. The conventional derivation of radiance obscures this 
point somewhat and it seems desirable to find an approach which makes 
this feature manifest from the outset.   
For the two systems considered here, and presumably this is true in general,
 the effect of the tunneling was to shift the distribution of any particles
 that were
 present before tunneling.  In the present case initial excitations were damped
because the final surface is rotated clockwise relative to the initial
surface.  A counterclockwise rotation would have led to amplification.
In \cite{Guth90} numerical investigations are quoted which show that the
rotation is always clockwise for the false vacuum bubble. One is led to
speculate whether this is a general phenomenon --- whether all tunneling
transitions lead to damping.  

%\appendix
%\input appendix

\end{document}